\documentclass[fleqn,usenatbib]{mnras}

\usepackage{newtxtext,newtxmath}

\usepackage[T1]{fontenc}

\DeclareRobustCommand{\VAN}[3]{#2}
\let\VANthebibliography\thebibliography
\def\thebibliography{\DeclareRobustCommand{\VAN}[3]{##3}\VANthebibliography}

\usepackage{graphicx}	
\usepackage{color, xcolor}
\usepackage{subfigure}

\title[Multi-zone model of blazars]{A multi-zone view on the multi-wavelength emission of blazars}

\author[R.-Y. Liu et al.]{
Ruo-Yu Liu$^{1,5}$,
Rui Xue$^{2}$,
Ze-Rui Wang$^{3}$,
Hong-Bin Tan$^{1,5}$,
Markus B\"ottcher$^4$
\\
$^{1}$School of Astronomy and Space Science, Nanjing University, Nanjing 210023, China; {\color{blue} \rm ryliu@nju.edu.cn}\\
$^{2}$Department of Physics, Zhejiang Normal University, Jinhua 321004, China; {\color{blue} \rm ruixue@zjnu.edu.cn}\\
$^{3}$College of Physics and Electronic Engineering, Qilu Normal University, Jinan 250200, People's Republic of China; {\color{blue}\rm zeruiwang@qlnu.edu.cn}\\
$^{4}$Centre for Space Research, North-West University, Potchefstroom, South Africa; {\color{blue} \rm Markus.Bottcher@nwu.ac.za}\\
$^{5}$Key laboratory of Modern Astronomy and Astrophysics (Nanjing University), Ministry of Education, Nanjing 210023, People's Republic of China\\}

\date{Accepted XXX. Received YYY; in original form ZZZ}

\pubyear{2022}

\begin{document}
\label{firstpage}
\pagerange{\pageref{firstpage}--\pageref{lastpage}}

\maketitle

\begin{abstract}
In this work, a time-dependent modeling is developed to study the emission properties of blazars in the {low} state. Motivated by various observations, we speculate and assume that numerous discrete radiation zones throughout the jet of a blazar contribute to the broadband emission. We model the temporal evolution of the electron spectrum in each emission zone taking into account the injection, cooling and escape of relativistic electrons. By doing so, we are able to calculate the multi-wavelength emission of each radiation zone. The observed emission of a blazar is then the superposition of the emission from all discrete radiation zones. We revisit the multi-wavelength spectral energy distributions, light curves and polarisation under the model, and discuss its potential to reproduce the flat radio spectra, the core-shift phenomena, the minute-scale gamma-ray variability, and the large polarisation-angle swings, which are difficult to explain under the conventional one-zone models simultaneously.
\end{abstract}



\section{Introduction}
As a class of active galactic nuclei \citep[AGNs;][]{1978PhyS...17..265B, 1995PASP..107..803U}, blazars are one of the most luminous and highly variable objects in the universe. A blazar's radiation is dominated by the non-thermal emission of its relativistic jet, which is Doppler boosted.  The spectrum spans from the radio to the $\gamma$-ray band and shows a double-hump structure. While the low-energy hump is generally believed to originate from the synchrotron radiation of primary relativistic electrons, the origin of high-energy hump is under debate. In leptonic models, the high-energy hump is attributed to the inverse-Compton (IC) scattering of the synchrotron photons emitted by primary electrons in the jet \citep[synchrotron-self Compton, SSC; e.g.,][]{1967MNRAS.137..429R, 1974ApJ...188..353J, 1985ApJ...298..114M} and/or the thermal photons from surrounding structures (external Compton, EC) such as broad-line region (BLR) and/or the dusty torus \citep[DT;][]{1993ApJ...416..458D, 2000ApJ...545..107B}. In hadronic models, proton synchrotron emission \citep[e.g.,][]{2000NewA....5..377A} or the emission from secondary particles generated in the interactions between ultra-relativistic protons and soft photons \citep[e.g.,][]{2017MNRAS.464.2213P} can explain the high-energy hump as well.

When studying the multi-wavelength spectral energy distribution (SED) and variability of blazars, the most commonly used model is the one-zone leptonic model \citep[e.g., ][]{1997A&A...320...19M, 2000ApJ...536..729L, 2002ApJ...581..127B, 2004ApJ...609..576B, 2004ApJ...617..113W, 2008ApJ...686..181F, 2009ApJ...692...32D, 2012ApJ...752..157Z, 2013ApJ...768...54B, 2014Natur.515..376G, 2014MNRAS.439.2933Y, 2017ApJ...851...33P, 2020ApJS..248...27T, 2021arXiv211203941T}. It is assumed that all the jet's non-thermal emission comes from a compact spherical region, which is also referred to as a ``blob''. Although one-zone leptonic models have had great success in fitting the multi-wavelength SEDs and explaining the variability at different wavebands, there are growing observational hints indicating the existence of more than one dissipation regions in jets of blazars. In the framework of the one-zone leptonic model, since the multi-wavelength emission is emitted by the same population of relativistic electrons, it is predicted that there are close correlations between light curves (LCs) at different energy bands. Such correlations have been indeed observed \citep{1985ApJ...298..114M, 2003heba.conf..173M, Marscher08, Marscher10, 2015ApJ...807...79H, 2017Ap&SS.362..189W, 2020PASJ...72...44Z}. However, in several blazars, variability patterns have been observed which are not correlated between different wavelength bands. \cite{2019ApJ...880...32L} reported that about 54.5\% of optical flares and 20\% of $\gamma$-ray flares are orphan events in their sample. At the same time, TeV orphan flares are sometimes observed \citep[e.g.,][]{2019A&A...623A.175M}. Such orphan flares strongly argue against the one-zone leptonic model \citep[for a review, see][]{2019Galax...7...20B}. Anti-correlated light curves between different bands observed from some blazars during some periods are also at odds with one-zone models \citep{2014A&A...571A..83P,2020ApJ...890...97A, 2021A&A...655A..89M}.

Short time-scale variability such as minute-scale variability has been observed in GeV and TeV observations \citep{2007ApJ...664L..71A, 2007ApJ...669..862A, 2016ApJ...824L..20A, 2018ApJ...854L..26S}. Such minute-scale variability implies that the dissipation region is compact, which means that if we believe the observed photons are all coming from only one small blob, the high-energy $\gamma$ rays are likely to be absorbed due to the internal $\gamma\gamma$ annihilation. However, few blazars show a feature of substantial attenuation in their $\gamma$-ray spectrum \citep{2018MNRAS.477.4749C}. Therefore, in order to reduce the number density of soft photons so as to make the dissipation region transparent to the high-energy $\gamma$-ray photons, the Doppler factor of the jet need be larger than 50 in one-zone models \citep[e.g.,][]{2008MNRAS.384L..19B}. However, such a high Doppler factor is rarely measured in observations \citep{2009A&A...494..527H}. This difficulty is also known as the Doppler-factor crisis \citep{2010ApJ...722..197L}.

In addition to the difficulties faced by the one-zone models described above, there is also direct observational evidence against the scenario of a single compact radiation zone dominating the jet's emission. For example, substructures like bright knots have been observed along the jet of several blazars \citep[e.g., OJ 287, ][]{2011ApJ...729...26M}. Besides, \cite{2016A&A...595A..54M} applied the wavelet-based image segmentation and evaluation analysis to the jet image of M~87 obtained from 43~GHz Very Long Baseline Array (VLBA) observations. Using the segmented wavelet decomposition (SWD) and intermediate wavelet decomposition (IWD) methods, the VLBA image of M~87 jet can be decomposed into numerous local regions of different proper motions, implying the existence of discrete substructures inside the jet. On the other hand, \cite{2019ApJ...871..143P} found significant offsets between the optical core and the radio core pinpointed with $Gaia$ and very long baseline interferometry (VLBI) in more than 1000 AGNs. Among them, the {\it Gaia} optical centroids of 85\% of BL Lacs are located farther to the nucleus than the VLBI radio centroid position (downstream offset). This result implies that extended bright optical jets are quite common in BL Lacs, and suggests that particle acceleration and energy dissipation do not occur only in one compact region, but also occur throughout the jet.

Furthermore, the flat radio spectrum, as one of the most stable features of blazars \citep[e.g.,][]{2011A&A...536A..15P}, cannot be reproduced in one-zone models. This is because the synchrotron self-absorption effect is inevitable in a compact radiation zone which is solely responsible for the intense emission of a blazar. One-zone models therefore predict a very hard spectrum in the radio band which is inconsistent with measurements \citep[e.g.,][]{2009A&A...501..879T, 2011ApJ...736..131A, 2017MNRAS.464..599D, 2021MNRAS.506.5764D, 2022A&A...657A..20B}. As a result, the radio spectral data is usually ignored in the SED fitting with one-zone models \citep[e.g.,][]{2009MNRAS.399.2041G, 2010MNRAS.402..497G,2013ApJ...771L...4C, 2021JHEAp..29...31P}. On the other hand, an extended, continuous jet model with a uniform conical structure was proposed by \citet[][hereafter, the BK jet model]{1979ApJ...232...34B}, which is demonstrated to be capable of accounting for the flat radio spectrum. Assuming that the magnetic energy is conserved at different distances from the central black hole, the flat radio spectrum can be reproduced by the superposition of synchrotron self-absorbed radiation from different parts along the jet. This kind of conical continuous jet model successfully predicts the radio core-shift phenomenon, which has been confirmed by observations \citep{2011Natur.477..185H, 2011A&A...532A..38S}.

While the BK continuous jet model can successfully explain the radio data, it does not explain some observational features of blazars such as the fast temporal variability, the discrete structure and so on. Indeed, some magnetohydrodynamic (MHD) simulations have suggested an episodic blazar jet, and the corresponding dynamical model has been studied \citep[e.g.,][]{2009MNRAS.395.2183Y, 2012ApJ...757...56Y, 2015RAA....15..207M, 2017MNRAS.468.2552L, 2020MNRAS.495.1549N, 2021ApJ...908..193M}. However, the radiation feature of such a jet model with multiple discrete emitting blobs is yet to be studied. Recently, \cite{2022PhRvD1053005W} proposed the so-called ``stochastic dissipation model'' to explain flares of different types (i.e., multiwavelength flares, orphan gamma-ray flares, orphan optical flares) in the same framework. The model suggests that there are numerous discrete radiation zones along the jet owing to certain dissipation events such as magnetic reconnection events. Most of the dissipation zones are comparatively weak and the superposition of their emission constitutes a {low-state} background emission or the low state of the blazar's emission, while flares would arise from intense dissipation events that could occur somewhere in the jet occasionally. Although blazar flares of different types can be consistently explained with this model, \cite{2022PhRvD1053005W} did not model the jet's emission in the {low} state and simply use a polynomial to represent its contribution.

The {low-state} emission of a blazar is actually important, as it may reflect the general properties of blazar jets. In fact, according to the physical picture of the stochastic dissipation model, there is no essential difference between the flaring state and the {low} state of the blazar, because the blazar's emission arises from the superposition of multiple discrete dissipation zones in either state. If one or a few dissipation zones in the jet become very active, they could lead to a sudden enhancement of the blazar's flux and produce a flare. 
Thus, studying the {low-state} emission formed by the superposition of numerous comparatively weak dissipation regions is also helpful to understand the flaring state. This motivates us to model the multi-wavelength SEDs and LCs of blazars in the {low} state based on the stochastic dissipation model, and try to interpret various aforementioned properties of blazar jets (such as the flat radio spectrum, and the minute-scale variability in the GeV-TeV band) in the same framework. The rest of this paper is structured as follows. In Section~\ref{model} we present a general description of this multi-zone model. We apply the model to revisit the multi-wavelength SED and LCs in Section~\ref{app}. In Section~\ref{DC}, we present our discussion and conclusions. Throughout the paper, we adopt the $\Lambda$CDM cosmology with $H_{\rm{0}}={69.6\, \rm{km~s^{-1} Mpc^{-1}}}$, $\Omega_{\rm{m}}=0.29$, $\Omega_{\rm{\Lambda}}=0.71$ \citep{2014ApJ...794..135B}.

\begin{figure}
\includegraphics[width=0.5\textwidth]{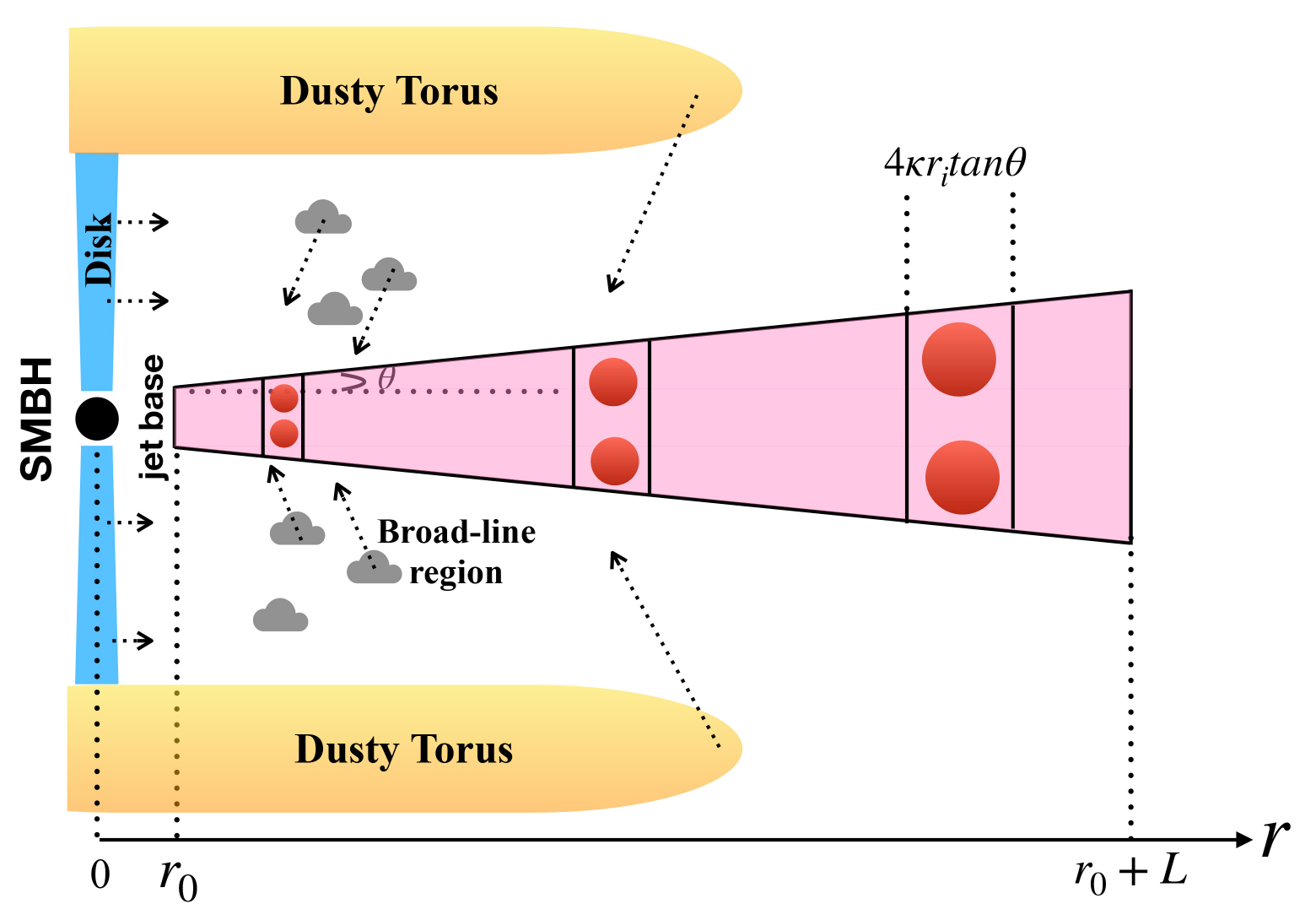}
\caption{A schematic illustration (not to scale) of the stochastic dissipation model. The conical jet is divided into numerous successional segments (pink trapezoids). Each segment contains several emitting blobs (crimson circles), which are supposed to be generated by random dissipation events along the jet. Electrons are accelerated in each blob and give rise to multiwavelength emission.
\label{fig:sketch}}
\end{figure}

\section{Model description}\label{model}
According to the VLBA observations of the M~87 jet \citep{2016A&A...595A..54M} and MHD simulations \citep[e.g.,][]{2009MNRAS.395.2183Y}, jets of AGNs could consist of many small blobs on different scales. This picture is supported by recent particle-in-cell simulations \citep{2016MNRAS.462.3325P, 2019MNRAS.482...65C}, which showed that quasi-spherical plasmoids with different sizes and electron numbers can be generated in the turbulence-triggered relativistic magnetic reconnection processes in the jet flow. MHD simulations also suggest that magnetic reconnection could energize particles in AGN jets locally \citep[e.g.,][]{2016MNRAS.462...48S, 2021ApJ...908..193M}. Based on these previous studies, the geometry of each blob in the jet is approximated as spherical in our work. Accelerated electrons are injected into each blob and emit photons through synchrotron and IC radiation when moving along the jet. We further assume that the jet is of a conical geometry \citep[e.g.,][]{2007ApJ...668L..27K} with a constant half-opening angle $\theta$, length $r_{\rm max}$ and distance between supermassive black hole (SMBH) and jet base $r_{0}$. We denote the jet's bulk Lorentz factor  by $\Gamma=(1-\beta^2)^{-1/2}$, where $\beta c$ is the jet speed at a viewing angle $\theta_{\rm obs}$ with respect to the observer's line of sight. Because of the beaming effect, the observed radiation is strongly boosted by a relativistic Doppler factor, $\delta_{\rm D}=[\Gamma(1-\beta \rm cos\theta_{\rm obs})]^{-1}$. In this work, assuming that $\theta_{\rm obs} \lesssim 1/\Gamma$ for blazars, we have $\delta \approx \Gamma$. A sketch of our model is shown in Fig.~\ref{fig:sketch}. We define $r$ as the distance from the SMBH along the jet axis, so that $r=0$ is defined as the position of the SMBH, $r=r_0$ as the position of jet base and $r=r_0+r_{\rm max}\simeq r_{\rm max}$ as the end of jet. To compute the jet's emission at different radii, we divide the jet into a number of $i_{\rm max}$ consecutive segments (the pink trapezoids in Fig.~\ref{fig:sketch}) with equal length in the logarithmic space, where the distance of the $i$th segment's far end from the SMBH is denoted by $r_i$. The radius of the blob generated in the $i$th segment in the comoving frame\footnote{Hereafter, quantities in the comoving frame are primed, and quantities in the rest frame are unprimed, unless specified otherwise.} is given by
\begin{equation}\label{eq:radius}
R'(r_i)=\kappa r_i\rm \tan\theta,
\end{equation}
where $\kappa$ is the ratio of blob's radius to its segment's radius. We assume the length of each segment to be twice the diameter of the blobs inside each segment itself, therefore we have the following relation
\begin{equation}
r_{i+1}=r_{i}(1+4\kappa \tan\theta),
\end{equation}
and
\begin{equation}
i_{\rm max}=\ln(r_{\rm max}/r_0)/\ln(1+4\kappa \tan\theta).
\end{equation}
Note that such a division is rather arbitrary and is simply for convenience of calculation.

\begin{figure*}
\centering
\subfigure{
\includegraphics[width=0.68\columnwidth]{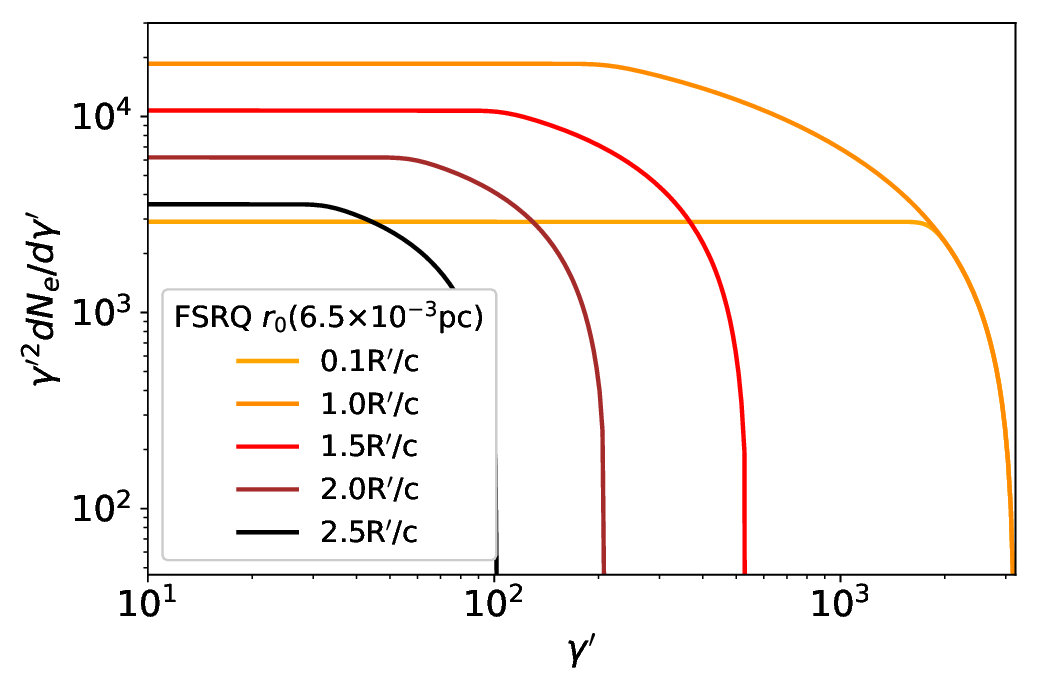}
}\hspace{-5mm}
\quad
\subfigure{
\includegraphics[width=0.68\columnwidth]{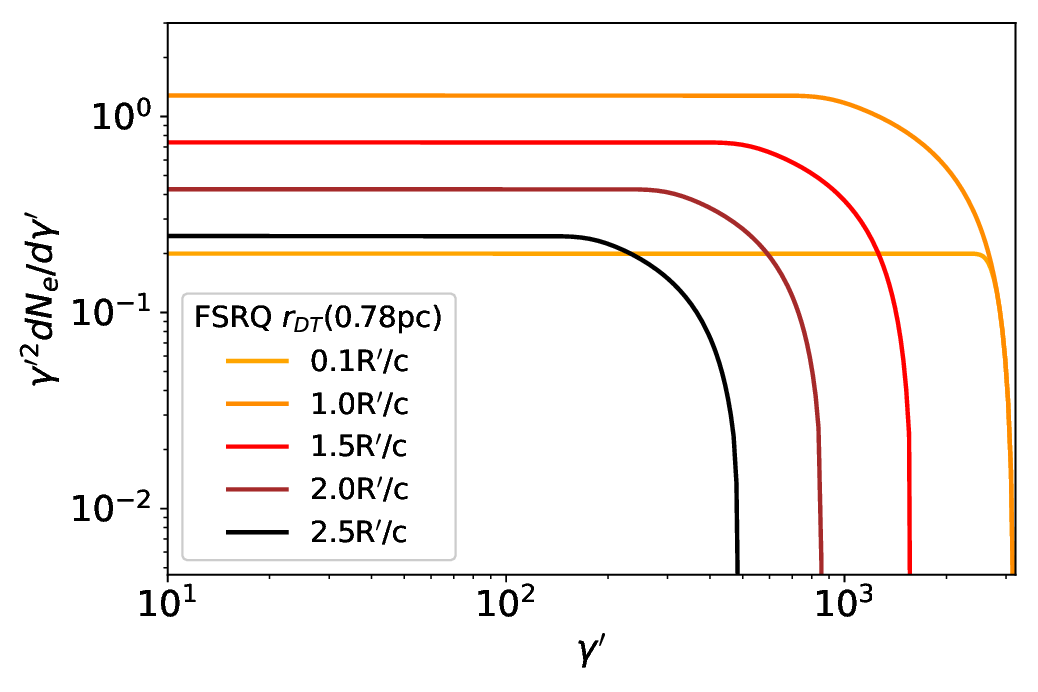}
}\hspace{-5mm}
\quad
\subfigure{
\includegraphics[width=0.68\columnwidth]{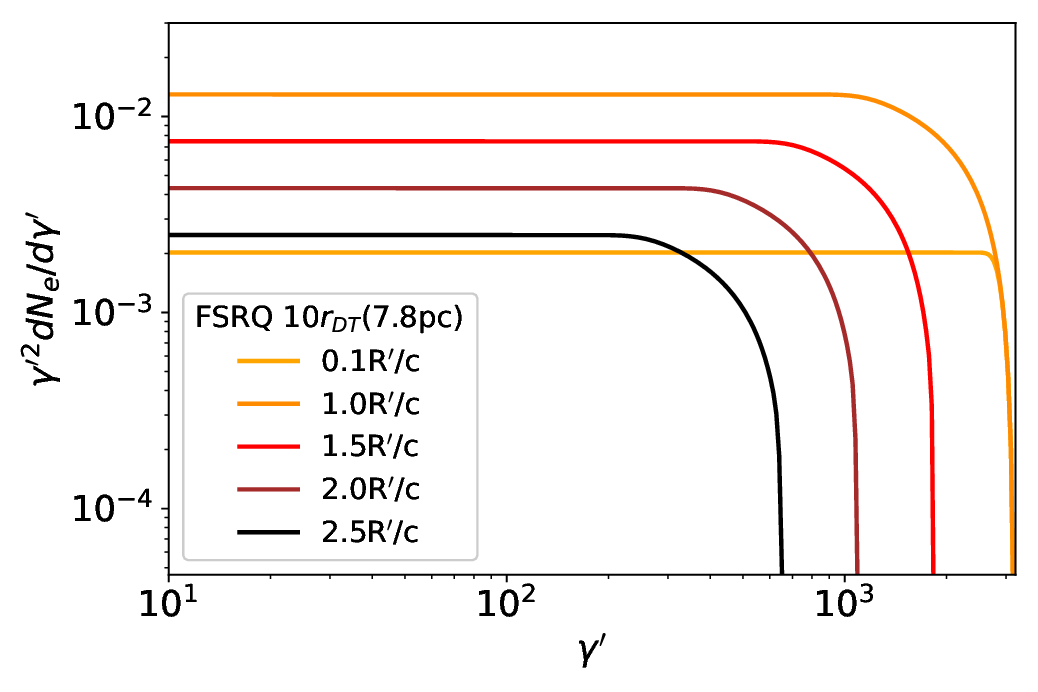}
}\hspace{-5mm}
\quad
\subfigure{
\includegraphics[width=0.68\columnwidth]{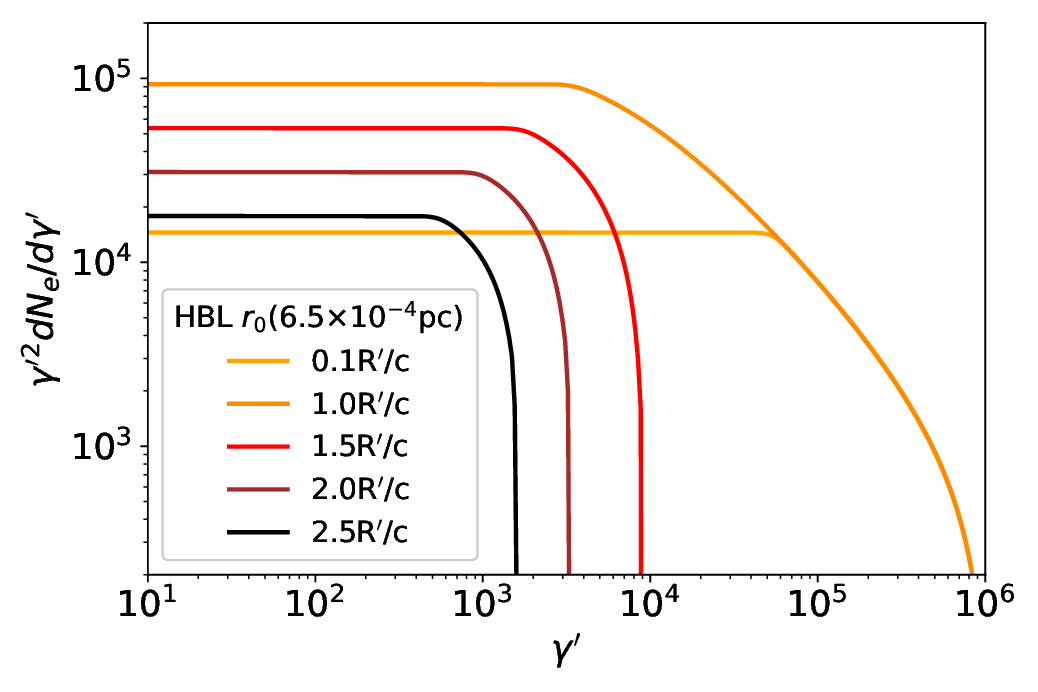}
}\hspace{-5mm}
\quad
\subfigure{
\includegraphics[width=0.68\columnwidth]{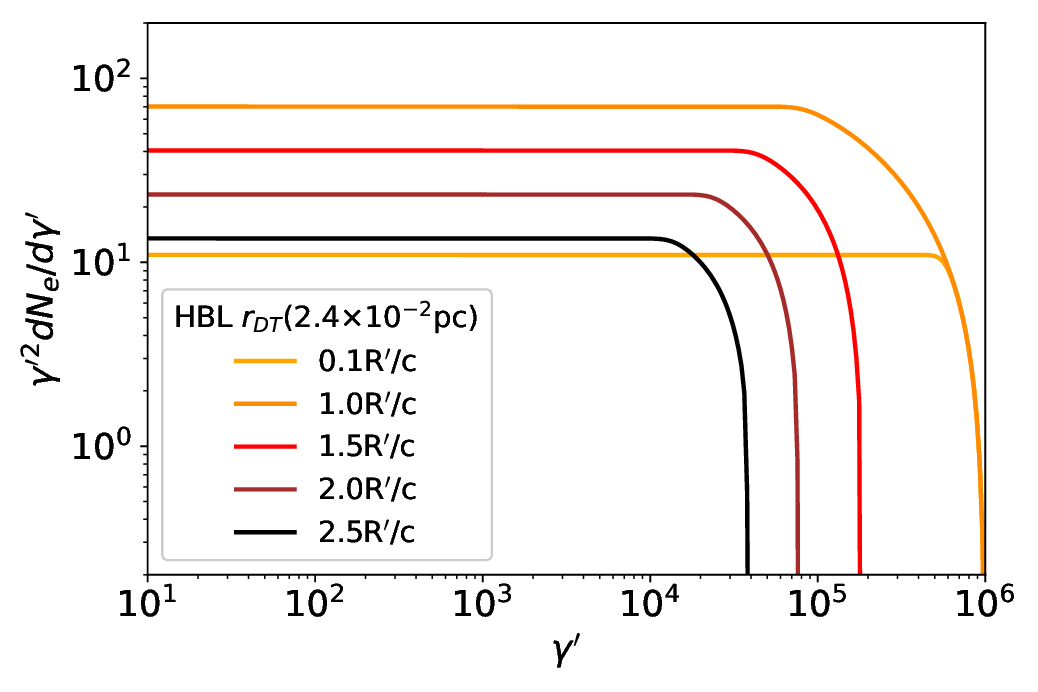}
}\hspace{-5mm}
\quad
\subfigure{
\includegraphics[width=0.68\columnwidth]{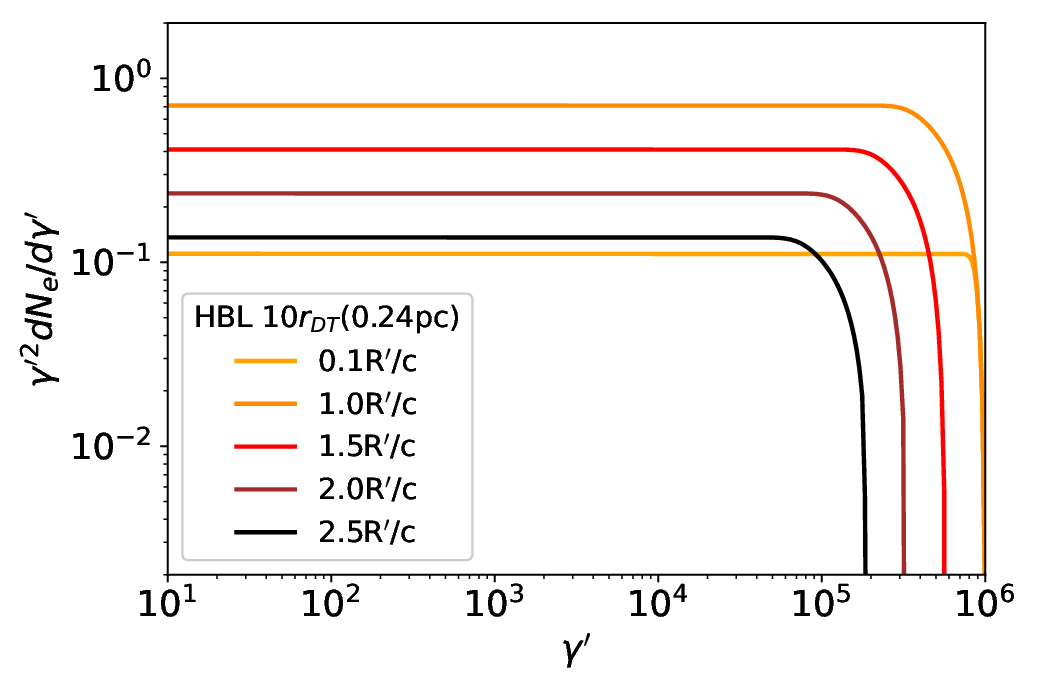}
}
\caption{The temporal evolution of electron distribution in a blob at the jet base, at the typical radius of the DT $r_{\rm DT}$, and beyond the DT for the benchmark FSRQ (three top panels) and for the benchmark HBL (three bottom panels), respectively. Parameters are the same as those shown in Table~\ref{parameters}. Curves of different colors represent the electron spectrum at different time after the blob formation as labeled in the legends.
\label{EED}}
\end{figure*}

\begin{figure*}
\centering
\subfigure{
\includegraphics[width=0.68\columnwidth]{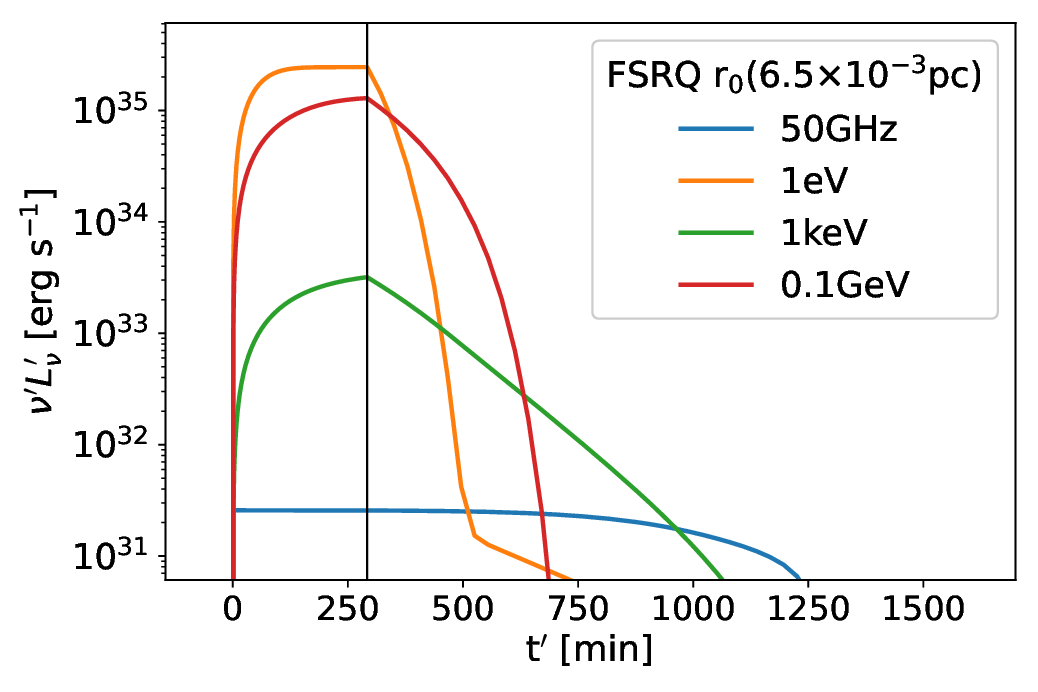}
}\hspace{-5mm}
\quad
\subfigure{
\includegraphics[width=0.68\columnwidth]{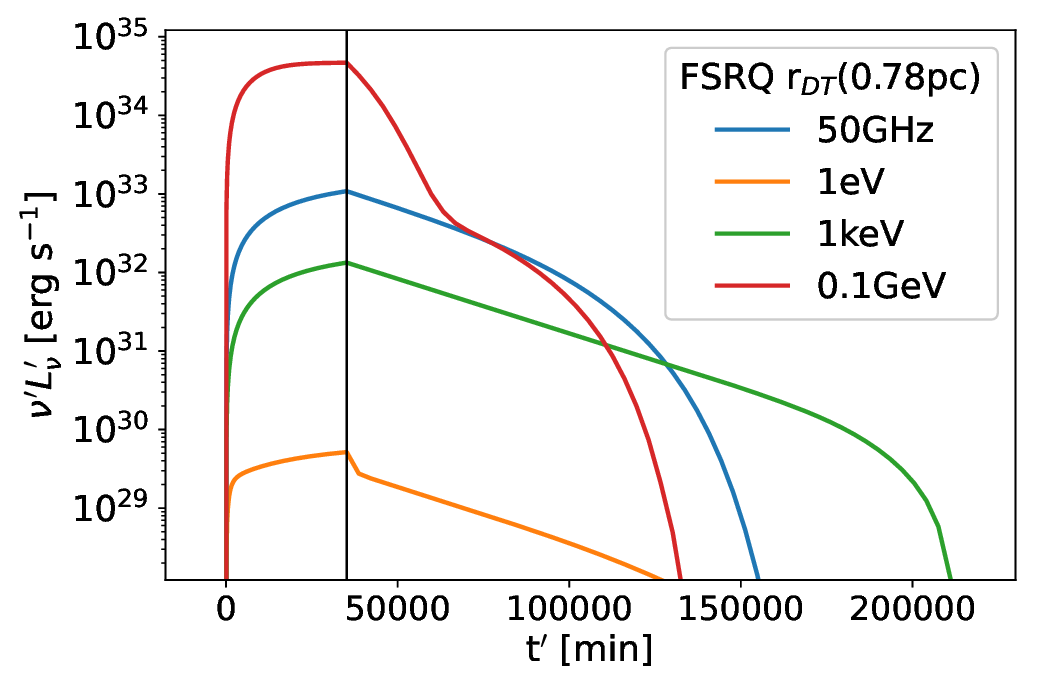}
}\hspace{-5mm}
\quad
\subfigure{
\includegraphics[width=0.68\columnwidth]{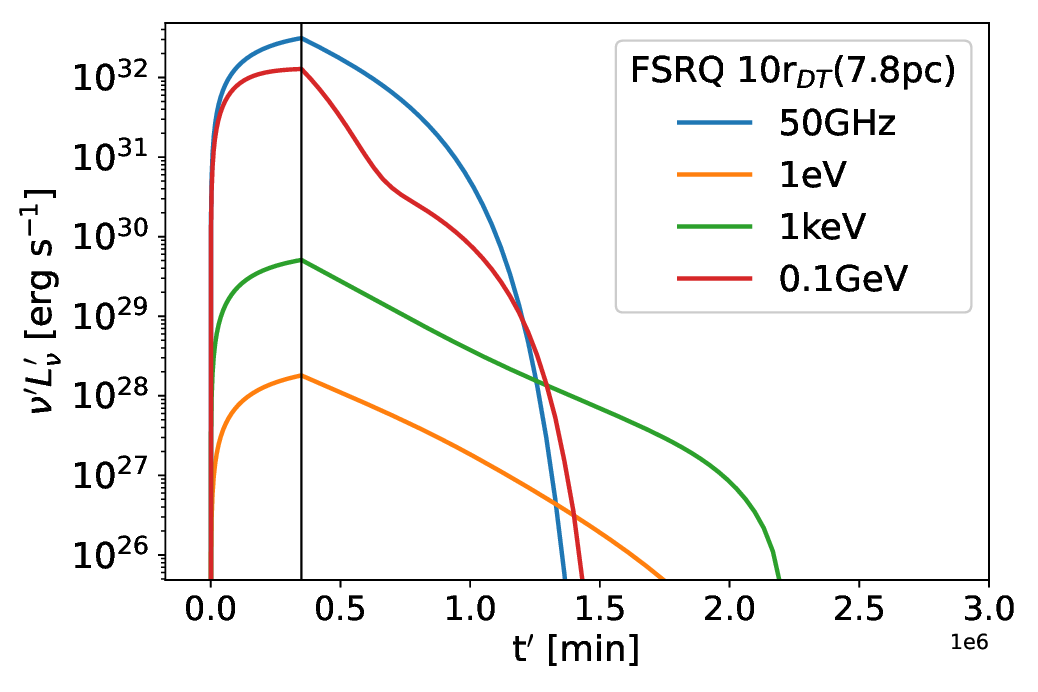}
}\hspace{-5mm}
\quad
\subfigure{
\includegraphics[width=0.68\columnwidth]{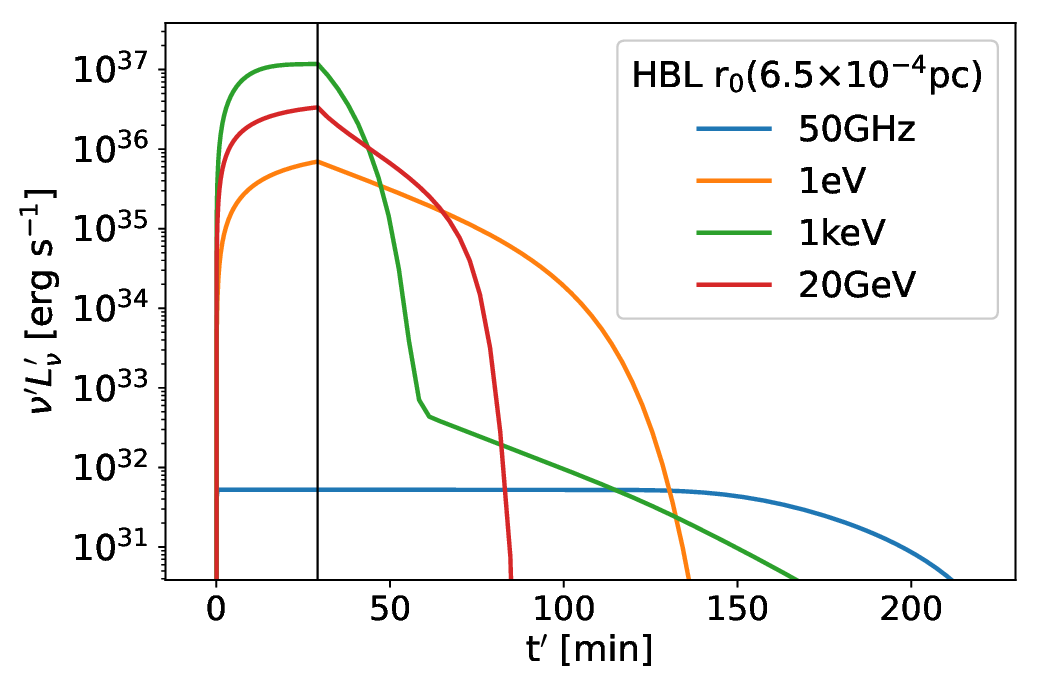}
}\hspace{-5mm}
\quad
\subfigure{
\includegraphics[width=0.68\columnwidth]{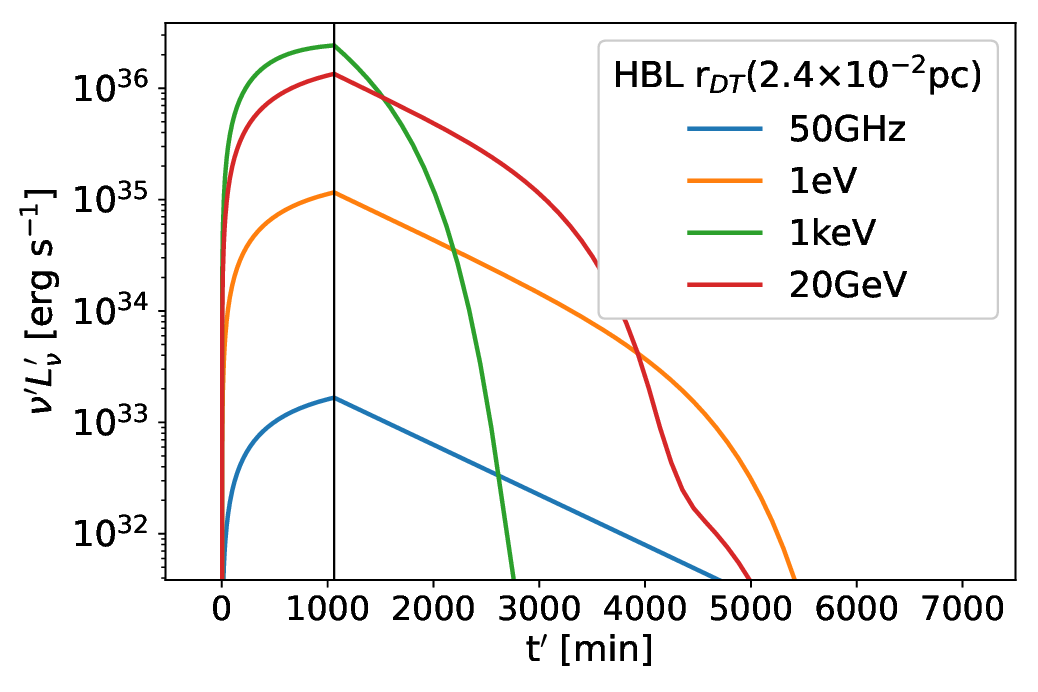}
}\hspace{-5mm}
\quad
\subfigure{
\includegraphics[width=0.68\columnwidth]{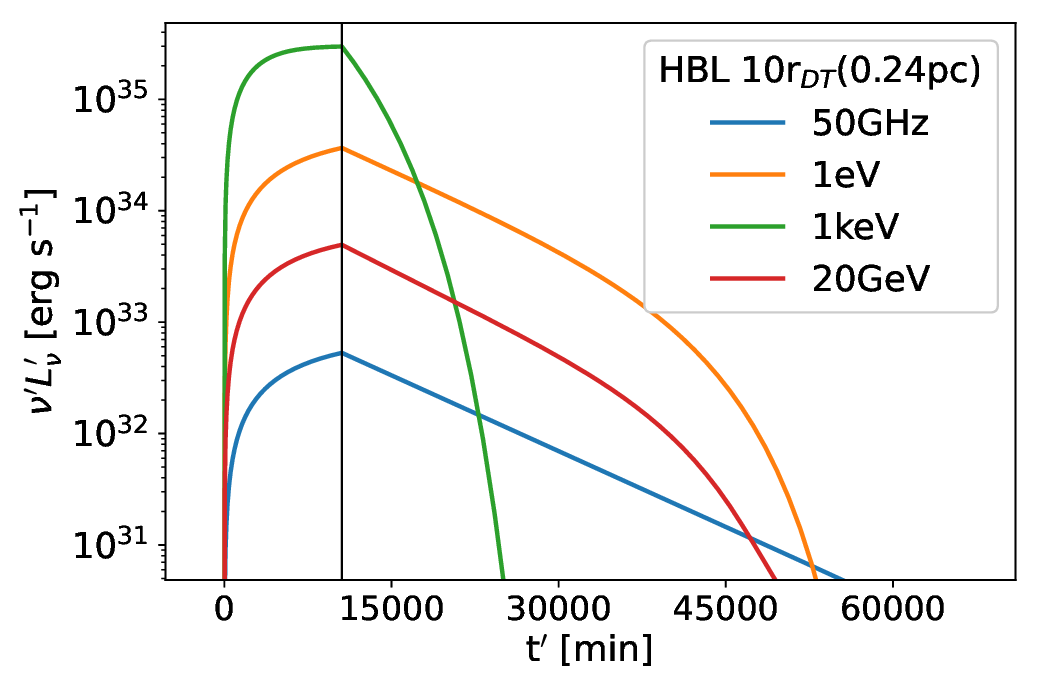}
}
\caption{Temporal evolution of fluxes (i.e., LCs) in radio, optical, X-ray and $\gamma$-ray bands emitted by a blob at the jet base, at the typical radius of the DT $r_{\rm DT}$, and beyond the DT for the FSRQ (three top panels) and the HBL (three bottom panels), respectively, in the comoving frame of the blob. The black vertical solid line represents the moment when the injection is stopped ($R'/\rm c$). Parameters are the same as those shown in Table~\ref{parameters}. 
\label{LC1}}
\end{figure*}

\begin{figure*}
\centering
\subfigure{
\includegraphics[width=0.68\columnwidth]{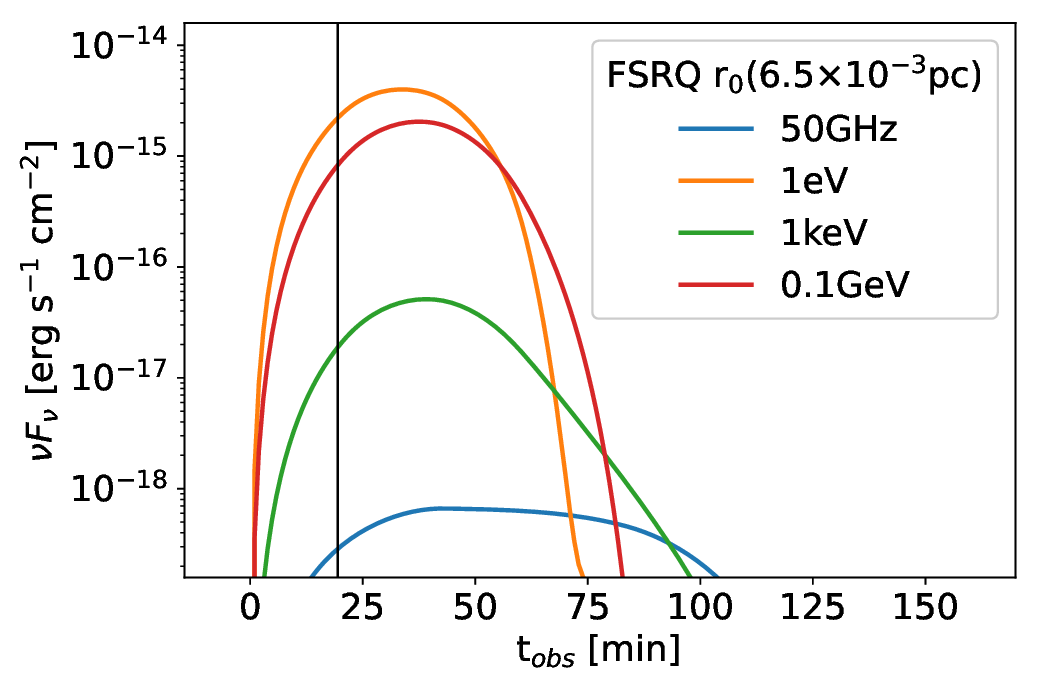}
}\hspace{-5mm}
\quad
\subfigure{
\includegraphics[width=0.68\columnwidth]{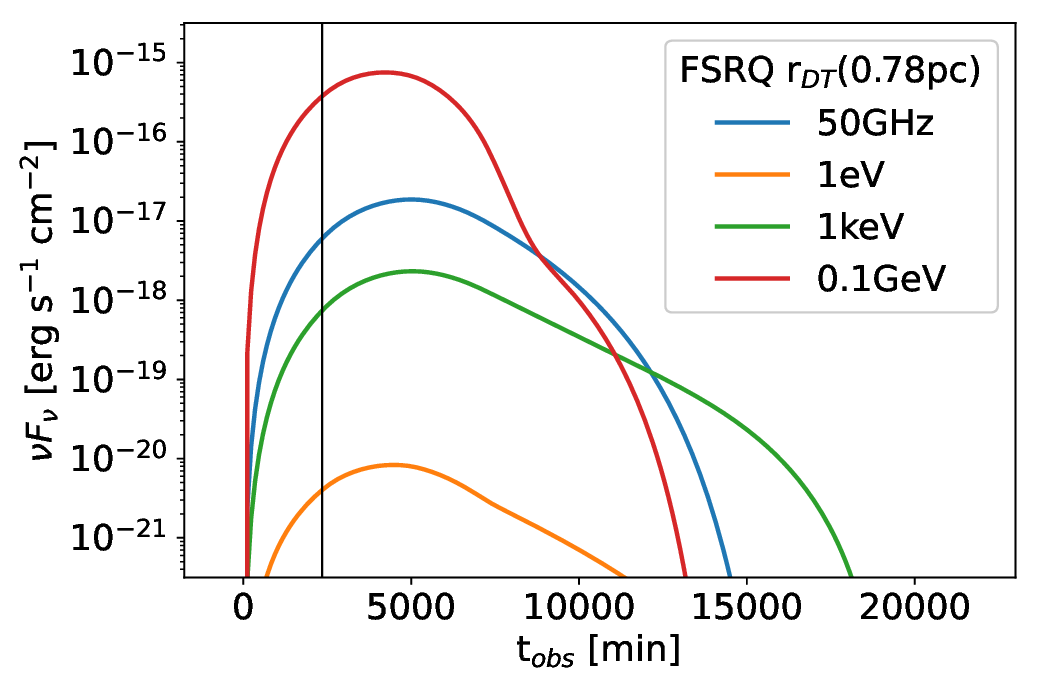}
}\hspace{-5mm}
\quad
\subfigure{
\includegraphics[width=0.68\columnwidth]{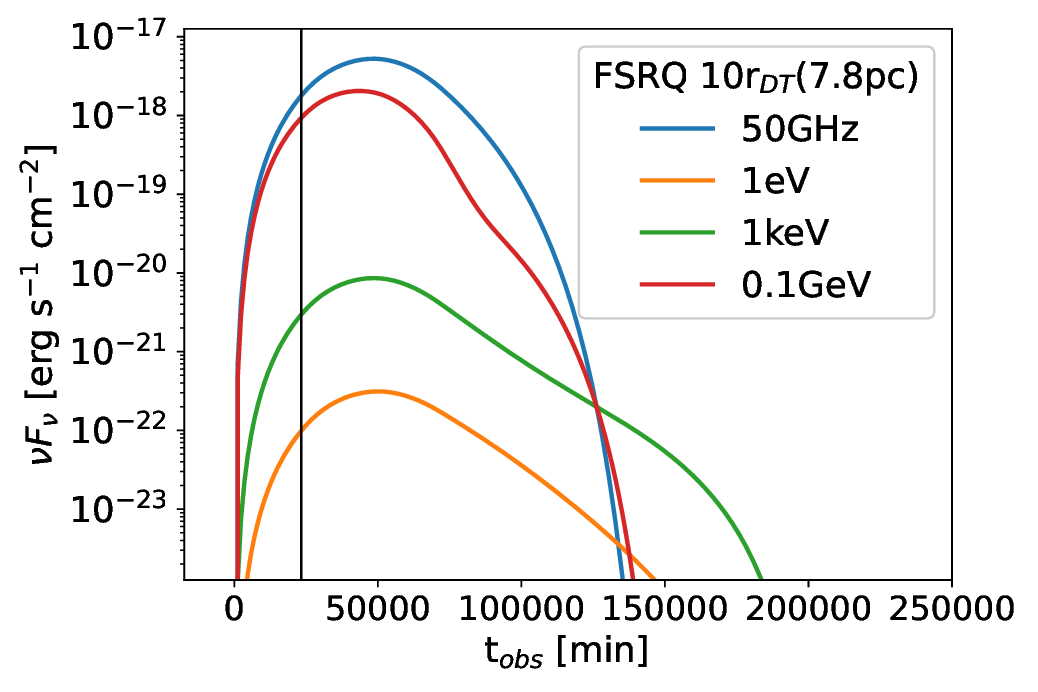}
}\hspace{-5mm}
\quad
\subfigure{
\includegraphics[width=0.68\columnwidth]{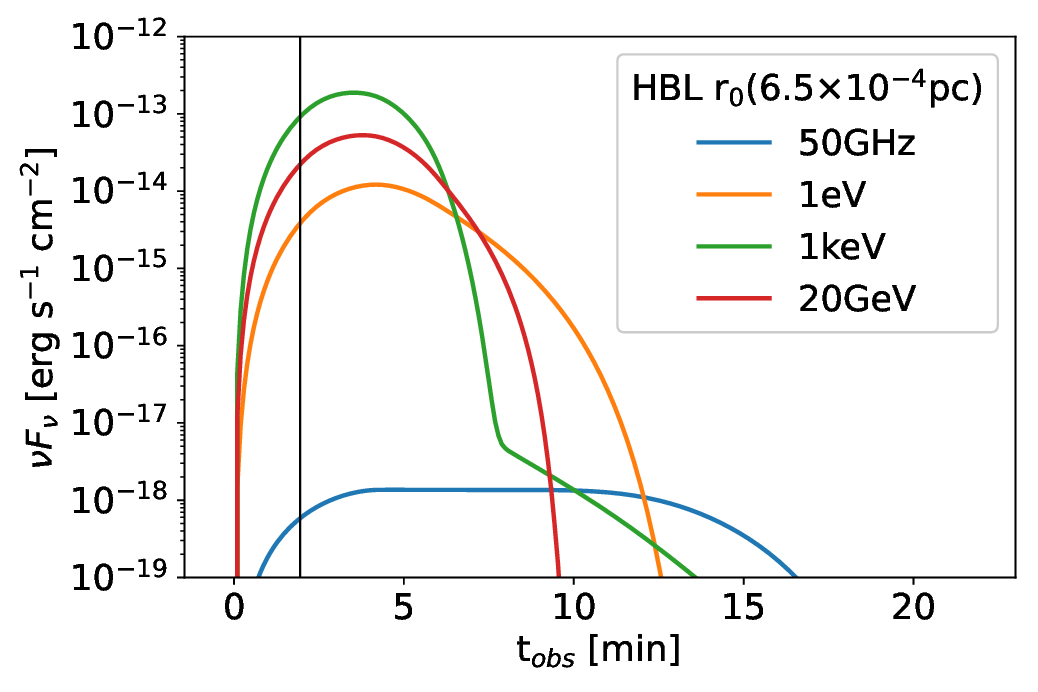}
}\hspace{-5mm}
\quad
\subfigure{
\includegraphics[width=0.68\columnwidth]{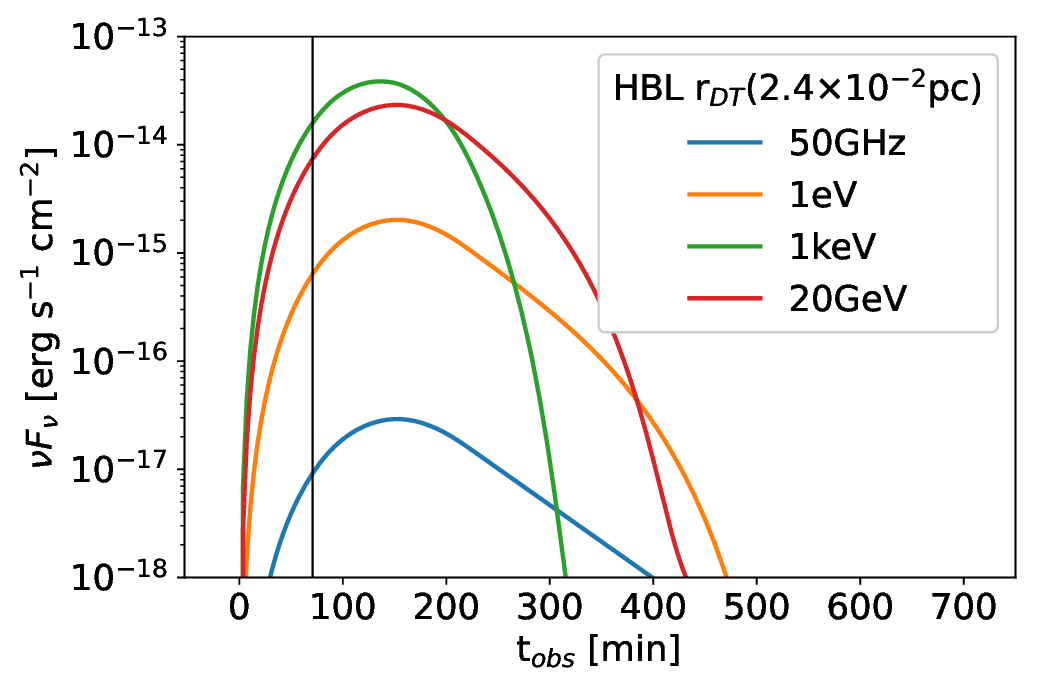}
}\hspace{-5mm}
\quad
\subfigure{
\includegraphics[width=0.68\columnwidth]{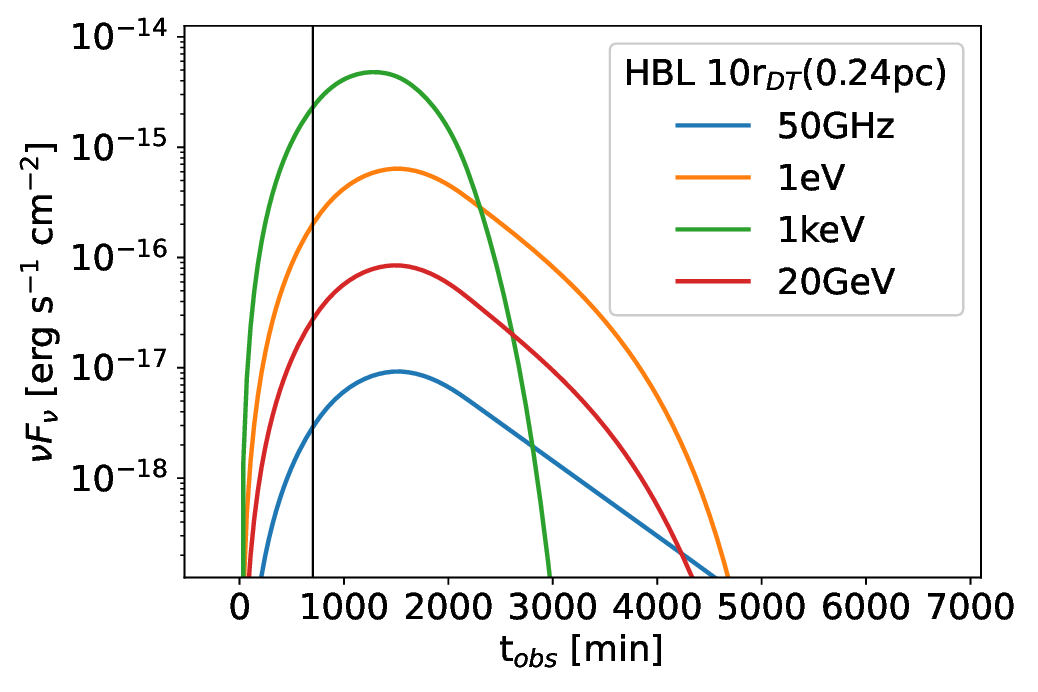}
}
\caption{{Same as Fig.~\ref{LC1}, but converted to the observer's frame with considering the different travelling time of the light emitted from different parts of the blob.}
\label{LC2}}
\end{figure*}

Following the basic assumption in the BK jet model \citep{1979ApJ...232...34B, 2012MNRAS.423..756P}, the power carried in the magnetic field $B(r_i)$ is conserved in each segment, therefore
\begin{equation}
B(r_i)=B(r_0)\frac{R'(r_0)}{R'(r_i)}.
\end{equation}
Through various acceleration mechanisms \citep[e.g.,][]{2020LRCA....6....1M}, like Fermi-type acceleration, electrons are accelerated and continuously injected into each blob with the injection time $t'_{\rm inj}$, which is assumed to be equal to the light crossing time $R'(r_i)/c$. For the sake of simplicity, the injected electron luminosity $L'_{\rm inj}$ and the shape of electron energy distribution (EED) for each blob are assumed to be the same, and the EED is a power-law distribution,
\begin{equation}
\dot{Q}'(\gamma', r_i)=\dot{Q}'_0(r_i)\gamma'^{-s},~ \rm for~\gamma'_{\rm min}<\gamma'<\gamma'_{\rm max},
\end{equation}
where $\gamma'_{\rm min}$ and $\gamma'_{\rm max}$ are the minimum and maximum electron Lorentz factors, and $\dot{Q}'_0(r_i)$ is a normalization constant in units of $\rm s^{-1}$. The normalization constant $\dot{Q}'_0(r_i)$ is determined by $\int \dot{Q}'(\gamma', r_i)\gamma' m_{\rm e} c^2d\gamma'= L'_{\rm inj}$, where $m_{\rm e}$ is the electron rest mass, and $c$ is the speed of light. The local temporal evolution of the electron distribution $N'(\gamma', t', r_i)$ in each blob is governed by the continuity equation
\begin{equation}
\begin{split}
\frac{\partial N'(\gamma', t', r_i)}{\partial t'}=&-\frac{\partial}{\partial \gamma'}[\dot{\gamma'}(\gamma', t', r_i)N(\gamma', t', r_i)]\\
&-\frac{N'(\gamma', t', r_i)}{t'_{\rm esc}(r_i)}
+\dot{Q}'(\gamma', r_i),
\end{split}
\end{equation}
where $t'_{\rm esc}(r_i)=\frac{10R'(r_i)}{c}$ is the escape timescale \citep{2017ApJ...843..109G}, $\dot{\gamma'}(\gamma', t', r_i)$ is the total energy loss rate, given by
\begin{equation}\label{cooling rate}
\dot{\gamma'}(\gamma', t', r_i)=-\frac{c\gamma'}{R'(r_i)}-b'(\gamma', t', r_i)\gamma'^2,
\end{equation}
where the first term on the right-hand side describes the adiabatic loss and the second term represents the energy losses due to synchrotron, SSC and EC radiation, i.e.,
\begin{equation}
\begin{split}
b'(\gamma', t', r_i) = \frac{4}{3}\sigma_{\rm T}\frac{1}{m_{\rm e}c}[U_{\rm B}(r_i)+F_{\rm KN}(\gamma', t', r_i)u'_{\rm soft}(t', r_i)],
\end{split}
\end{equation}
where $\sigma_{\rm T}$ is the Thomson cross-section and $u'_{\rm B}(r_i) = B^2(r_i)/8\pi$ is the magnetic field energy density,
\begin{equation}
\begin{split}
F_{\rm KN}(\gamma', t', r_i)=\frac{9}{u'_{\rm soft}(t', r_i)}\int^{\infty}_0d\epsilon~\epsilon~n'_{\rm soft}(\epsilon, t', r_i) \\
\times \int^1_0 dq\frac{2q^2\textmd{ln}~q+q(1+2q)(1-q)+\frac{q(\omega q)^2(1-q)}{2(1+\omega q)}}{(1+\omega q)^3}
\end{split}
\end{equation}
is a numerical factor accounting for Klein-Nishina effects \citep{2010NJPh...12c3044S}, where $\epsilon$ is the energy of target photons, $n'_{\rm soft}(\epsilon, t', r_i)$ is the number density distribution of target photons, $\omega=4\epsilon \gamma'/(m_{\rm e}c^2)$, and $u'_{\rm soft}(t', r_i)=u'_{\rm blob}(t', r_i)+u'_{\rm seg}(r_i)+u'_{\rm ext}(r_i)$ is the energy density of target photons. $u'_{\rm blob}(t', r_i)$ is the energy density of synchrotron photons {from the blob itself. $u'_{\rm seg}(r_i)$ accounts for the radiation from other neighboring radiation zones, i.e., blobs in the same segment. Therefore, $u'_{\rm seg}$ depends on the number of blobs in certain segment $N_i$, which will be addressed in the next section. Note that this term is not evolved with time since we consider the {low} state of the jet in this work. $u'_{\rm ext}(r_i)$ is the energy density of external photons.} In the environment of a blazar jet, external photons are mainly from BLR and DT, and their energy densities in the comoving frame of a blob, which is located at a distance $r$ from the SMBH, can be approximately written as \citep{2012ApJ...754..114H}
\begin{equation}
u'_{\rm BLR}(r_i) \approx \frac{\xi_{\rm BLR} \Gamma^2L_{\rm D}}{4\pi r_{\rm BLR}^2c[1+(r_i/r_{\rm BLR})^3]}
\end{equation}
\begin{equation}
u'_{\rm DT}(r_i) \approx \frac{\xi_{\rm DT} \Gamma^2L_{\rm D}}{4\pi r_{\rm DT}^2c[1+(r_i/r_{\rm DT})^4]},
\end{equation}
where $\xi_{\rm BLR}=0.1$ and $\xi_{\rm DT}=0.1$ are the fractions of the disk luminosity $L_{\rm D}$ reprocessed into the BLR and DT radiation, respectively, $r_{\rm BLR} = 0.1(L_{\rm D}/10^{46}\rm erg \,  s^{-1})^{1/2}$~pc and $r_{\rm DT} = 2.5(L_{\rm D}/10^{46}\rm erg \,  s^{-1})^{1/2}$~pc are the characteristic distances where such reprocessing takes place \citep{2008MNRAS.387.1669G}. The energy spectrum of the BLR and DT radiation is taken to be that of a blackbody with a peak at $2\times10^{15}\Gamma$ Hz \citep{2008MNRAS.386..945T} and $3\times10^{13}\Gamma$ Hz \citep{2007ApJ...660..117C} in the comoving frame, respectively. The corresponding cooling timescales calculated as $1/b'(\gamma', t', r_i)\gamma'$ are shown in Fig.~\ref{tcool}. After solving for the temporal evolution of $N'(\gamma', t', r_i)$ (Fig.~\ref{EED}), the LC of a single blob $\nu L_{\nu}'(E', t', r_i)$ at a distance $r_i$, consisting of the synchrotron, SSC and EC radiation (Fig.~\ref{LC1}), can be calculated using the \texttt{NAIMA} python package\footnote{https://naima.readthedocs.io/en/latest/} \citep{2015ICRC...34..922Z}. Note that due to the adiabatic cooling, the radiation of a blob will quickly decline after the cessation of particle injection. Given both the injection timescale and the adiabatic cooling timescale to be $R'/c$ in the comoving frame, the duration of the active phase of a blob is approximately estimated by $\tau_{\rm blob}'=2R'/c$ in the comoving frame or $\tau_{\rm blob}=2R'/c\delta_{\rm D}$ in the observer's frame. The comoving-frame LCs emitted from blobs at different distances are shown in Fig.~\ref{LC1}. 
Then, we can convert the LC into the flux received by the observer, by $t=t'/\delta_{\rm D}$, $E=\delta_DE'$ (here we neglect the cosmological redshift). We also take into account the different travelling time of the light emitted from different parts of the blob at the same moment, assuming electrons are homogeneously distributed inside the blob. This results in smoothing of the LC received by the observer. The overall conversion can be given by
\begin{equation}
\begin{split}
    \nu F^{\rm b}_\nu(E,t,r_i)=&\frac{(2R')^{-1}\delta_{\rm D}^4}{4\pi D_L^2}\int_0^{2R'} \nu L'_\nu\left[E/\delta_D, \delta_{\rm D}(t-x/c), r_i\right]dx,  \\
    &{\rm with}~L_\nu'(t'<0)=0,
\end{split}
\end{equation}
where the expected LCs in the observer's frame are shown in Fig.~\ref{LC2}.

{We note that the obtained LCs are based on the assumption of $t_{\rm inj}'=R'/c$. This may be true if the dissipation is caused by, for example, a mildly relativistic shock induced by the internal collision. Of course, the particle injection timescale could be longer or shorter, and not necessarily equal to the light crossing time of the blob. For a shorter injection timescale, the LC will be more or less the same because the variability is subject to the light travelling time. If the injection timescale is longer, the rising phase of the LC of each blob will be longer and hence the superimposed LC of the entire jet would be less variable. For reference, we compare the SED and LCs between the case of $t_{\rm inj}'=R'/c$ and the case of $t_{\rm inj}'=2R'/c$ in Appendix~\ref{appB} (see Fig.~\ref{fig:sed_tinjcompare} and \ref{fig:LC_tinjcompare}).} 


\begin{figure}
\centering
\subfigure{
\includegraphics[width=1\columnwidth]{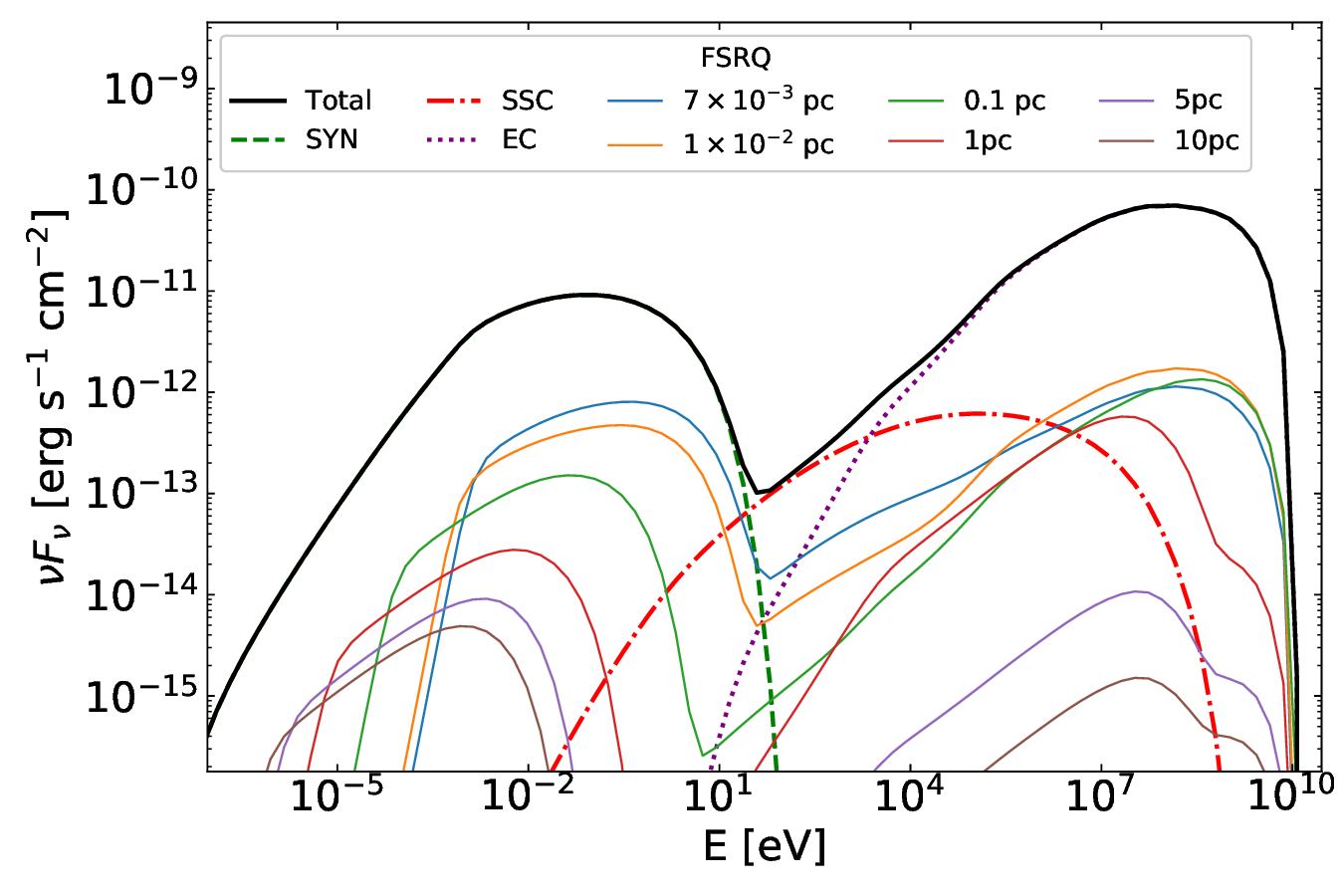}
}
\quad
\subfigure{
\includegraphics[width=1\columnwidth]{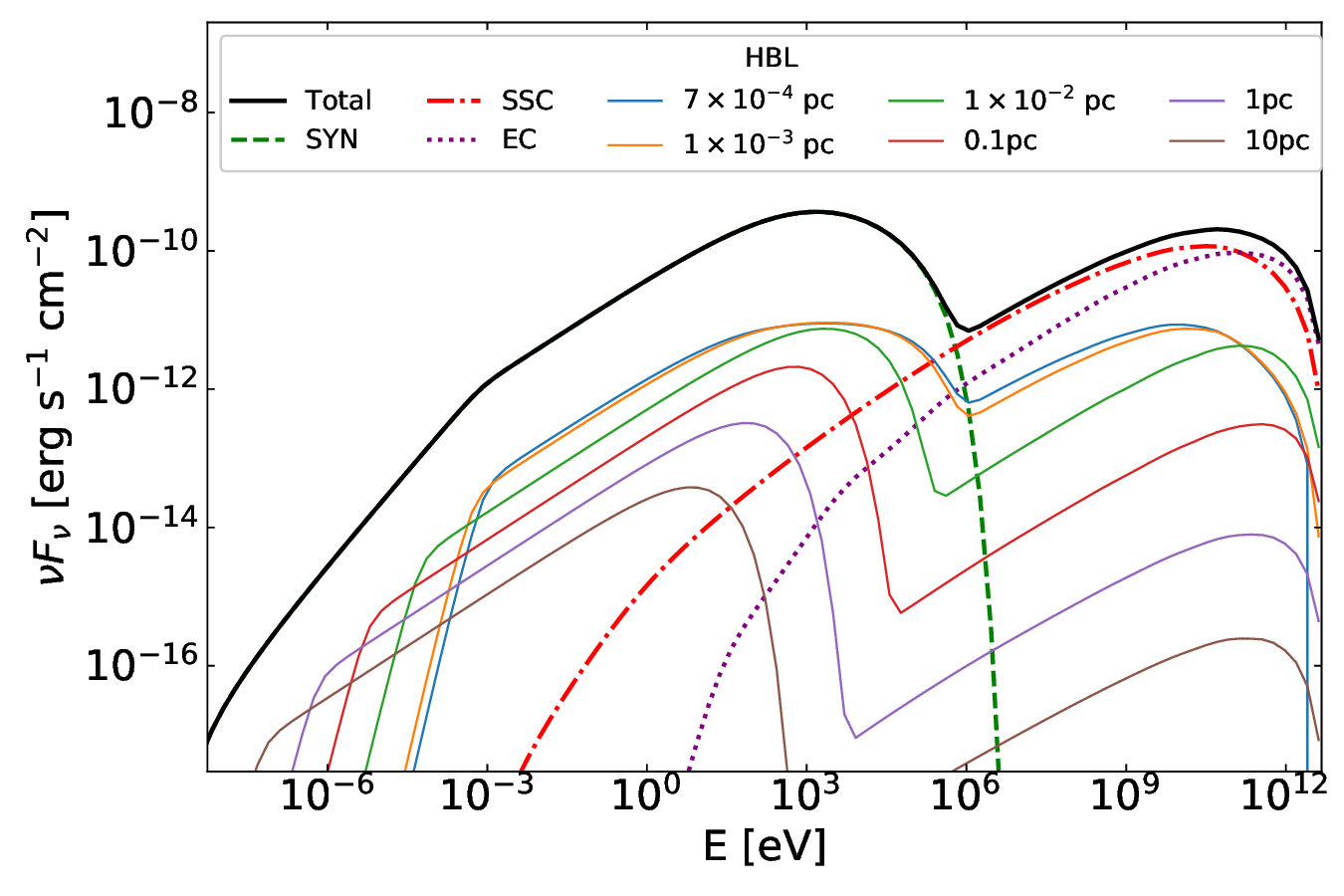}
}
\caption{Sum of the average SEDs of FSRQ (upper panel) and HBL (lower panel) from all blobs along the jets, as shown with thick solid black curves. The dashed green, dot-dashed red and dotted purple curves represent sum of the average synchrotron emission, SSC emission and EC emission, respectively. Thin solid colored curves represent the average contribution from some representative segments at different distances $r$.
\label{ave}}
\end{figure}

\section{Application}\label{app}
Let us assume that blobs could be generated at any distance of the jet, and define $p(r)$ to be the probability of a dissipation event occurring at unit time (in the rest frame) and unit jet length at a distance $r$. As such, we may expect that $N$ blobs are generated in a period of time $T$ or a total dissipation rate $\dot{N}$ over the jet, i.e.,\footnote{As shown in Fig.~\ref{fig:sketch}, the end of jet is located at $r_0+L$. Since the jet length $L$ is much larger than the distance between jet base and SMBH $r_0$, $r_0+L\approx L$.}
\begin{equation}\label{eq:generate_rate}
\dot{N}\equiv N/T=\int_{r_0}^{r_{\rm max}} p(r)dr,
\end{equation}
The form of $p(r)$ is probably dependent on the microphysics (such as plasma instabilities) and macrophysics (such as internal collisions) of the jet. We here simply parameterise it as
\begin{equation}
p(r)=Ar^{-\alpha}
\end{equation}
with
\begin{equation}
A=\left\{
\begin{array}{ll}
\dot{N}(1-\alpha)(r_{\rm max}^{1-\alpha}-r_0^{1-\alpha})^{-1}, ~\alpha \neq 1\\
\dot{N}(\ln(r_{\rm max}/r_0))^{-1}, \hspace{35pt}  \alpha = 1
\end{array} \right.
\end{equation}
being the normalization and $\alpha$ being a model parameter.

By doing so, we may {depict the number of blobs appearing in the $i$th segment during an observational period $T$ by discretizing Eq.~\ref{eq:generate_rate} as}
\begin{equation}\label{Ni}
N_i=Tp(r_i)(r_{i+1}-r_i).
\end{equation}
Here we have implicitly assumed that $p(r_i)$ is independent of the observation time, {i.e., each dissipation event occurs randomly at a uniform probability over time in the observer's frame. We then employ the Monte-Carlo method to generate the starting time of the $j$th blob in the $i$th segment, as denoted by $t_{i,j}$, in the timeline of a total duration $T$, where $j=1, 2, ..., N_i$. Obviously, we need to repeat the above process separately for each segment, i.e., $i_{\rm max}$ times in total.} After completion, we can obtain the expected LCs of the blazar by
\begin{equation}
\nu F_\nu(E, t)=\sum_{i=1}^{i_{\rm max}}\sum_{j=1}^{N_i}\nu F_\nu^{\rm b}(E, t, r_i)\Theta(t-t_{i,j}),
\end{equation}
Given $T$ a sufficiently long time, the {low-state} spectrum of the jet can be given by
\begin{equation}
    \overline{\nu F_\nu}(E)=T^{-1}\int_0^T \nu F_\nu(E,t)dt,
\end{equation}
while the contribution of the $i$th segment to the flux reads
\begin{equation}\label{eq:flux_seg}
\overline{\nu F_\nu^{\rm s}}(E, r_i)=T^{-1}\sum_{j=1}^{N_i}\int_0^{T} \nu F_{\nu}(E,t, r_i)\Theta(t-t_{i,j})dt.
\end{equation}
With Eq.~(\ref{eq:flux_seg}), the energy density of the collective radiation of all the blobs in a certain segment $u'_{\rm seg}(r_i)$ can be obtained by $u'_{\rm seg}(r_i)\approx \overline{\nu F_\nu^{\rm s}}(E, r_i)D_{\rm L}^2/(S(r_i)c\delta_{\rm D}^4)$, where $D_L$ is the luminosity distance, and $S(r_i)=\pi (r_i^2+r_{i+1}^2)\tan^2\theta+\pi(r_{i+1}^2-r_{i}^2)\tan\theta/\cos\theta$ represents the surface area of each segment. Considering Eq.~\ref{eq:radius}, we have approximately $S(r_i)\sim 2\pi R'(r_i)^2/\kappa^2$. On average, the number of blobs active in the $i$th segment at the same time can be estimated by $n_i\approx p(r_i)(r_{i+1}-r_i)\tau_{\rm blob}(r_i)=8p(r)R'(r_i)^2/c\delta_D$. The ratio $u'_{\rm seg}/u'_{\rm b}\sim n_i\kappa^2$ determines the relative importance of photons emitted by nearby blobs and photons emitted by the blob itself as the target for the IC emission of the blob. 

As a general study, instead of reproducing the SEDs of specific objects, we employ the model to generate the {low-state} SEDs and LCs of a typical flat-spectrum radio quasar \citep[FSRQ,][]{1995PASP..107..803U} and a typical high-synchrotron-peaked BL Lac object \citep[HBL,][]{2010ApJ...716...30A}, respectively. In the modeling, the redshifts of two objects are taken as that of 3C 279 ($z=0.536$) and Mrk 421 ($z=0.03$), which are the representative of an FSRQ and an HBL respectively.

It should be noted that the number of blobs $N$ generated during a period $T$ cannot be too high, since otherwise the total volume of blobs would be larger than the jet itself. The filling factor of the jet $\lambda${, which represents the ratio of the volume of all blobs to the volume of the jet,} can be given by
\begin{equation}\label{fill}
\lambda\approx \frac{\int \tau_{\rm blob}p(r)\frac{4}{3}\pi R'(r)^3/\delta_D~dr}{(1/3)\pi \tan^2\theta r_{\rm max}^3}.
\end{equation}
{For simplicity, here we neglect the possibility of two (or more) blobs colliding with each other.} When setting the initial free parameters, we ensure that the filling factor $\lambda$ does not exceed unity. The value of $\lambda$ would affect the variability in LCs. A study of the variability features with different $\lambda$ will be shown in Section~\ref{doppler}. {For reference, if we assume that the blazar's emission mostly originates from a single blob in the jet (i.e., one-zone models), we may find the filling factor to be $\lambda \sim 10^{-12}(R'/10^{16}{\rm cm})^{3}(r_{\rm max}/100{\rm pc})^{-3}(\theta/5^\circ)^2(\delta_D/10)^{-1}$.}

\begin{table*}
\begin{minipage}[t][]{\textwidth}
\caption{Summary of Model Parameters.}
\label{parameters}
\begin{tabular}{cccc}
\hline\hline
Free parameters	&	FSRQ	&	HBL	&	Notes	\\
\hline							
$r_{\rm max}$ (pc)	&	200	&	200	&	Jet length	\\
$\kappa$	&	0.3	&	0.3	&	 Ratio of blob's radius to its segment's radius 	\\
$\dot{N}~\rm (s^{-1})$	&	$0.13$	&	$0.4$	&	Blob generation rate of the entire jet	\\
$\alpha$	&	1.9	&	1.9	&	 Index of the dissipation probability $p(r)$	\\
$r_0$ (pc)	&	$6.5\times10^{-3}$	&	$6.5\times10^{-4}$	&	 Position of jet base	\\
$\theta~(\circ)$	&	5	&	5	&	Jet's half-opening angle	\\
$\Gamma$	&	15	&	15	&	Jet's bulk Lorentz factor \\
$L_{\rm D}$ (erg/s)	&	$1\times10^{45}$	&	$1\times10^{42}$	&	Disk luminosity	\\
$B(r_0)$ (G)	&	8	&	4	&	Magnetic field at jet base	\\
$L'_{\rm inj}$~(erg/s)	&	$5\times10^{39}$	&	$5\times10^{38}$	&	Injection electron luminosity	\\
$s$	&	2	&	2	&	The spectral index of electron energy distribution	\\
$\gamma'_{\rm min}$	&	1	&	1	&	Minimum electron Lorentz factor	\\
$\gamma'_{\rm max}$\footnote{The difference between $\gamma'_{\rm max}$ for the FSRQ and for the HBL is required to reproduce the different peak frequencies of the synchrotron humps in their SEDs.  Such a result may imply that there is a difference in the particle acceleration efficiency between FSRQ and HBL. Alternatively, one may employ a broken-power-law function for the electron distribution of the sources and tune the break energy to reproduce the observed peak frequency. This is similar to that often adopted in one-zone models, although the physical origin of the spectral break still remains unclear.}	&	3.2$\times10^{3}$	&	1$\times10^6$	&	Maximum electron Lorentz factor	\\
\hline							
Derived/Fixed parameters	&		&		&		\\
\hline
$\lambda$	&	18\%	&	7\%	&	Jet filling factor given by Eq.~\ref{fill}	\\
$\delta_{\rm D}$  &    $\Gamma$     &    $\Gamma$  &  Doppler factor\\
$\theta_{\rm obs}$   & $\approx 1/\Gamma$ & $\approx 1/\Gamma$  &   Angle between the observer's LOS and the jet's axis\\
\hline
\end{tabular}
\end{minipage}
\label{tab:parameters}
\end{table*}

\subsection{Multiwavelength SEDs of FSRQ and HBL}\label{sec:sed}
\begin{figure}
\centering
\subfigure{
\includegraphics[width=1\columnwidth]{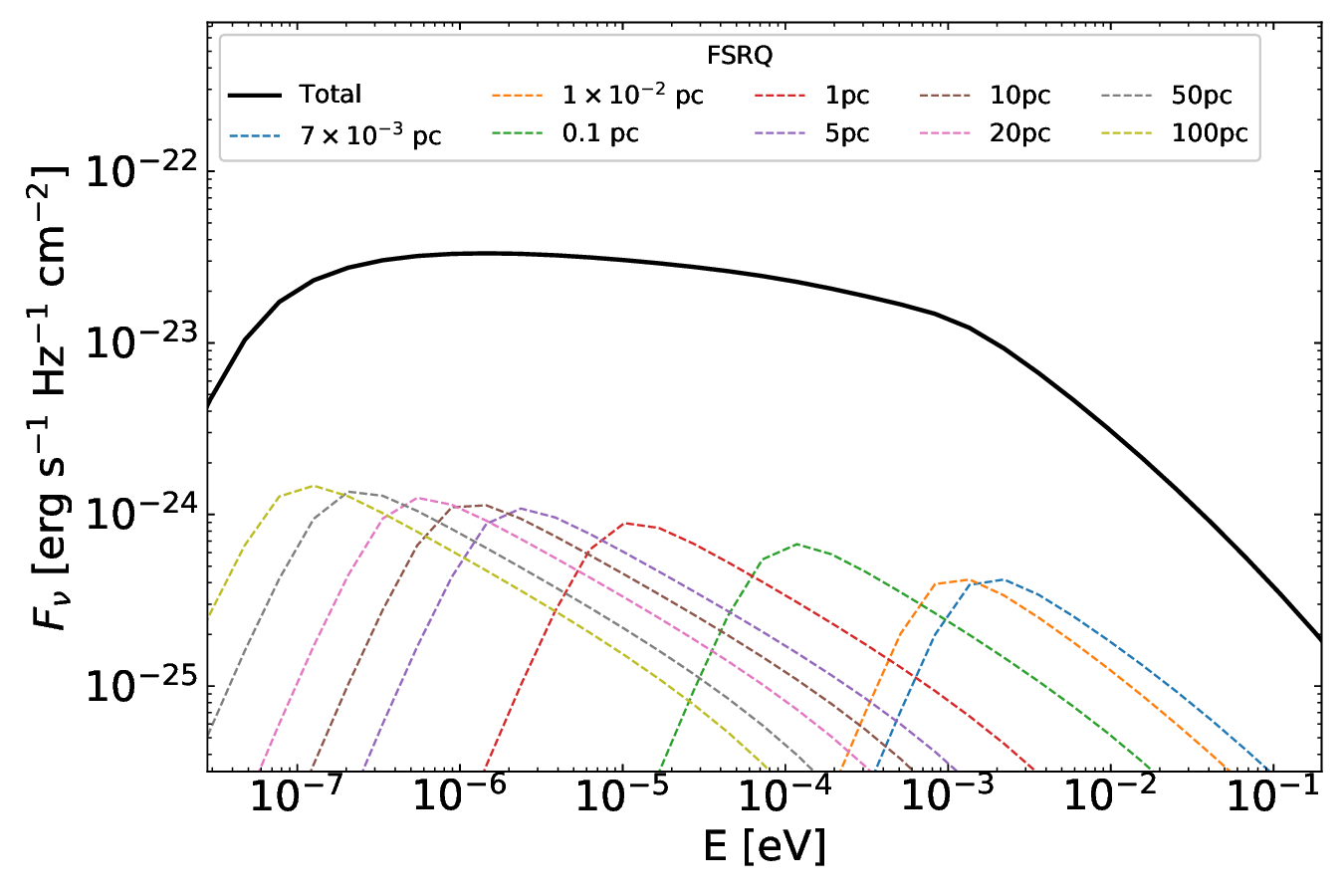}
}
\quad
\subfigure{
\includegraphics[width=1\columnwidth]{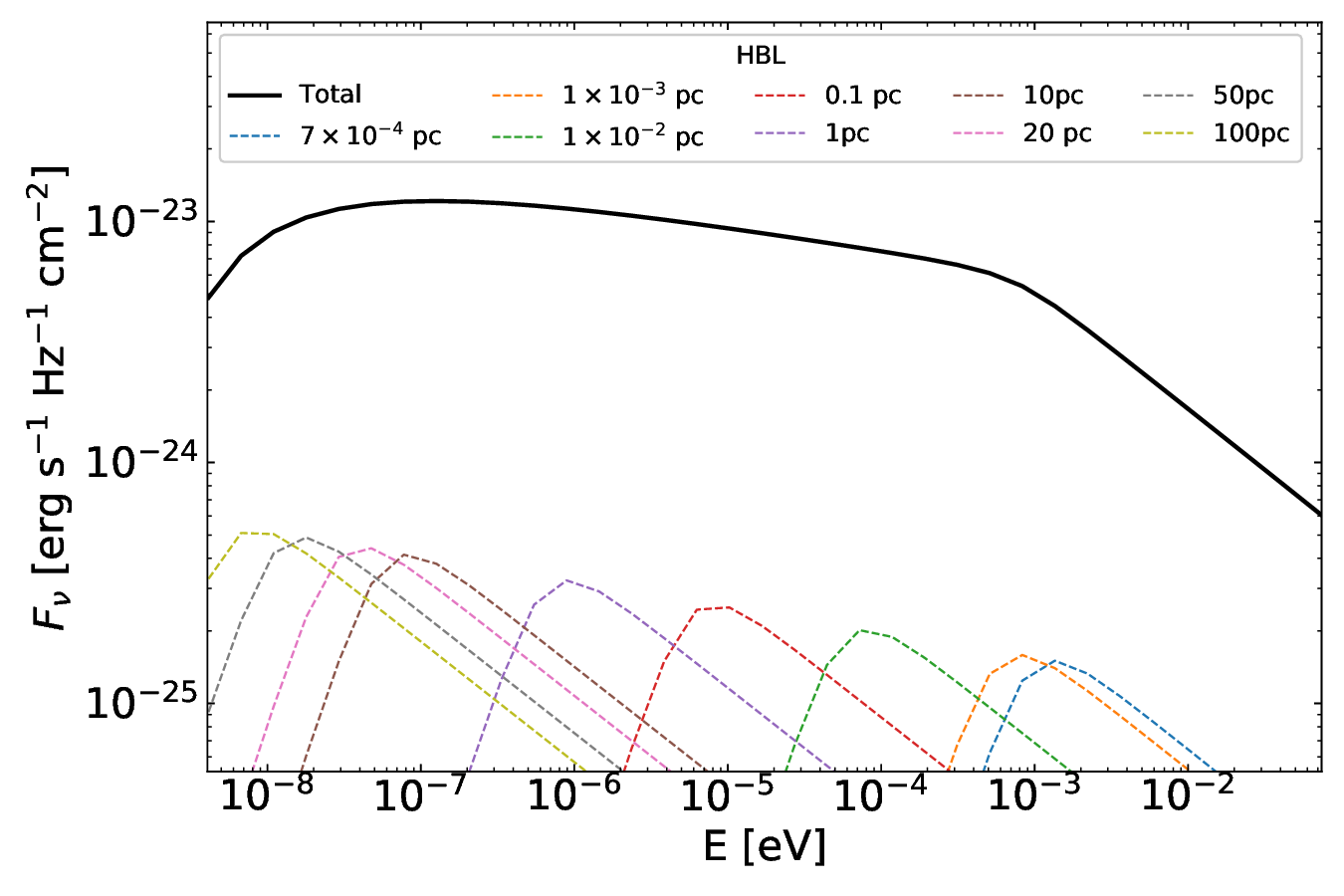}
}
\caption{This figure shows the predicted average flat radio spectra (the solid black curve) of an FSRQ (upper panel) and an HBL (lower panel) explained as the superposition of synchrotron emission from the blobs at different distances along the jet (the dashed colored curves).
\label{flat}}
\end{figure}

It is well known that the flat radio spectrum is one of the most remarkable and common features of blazars. Therefore, it is a critical test for the model to reproduce the flat radio spectrum. In our model, the jet's emission is constituted by the superposition of blobs at different distance, and thus the spectrum is dependent on the form of $p(r)$ (or more specifically, the parameter $\alpha$ for a power-law form). It can be shown analytically that the predicted spectral index of the radio spectrum $F_\nu \propto \nu^{\alpha-2}$ (see Appendix~\ref{appA}), as also supported by our numerical calculation. While setting $\alpha=2$, a flat radio spectrum can be generated.  \cite{2011A&A...536A..15P} suggested that the mean value of the spectral index of the radio spectrum of blazars is about $-0.1$, implying $\alpha \simeq 1.9$ in our model. We therefore employ $\alpha=1.9$ as the benchmark parameter here and discuss the cases with other values of $\alpha$ in the Appendix~\ref{fig:sed_alpha}. In Fig.~\ref{ave}, we present the typical SEDs of an FSRQ and an HBL in the {low} state predicted by our model, where the model parameters are given in Table~\ref{parameters}. For an FSRQ (the upper panel of Fig.~\ref{ave}), as a typical low synchrotron peaked \citep[LSP,][]{2010ApJ...716...30A} blazar, the peak frequency of its low-energy hump is lower than $10^{14}~\rm Hz$, while the high-energy hump peaks around 0.1~GeV and is dominated by EC radiation. For an HBL (lower panel of Fig.~\ref{ave}), its low-energy hump peaks beyond $10^{17}~\rm Hz$, while the high-energy hump peaks around 20~GeV and is dominated by SSC radiation. In Fig.~\ref{flat}, we zoom into the radio band of the spectrum, in the form of $F_\nu$, to show that the flat radio spectra of FSRQ and HBL are well reproduced. The contributions of blobs at different distances in the jet are decomposed for reference, as shown with the dashed colored curves. It is clear that the turnover frequencies of the synchrotron radiation of blobs vary with the distance to the SMBH. For a blob closer to the SMBH, its magnetic field is higher and the blob size is smaller, so the synchrotron self-absorption is stronger, truncating the radio spectrum at a higher frequency, which is also referred to as the core-shift phenomenon. Therefore, in terms of the interpretation to the flat radio spectra and the core-shift phenomenon, our model is basically the same as the standard BK jet model  \citep[e.g.,][]{2019ApJ...870...28F}. The only difference is that particles are all injected from the jet base and advected to larger distance in the BK jet model. When considering adiabatic losses, re-acceleration of electrons along the entire jet must be introduced to the BK model to balance the energy loss of electrons \citep{2013MNRAS.429.1189P, 2015MNRAS.453.4070P, 2019MNRAS.485.1210Z}, or otherwise a specific geometry of the jet need be introduced \citep{2006MNRAS.367.1083K}. In our model, particles are accelerated locally at different positions of the jet.

\begin{figure}
\centering
\subfigure{
\includegraphics[width=1\columnwidth]{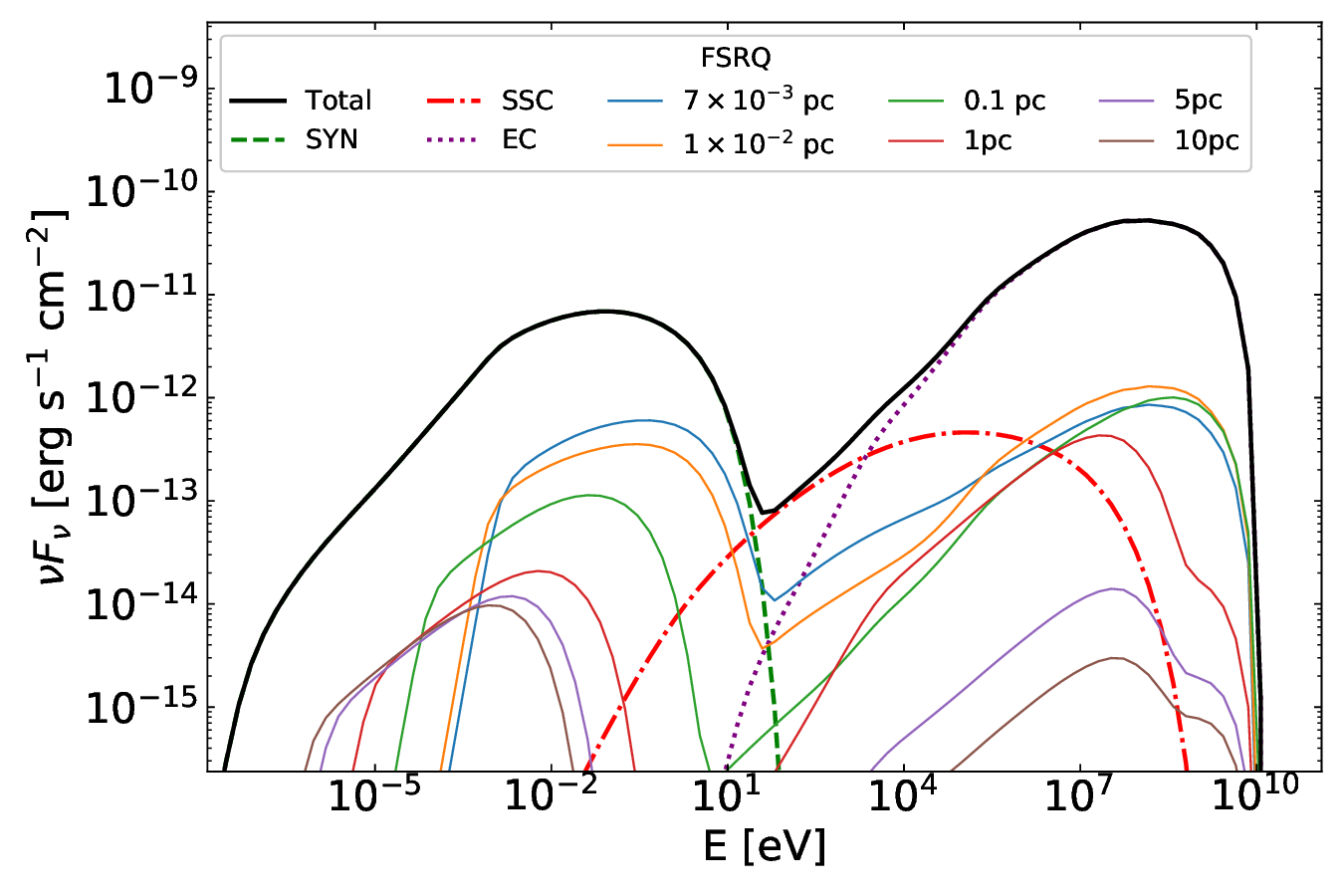}
}
\quad
\subfigure{
\includegraphics[width=1\columnwidth]{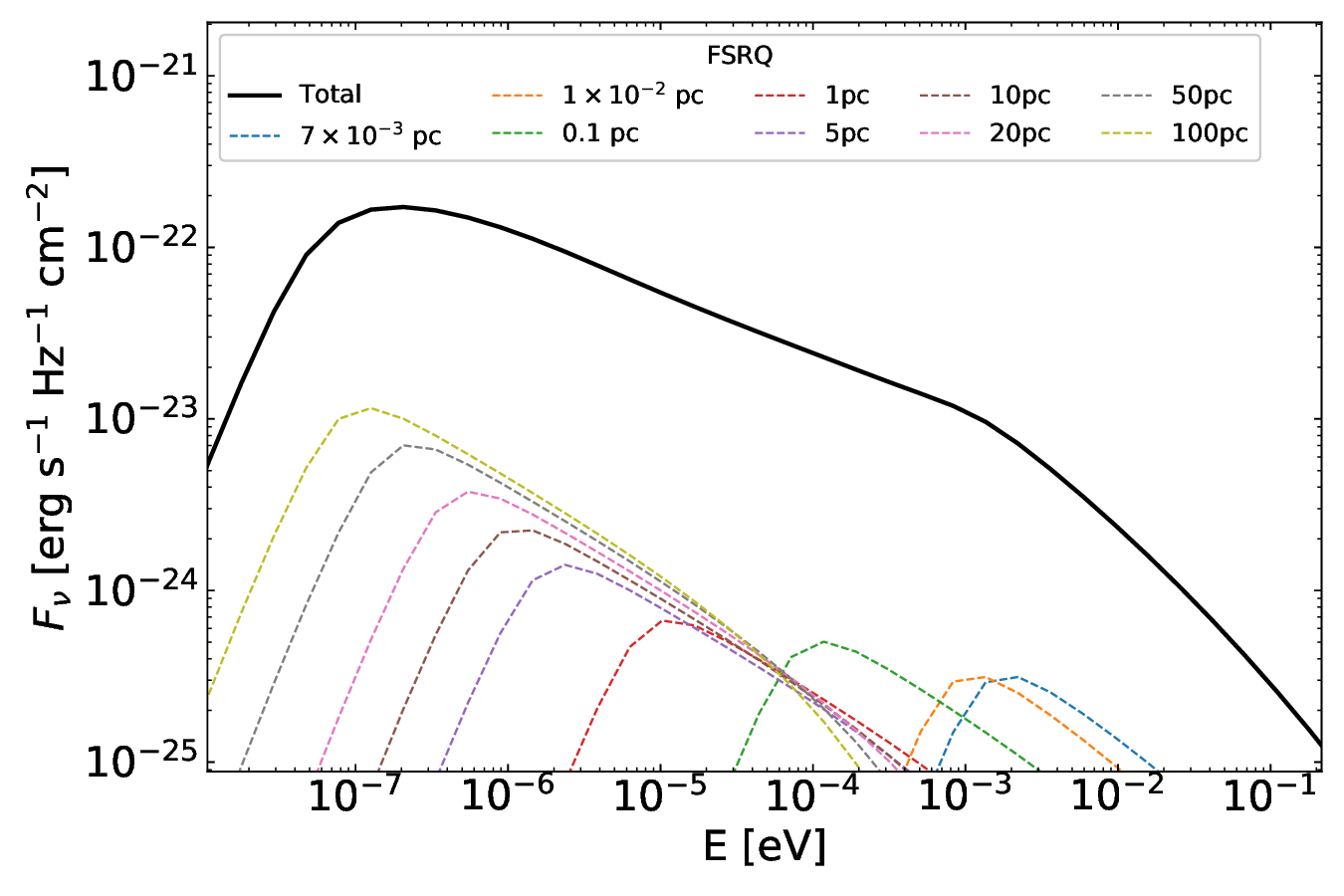}
}
\caption{The SED (upper panel) and radio spectrum (lower panel) of an FSRQ when setting $\alpha=1.9$ for $r<\rm 2~pc$ and $\alpha=1.3$ for $r\geqslant \rm 2~pc$. The other parameters are same as shown in Table~\ref{parameters}.
\label{break}}
\end{figure}

We also note that the radio spectra of some blazars become steep again at a certain frequency $\nu_b$, which is generally below $0.1-1\,$GHz, e.g. 3C 279 (see Fig.~7.5 of \citealt{Marscher10}) and 3C 454.3 (see Fig.~16 of \citealt{2011A&A...536A..15P}). In order to reproduce this feature in the radio spectrum, we may introduce a break in $\alpha$ at a certain $r$ where the synchrotron self-absorption (SSA) frequency is equal to $\nu_b$. The break of $\alpha$ is not unreasonable, because both MHD simulations and the analytical solution indicate that although a jet is usually magnetically dominated near the SMBH, the kinetic energy becomes equal or dominant at large distance \citep[e.g.][]{2019MNRAS.490.2200C,2021ApJ...906..105C}. As a result, the dissipation mechanism could be changed and the dissipation probability distribution may change subsequently. Here we take the {FSRQ} as an example. If assuming the radio spectrum changes from flat to steep at $5\times10^{-6}~\rm eV$, we can set the index of $p(r)$ after 2~pc as $\alpha=1.3$. The corresponding SED and radio spectrum are shown in Fig.~\ref{break}. 

\subsection{Flux profile of the jet}
\begin{figure}
\centering
\subfigure{
\includegraphics[width=1\columnwidth]{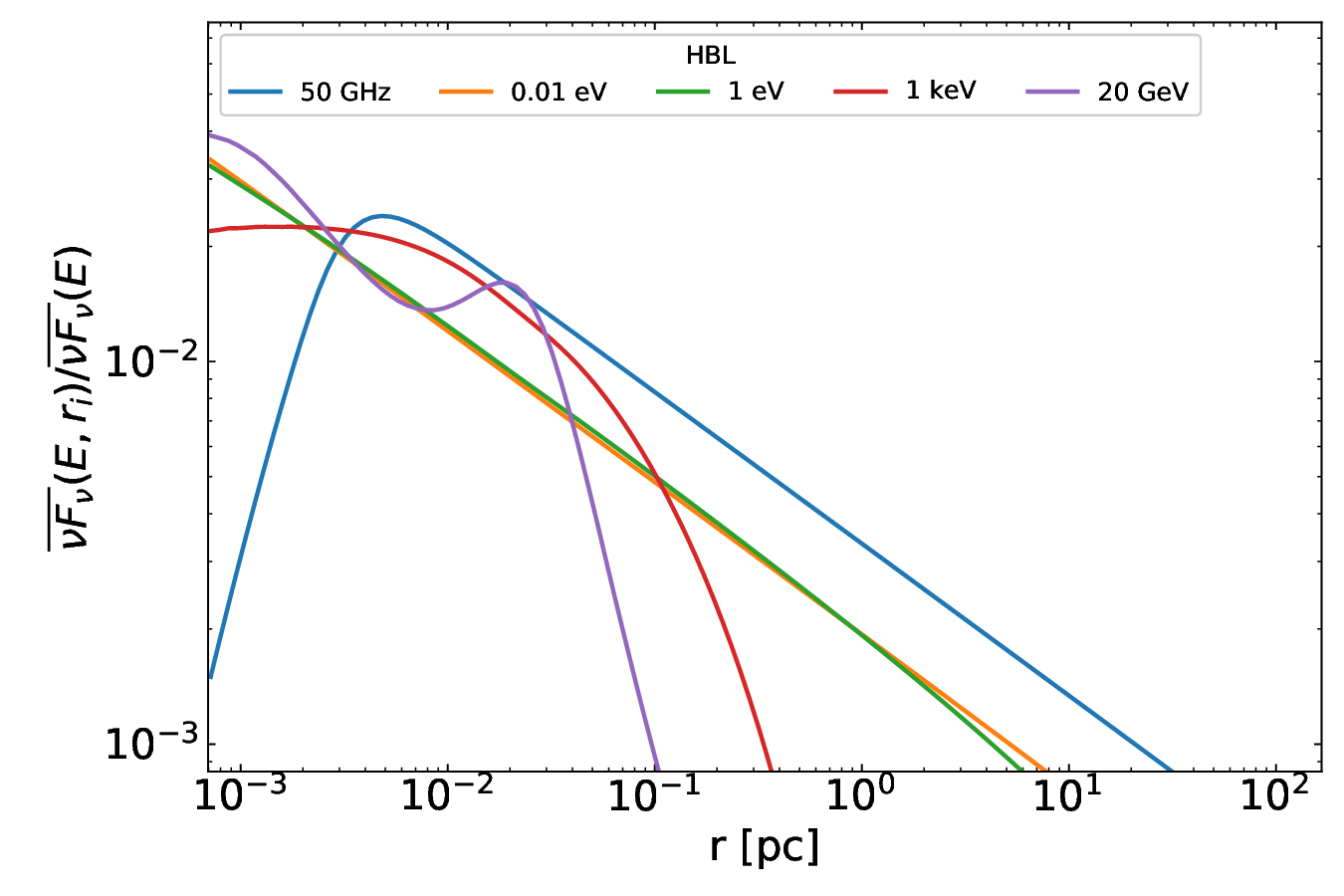}
}
\quad
\subfigure{
\includegraphics[width=1\columnwidth]{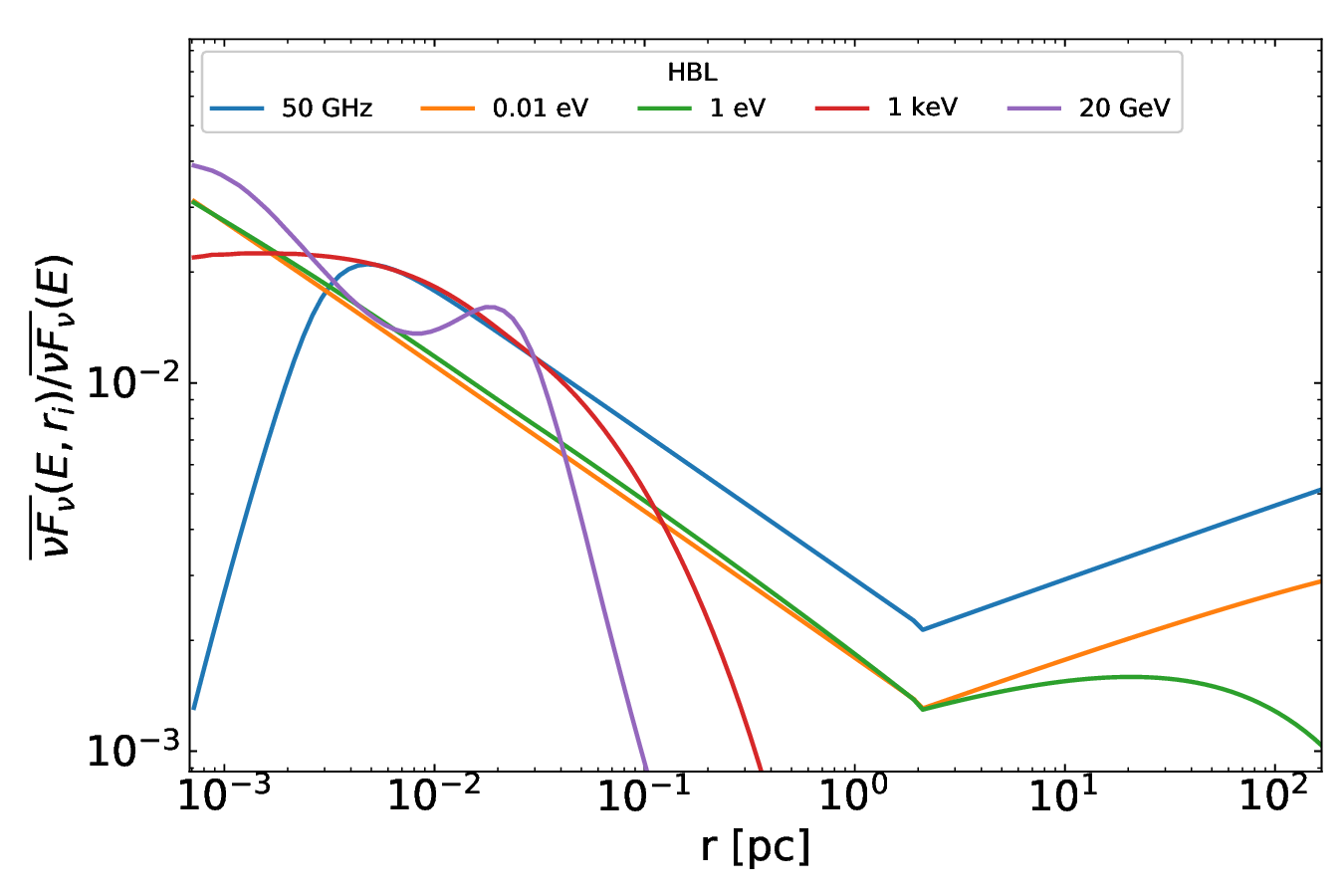}
}
\caption{The ratio of the average contribution of the $i$th segment $\overline{\nu F_\nu}(E, r_i)$ to the average contribution of the whole jet $\overline{\nu F_\nu}(E)$ of an HBL as a function of $r$ at 50~GHz, 0.01~eV, 1~eV, 1~keV and 20~GeV bands, respectively. The used parameters are the same as shown in Table~\ref{parameters}. The upper panel show the result when setting $\alpha=1.9$ along the whole jet and the lower panel show the result when setting $\alpha=1.9$ for $r<\rm 2~pc$ and $\alpha=1.3$ for $r\geqslant \rm 2~pc$.
\label{fig:LCr}}
\end{figure}

In our model, blobs are distributed along the jet and hence the jet's radiation is naturally extended, but with different profiles in different energy bands. In the upper panel of Fig.~\ref{fig:LCr}, we take the HBL shown in Figs.~\ref{ave} as an example to show the radiation profiles of the jet at five different energies, i.e., the ratio of the $i$th segment's flux $\overline{\nu F_\nu}(E, r=r_i)$ to the total flux of the jet $\overline{\nu F_\nu}(E)$ as the function of $r$. It can be seen that the optical (synchrotron) radiation from different segments gradually decreases since the magnetic field in the blob $B(r)$ decreases with the increasing $r$. The X-ray and gamma-ray profiles decrease quite rapidly, however, due to different reasons. The rapid decline of the X-ray profile is not only due to the decreasing magnetic field but also caused by the maximum synchrotron energy radiated by accelerated electrons falling below the X-ray band.  The decline of the $\gamma$-ray profile is due to the decrease of the target radiation density and the latter resulting from the combination of a lower synchrotron luminosity and a larger blob size with increasing $r$. The hump around $0.01-0.1\,$pc in the gamma-ray profile is due to the contribution of the DT radiation.  In the lower panel of Fig.~\ref{fig:LCr}, we set a break in $p(r)$ with $\alpha$ changing from 1.9 to 1.3 at $r\geqslant \rm 2~pc$, corresponding to case shown in Fig.~\ref{break}. The radio profile first rises and then falls. This is because the radio emission in the blobs near the jet base will be absorbed due to the SSA process. As $r$ increases, the SSA optical depth decreases gradually, and at a certain position ($4\times10^{-3}~\rm pc$ in Fig.~\ref{fig:LCr}), the emitted radio flux reaches its peak when the SSA optical depth $\tau_{\rm SSA}$ decreases to unity. Since $\tau_{\rm SSA}\propto (r\nu)^{-(s+4)/2}$ (Eq.~\ref{eq:ssa}), we may expect the shift of radio cores at different frequencies as $r_{\rm core}\propto \nu^{-1}$, which is consistent with radio measurement \citep{2009MNRAS.400...26O, 2011Natur.477..185H, 2011A&A...532A..38S}.



\subsection{Variability and the Doppler-factor crisis}\label{doppler}

\begin{figure*}
\centering
\subfigure{
\includegraphics[width=1.53\columnwidth]{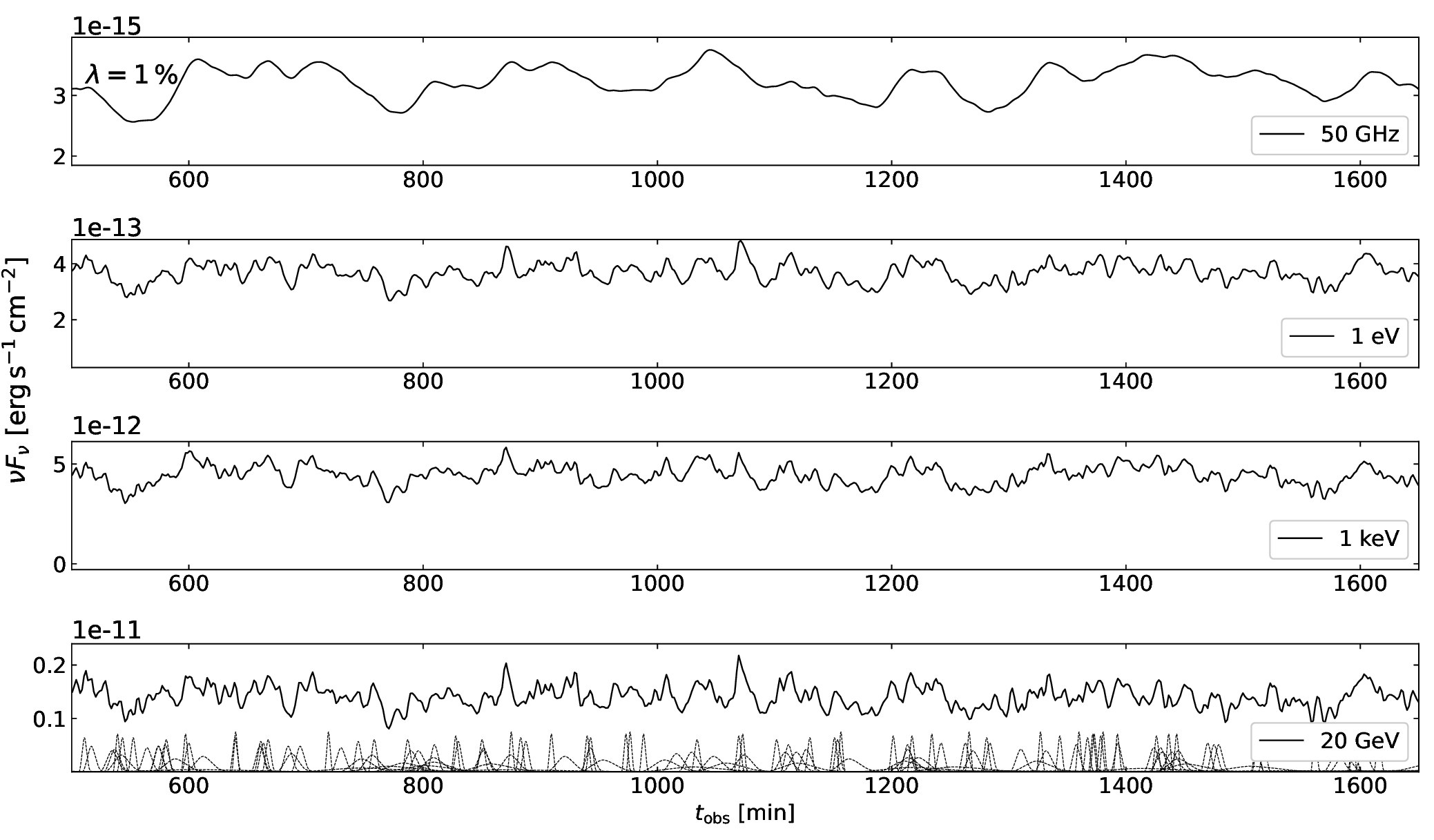}
}
\subfigure{
\includegraphics[width=1.53\columnwidth]{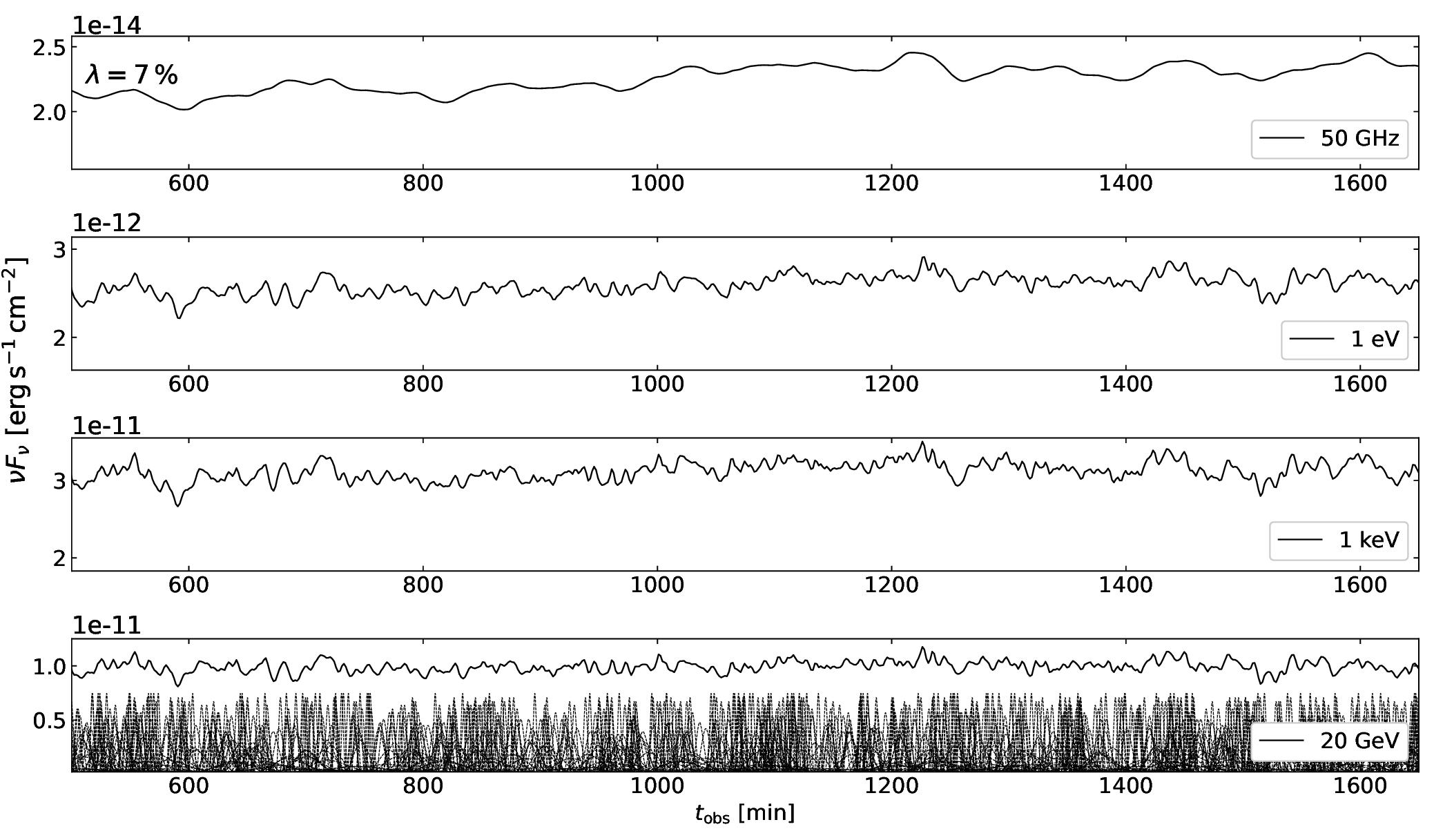}
}
\subfigure{
\includegraphics[width=1.53\columnwidth]{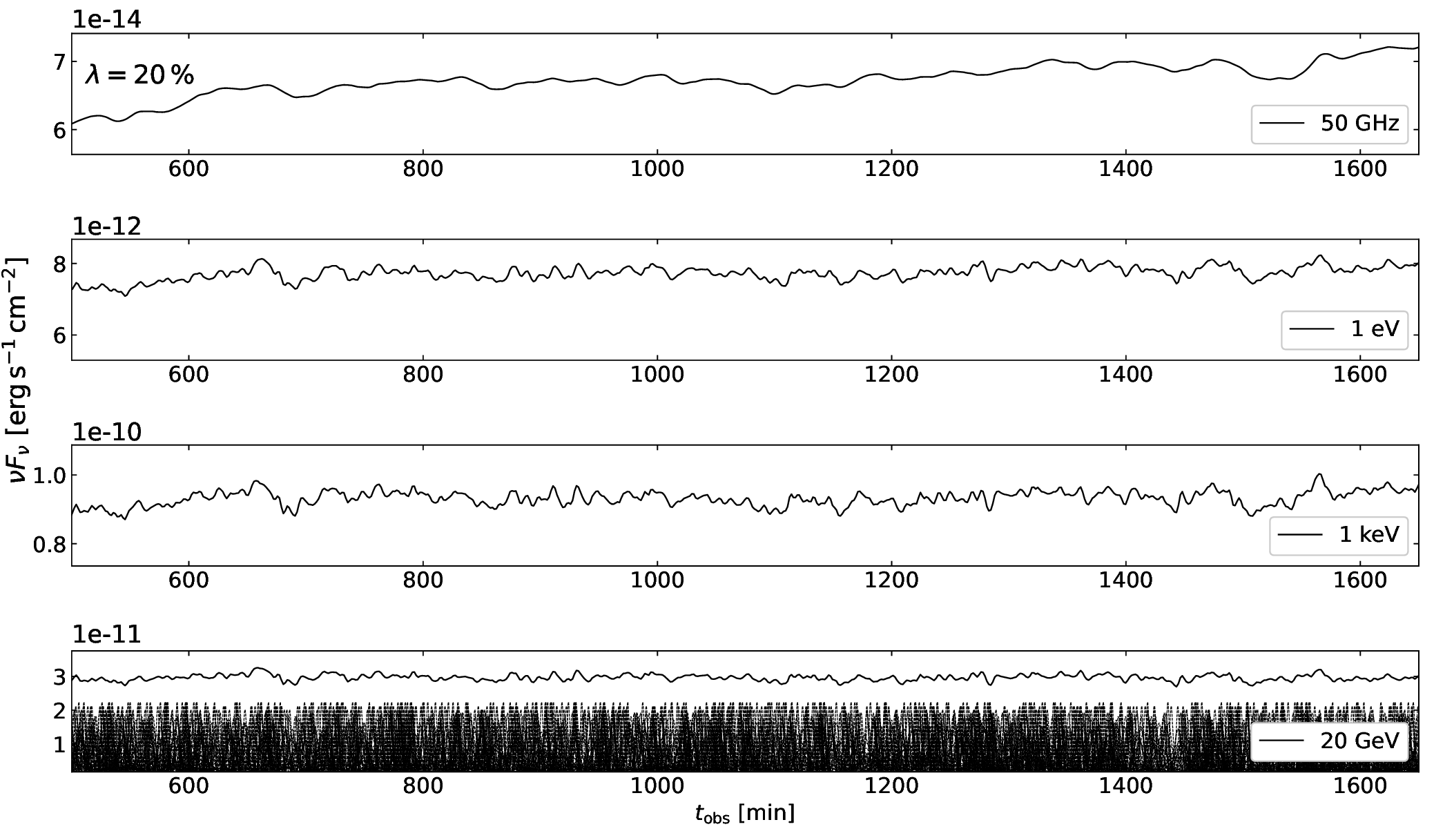}
}
\caption{The multi-wavelength LCs of the HBL with different filling factor $\lambda$. The parameters are the same as those shown in Table~\ref{parameters}. The black thin dotted curves in the subfigures of 20 GeV LCs represent the LCs from each blob at different segments. Note that the number of dashed curves is one-thirtieth of the number actually considered in the modeling. In addition, for the case of $\lambda=1\%$, the flux of each dashed line is amplified by a factor of 10 artificially for visibility; for the case of $\lambda=7\% \rm~and~20\%$, the flux of each dashed line is amplified by a factor of 100.
\label{LC-mw}}
\end{figure*}


\begin{figure}
\centering
\includegraphics[width=1\columnwidth]{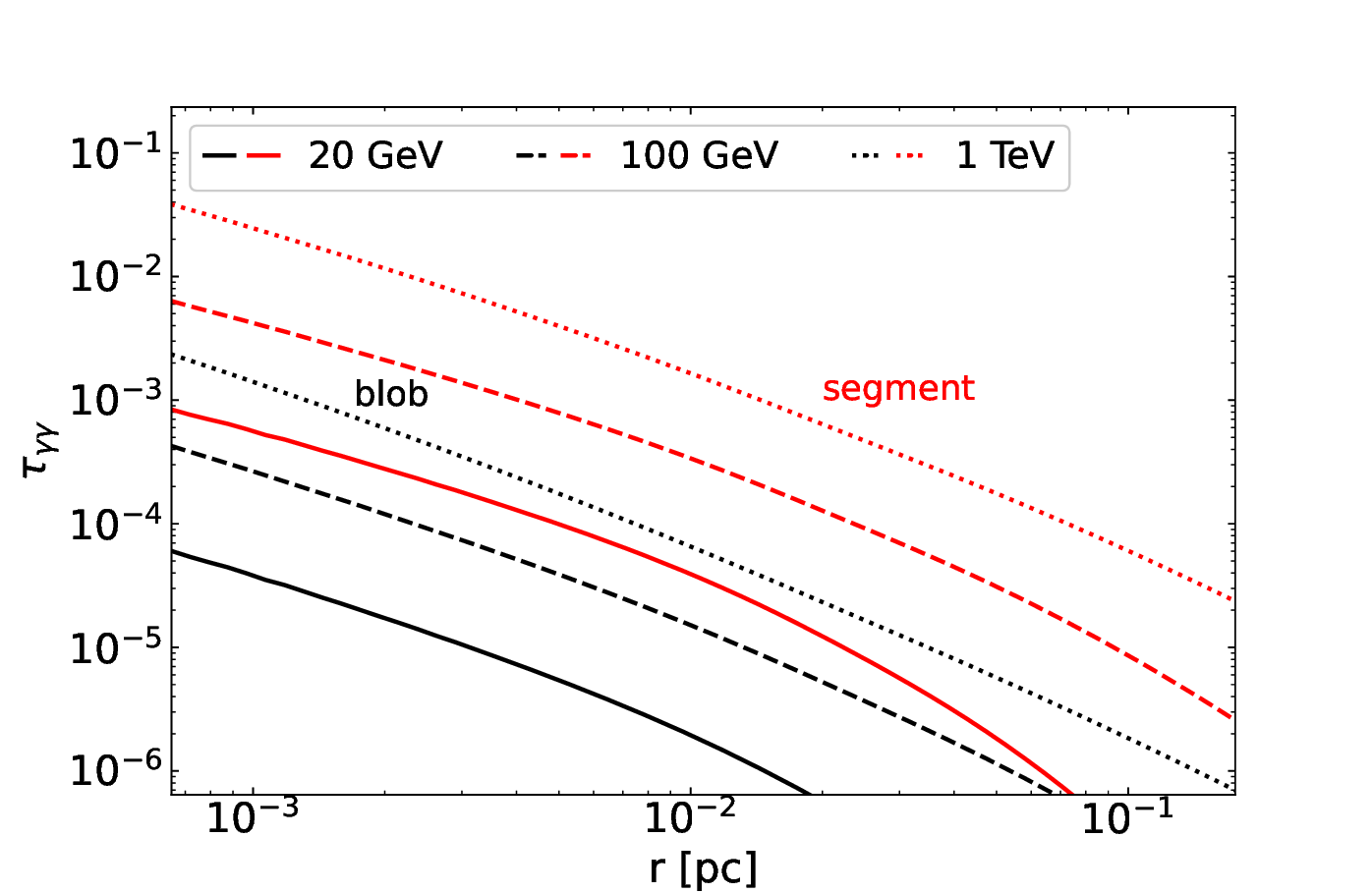}
\caption{The internal $\gamma \gamma$ opacity as a function of distance from the jet base. The black and red curves show the opacities in a blob and a segment, respectively. The dotted, dashed and solid curves represent the opacities at 1 TeV, 100 GeV and 20 GeV, respectively. Parameters are based on the benchmark HBL shown in Table~\ref{tab:parameters}.
\label{tau}}
\end{figure}

In one-zone models, the opacity for internal $\gamma\gamma$ absorption $\tau_{\gamma\gamma}$ increases as the Doppler factor $\delta_{\rm D}$ and the blob radius $R'$ decrease \citep[i.e. $\tau_{\gamma\gamma}\propto \delta_{\rm D}^{-4} {R'}^{-1}$,][]{2008ApJ...686..181F, 2019ApJ...871...81X}. Since the blob radius is limited by the observed minute-scale variability timescale $t_{\rm var}^{\rm obs}$ at a level of $R'\leq ct_{\rm var}^{\rm obs}\delta_{\rm D}$, an extremely large $\delta$ has to be introduced to make the dissipation zone transparent to GeV-TeV photons, as we have introduced in the beginning. As a result, the explanation of the minute-scale variability of several TeV HBLs, such as Mrk 501 \citep{2008MNRAS.384L..19B}, is probably problematic in the one-zone models. Note that minute-scale variability is also observed around 10\,GeV for FSRQs \citep{2016ApJ...824L..20A, 2018ApJ...854L..26S}. However, FSRQs do not always require an extremely large Doppler factor, since the target photons for the $\gamma\gamma$ absorption are sometimes in the valley between the two humps of the SEDs. In the following, we only show the internal $\gamma\gamma$ opacity of the HBL in our multi-zone model.
 
In our modeling, the jet's emission arises from the superposition of numerous blobs of different radii. To interpret the minute-scale variability, the size of the blobs contributing to the rapidly varying gamma-ray emission need to be {$R'<2.7\times 10^{14}\,$cm} for a Doppler factor of {$\delta_{\rm D}=15$}. Assuming the jet's opening angle to be $\theta=5^\circ$, the corresponding distance of the blob from the SMBH can be given by {$r<3\times10^{-3}(\delta_{\rm D}/15)(\theta/5^\circ)^{-1}(\kappa/0.3)^{-1}~\rm pc$}. Such a small distance is consistent with the result shown in Fig.~\ref{ave}, where we can see that the gamma-ray emission is mainly contributed by blobs at small distance. {Note that the blazar jet's opening angle could be as small as $1^\circ$ or so, according to the study of \citet{Clausen-Brown13} with the MOJAVE VLBI survey. For $\theta=1^\circ$, the blob size would be 5 times smaller at the same distance if all the other parameters are kept the same. We may then expect the presence of faster variability in the LC. On the other hand, if we put the distance of all the blobs 5 times larger than the case of $\theta=5^\circ$, we would get the same blob size. In this case, the radiation property of each blob would be the same if the magnetic field in the blob is also kept the same (i.e., via increasing $B(r)$ by 5 times with respect to the case of $\theta=5^\circ$).}

The target radiation field relevant for the internal $\gamma\gamma$ absorption in a blob is composed of two components with one being the radiation emitted by the blob itself $\tau_{\gamma\gamma}^{\rm blob}$ and the other being the radiation from other blobs in the current segment $\tau_{\gamma\gamma}^{\rm seg}$, which can be estimated, respectively \citep{2008ApJ...686..181F},
\begin{equation}\label{opacity-blob}
\tau_{\gamma\gamma}^{\rm blob}\approx f_{\rm b}(r,E)\frac{\sigma_{\gamma\gamma}D_{\rm L}^2E_{\gamma}\overline{\nu F_\nu}(E)}{2m_{\rm e}^2c^5R'\Gamma^5},
\end{equation}
and
\begin{equation}\label{opacity-seg}
\tau_{\gamma\gamma}^{\rm seg} \approx f_{\rm s}(r,E)\frac{\sigma_{\gamma\gamma}D_{\rm L}^2E_{\gamma}\overline{\nu F_\nu}(E, r)}{2m_{\rm e}^2c^5(R'/\kappa)\Gamma^5},
\end{equation}
where $E_\gamma$ is the energy of high-energy photons, $\sigma_{\gamma\gamma}\approx 1.68\times10^{-25}~\rm cm^2$ is the peak of the $\gamma \gamma$ pair-production cross section, and we used here the $\delta$-approximation for the cross section so that $EE_\gamma=2\delta_{\rm D}^2m_{\rm e}^2c^4$. The gamma-ray opacity of $\tau_{\gamma\gamma}^{\rm blob}$ and $\tau_{\gamma\gamma}^{\rm seg}$ for the benchmark HBL are shown in Fig.~\ref{tau}.

{In Eq.~(\ref{opacity-blob}), $f_{\rm b}(r,E)$ represents the fraction of the contribution of a single blob at distance $r$ to the blazar flux at energy $E$. If we set $f_{\rm b}=1$, the formula returns to that in one-zone models where the single blob contributes all the observed emission.} The value of $f_{\rm b}$ is generally much smaller than unity as long as the filling factor $\lambda$ is not too small. Taking the case of the HBL shown in Fig.~\ref{ave} for example, we find {$f_{\rm b}\approx 2.7\times10^{-5}\lambda^{-1}$} at the jet base (i.e., $r=r_0$) for X-ray photons with {$E\simeq 5.9\,$keV} (which absorb 20\,GeV photons), leading to 2500 times smaller opacity compared to that in the one-zone model {for $\lambda=7\%$}. For our benchmark parameters, we have $R'_0 = 5.2\times10^{13}~\rm cm$ for the radius of the blob at the jet base, {$\overline{\nu F_\nu}(E)\approx 3.2\times10^{-10}~\rm erg~s^{-1}~cm^{-2}$} for the flux of target photons of 5.9\,keV. One then can obtain the internal $\gamma \gamma$ absorption opacity of the blob at jet base to be {$\tau_{\gamma \gamma}^{\rm blob}(r_0)\approx 6.8\times10^{-5}$ for photons of $E_{\gamma}=20~\rm GeV$}. 

As for the opacity provided by the segment, i.e., Eq.~(\ref{opacity-seg}), $f_{\rm s}(r,E)$ represents the fraction of the blazar's emission contributed by the segment at distance $r$. {We can obtain the internal $\gamma \gamma$ opacity in the segment at the jet base to be $\tau_{\gamma\gamma}^{\rm seg}\simeq 10^{-3}$} for $E_\gamma=20$\,GeV photons with benchmark parameters, where the flux of target photons is  {$\overline{\nu F_\nu}(E, r_0)\approx 8.8\times10^{-12}~\rm erg~s^{-1}~cm^{-2}$} and the transverse radius of the segment is $R'_0/\kappa=1.7\times 10^{14}\,$cm.
We see that both of the opacities are much smaller than unity without invoking an extremely large Doppler factor. The reason is as follows: for the opacity in an individual blob, the luminosity is much smaller than that in one-zone models since each blob only contributes a small fraction of the entire jet's emission; for the opacity in a segment, the luminosity is closer to that in one-zone models but the size of the radiation zone is the jet's transverse radius instead of the radius of the compact blob. 

The filling factor $\lambda$ is an important quantity. A larger filling factor means that the jet's emission is shared among more blobs. Hence, the target radiation density inside each blob is low, leading to a smaller opacity, and vice versa. On the other hand, from the perspective of LCs, a
large filling factor could smooth out the variability due to the superposition of LCs of various blobs. To see the influence of different $\lambda$ on LCs, we compare the LCs of radio, optical, X-ray and $\gamma$-ray emission of the HBL, with different values of $\lambda$ in  Fig.~\ref{LC-mw}. As expected, the LC becomes more variable as $\lambda$ decreases. For $\lambda=1\%$, minute-scale variability in the 20\,GeV band with flux doubling can appear; for $\lambda=7\%$, we still can observe minute-scale variability, but the amplitude of the variation is generally less than a factor of 2; for $\lambda=20\%$, the LC becomes quite smooth.

It is worth mentioning the famous minute-scale flare of PKS 2155-304 in the TeV gamma-ray band with a ten-fold flux enhancement reported by \cite{2007ApJ...664L..71A}. {\cite{2008MNRAS.384L..19B} suggested that for a variability timescale of 5 minutes at TeV energies and a target low-energy radiation of luminosity $10^{46}~\rm erg~s^{-1}$, the required minimum Doppler factor is at least 50 in order to make the radiation zone transparent to 1~TeV photons in the one-zone model.} If we fix the Doppler factor of the blazar jet to be 30 based on observations \citep{2009A&A...494..527H}, we find that the opacity can be smaller than unity as long as $\lambda$ is larger than 0.5\%  in our model, i.e., no need to introduce an extremely high Doppler factor. {The reason for this difference can be understood as follows: in the one-zone model considered in previous literature, the size of the radiation zone must be very small to account for the 5-minute variability. Since all the blazar’s radiation is emitted from this compact radiation zone in the one-zone model by definition, the photon density is so intense that the TeV gamma-ray opacity is far greater than unity unless the Doppler factor is as high as 50 or larger. In our model, the radiation zones of the blazar spread in a more extended region along the jet and hence the photon density in the TeV gamma-ray flare zone is much smaller. This is reflected by the factor $f_{\rm b}$ and $f_{\rm s}$ in Eq.~\ref{opacity-blob} and Eq.~\ref{opacity-seg} respectively. Therefore, the TeV gamma-ray opacity may be smaller than unity while keeping a relatively small Doppler factor.}

{On the other hand, however, the challenge is to reproduce the ten-fold flux enhancement of TeV gamma-rays on minute timescales. From Fig.~\ref{LC-mw}, we do not see a ten-fold flux variation in our current model setup, which aims for blazars'  emissions in the {low} state. However, the same idea may nevertheless be applied to the minute-scale TeV flare, for example, by considering a significantly enhanced electron injection luminosity (possibly via an intense dissipation event occurs at the jet base, e.g., \citealt{2017ApJ...841...61A} or blobs passing through re-collimation shocks in the jet, e.g., \citealt{2009ApJ...699.1274B, 2016A&A...588A.101F}) or an increase in the Doppler factor of a blob or a few blobs (probably due to the random proper motion of the blob) at the jet base within a short period of time. The synchrotron radiation of electrons at the jet base may be suppressed due to the strong radiation field at the jet base. As such, it will not violate the non-detection of a simultaneous minute-scale optical/X-ray flare and the suppressed synchrotron radiation will not provide a strong opacity for gamma ray absorption. A detailed study on the application of the minute-scale TeV flare is beyond the scope of this work and will be investigated in a separate paper.}

\subsection{Polarization Variability}\label{polarization}
Blazars have been observed to be linearly polarized in the radio \citep[e.g.,][]{1992ApJ...399...16A, 2005AJ....130.1418J} and optical bands \citep[e.g.][]{1980ARA&A..18..321A, 1988ApJ...333..666I} for many decades, with variation in both degree of polarization (DoP) and the position angle (PA). Polarization, as a common feature of blazars, is therefore another important observable that can be used to diagnose the jet physics.

We investigate the linear polarization of the blazar jet predicted by our model, which is an outcome of the superposition of different blobs in the jet. Following \citet{2014ApJ...780...87M}, the DoP from many radiation zones can be given by:
\begin{equation}\label{DOP}
\Pi=\frac{\sqrt{\left({\sum_{i=1}^{i_{\rm max}}\sum_{j=1}^{N_i}{Q_{i,j}}}\right)^2+\left(\sum_{i=1}^{i_{\rm max}}\sum_{j=1}^{N_i}{U_{i,j}}\right)^2}}{F_{\nu}},
\end{equation}
and the PA is
\begin{equation}\label{PA}
\chi=\frac{1}{2}{\arctan{\frac{\sum_{i=1}^{i_{\rm max}}\sum_{j=1}^{N_i}{U_{i,j}}}{\sum_{i=1}^{i_{\rm max}}\sum_{j=1}^{N_i}{Q_{i,j}}}}},
\end{equation}
where $Q_{i,j}=F_{i,j}\cos{2\chi_{i,j}}$ and $U_{i,j}=F_{i,j}\sin{2\chi_{i,j}}$ are the Stokes parameters for each blob. $\chi_{i,j}$ is the PA of the electric field vector in the observer's plane of the sky measured from some reference direction for each blob. It can be given by \cite{2003ApJ...597..998L}
\begin{equation}\label{cosPA}
\cos{\chi_{i,j}}=\boldsymbol{e}\cdot(\boldsymbol{n}\times\boldsymbol{l}),
\ \sin{\chi_{i,j}}=\boldsymbol{e}\cdot\boldsymbol{l},
\end{equation}
where $\boldsymbol{e}$ is the unit vector along the direction of the electric field of a linearly polarized electromagnetic wave:
\begin{equation}\label{PA:e}
\boldsymbol{e}=\frac{\boldsymbol{n}\times\boldsymbol{q'}}{\sqrt{q'^{2}-(\boldsymbol{n}\cdot\boldsymbol{q'})^{2}}},
\end{equation}
where 
\begin{equation}\label{PA:q}
    \boldsymbol{q'}=\boldsymbol{B'}+\boldsymbol{n}\times(\boldsymbol{v}\times\boldsymbol{B'})-\frac{\Gamma}{1+\Gamma}(\boldsymbol{B'}\cdot\boldsymbol{v})\boldsymbol{v}
\end{equation}
We establish a Cartesian coordinate with the jet direction aligned with the z-axis and define the x-axis so that the observer's line of sight to the blazar is in the x-z plane. Then, we have the bulk velocity of the jet to be $\boldsymbol{v}=\beta\{0,\,0,\,1\}$, the unit vector in the direction of the line of sight to be $\boldsymbol{n}=\{\sin{\theta_{\rm obs}},\,0,\,\cos{\theta_{\rm obs}}\}$, and the unit vector normal to the plane determined by $\boldsymbol{n}$ and the direction of the projection of the z-axis to the plane of the sky to be $\boldsymbol{l}=\{0,\,1,\,0\}$. In our case, we assume that the magnetic field of the jet consists of two components, with one being the spiral magnetic field along the entire jet, and the other being the local magnetic field with a random orientation in each blob. The field strength of the two components are assumed to be equal. The unit vector of the ordered spiral component can be characterized by $\boldsymbol{B'_{1}}=\{-\sin{\psi'}\sin{\phi'},\,\sin{\psi'}\cos{\phi'},\,\cos{\psi'}\}$. We assume that the magnetic field pitch angle $\psi'=45^{\circ}$ (the angle between the jet's axis and the magnetic field) and the azimuthal angle in cylindrical coordinate system {$\phi'=q\times 360^{\circ}\left(r_{i}-r_{0}\right)/L+\Delta\phi'$ in the following calculation, where the factor $q$ means that the spiral magnetic field component rotates $q$ turns within the range of the jet of length $L$ which is arbitrarily chosen and can be changed. $\Delta\phi'$ is the phase of the toroidal component in a certain transverse plane of the jet. We assume that a blob can appear at any azimuthal direction with respect to the jet axis with an equal probability, and thus $\Delta\phi'$ is simply random in $0-360^\circ$, making $\phi'$ a random quantity in our calculation.} The unit vector of the random magnetic field component in certain blob can be written as $\boldsymbol{B'_{2}}=\{\sin{\tilde{\theta}}\cos{\tilde{\phi}},\,\sin{\tilde{\theta}}\sin{\tilde{\phi}},\,\cos{\tilde{\theta}}\}$, where $\tilde{\phi}$ is uniformly random in the range $(0,\,2\pi)$ and $\cos{\tilde{\theta}}$ is uniformly random in the range $(-1,\,1)$. Then the unit vector of the magnetic field of the jet $\boldsymbol{B'}=\{B_{\rm x}',\,B_{\rm y}',\,B_{\rm z}'\}=\left(\boldsymbol{B'_{1}}+\boldsymbol{B'_{2}}\right)/\Vert{\boldsymbol{B'_{1}}+\boldsymbol{B'_{2}}}\Vert$.

\begin{figure}
\centering
\subfigure{
\includegraphics[width=1\columnwidth]{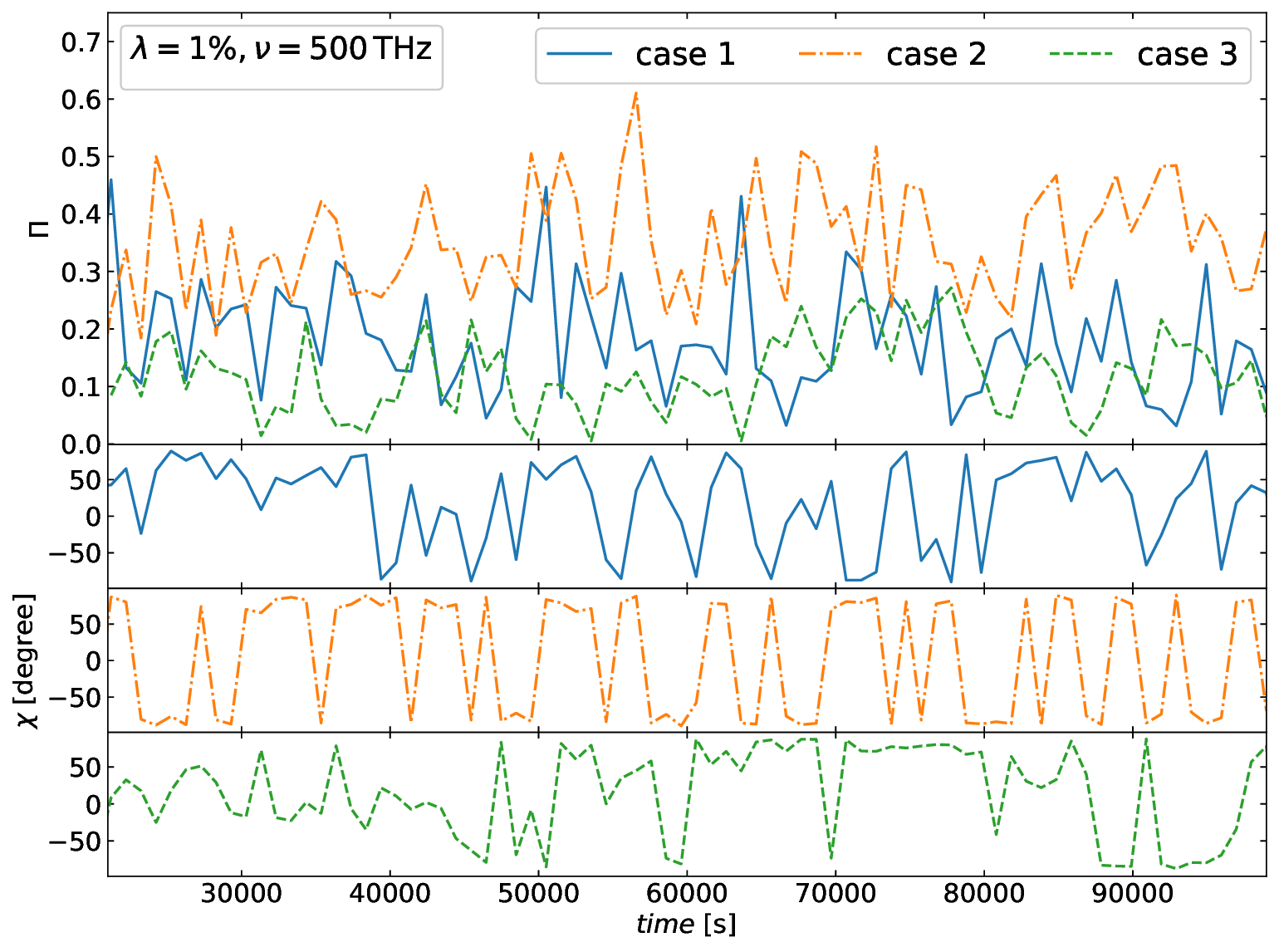}
}
\subfigure{
\includegraphics[width=1\columnwidth]{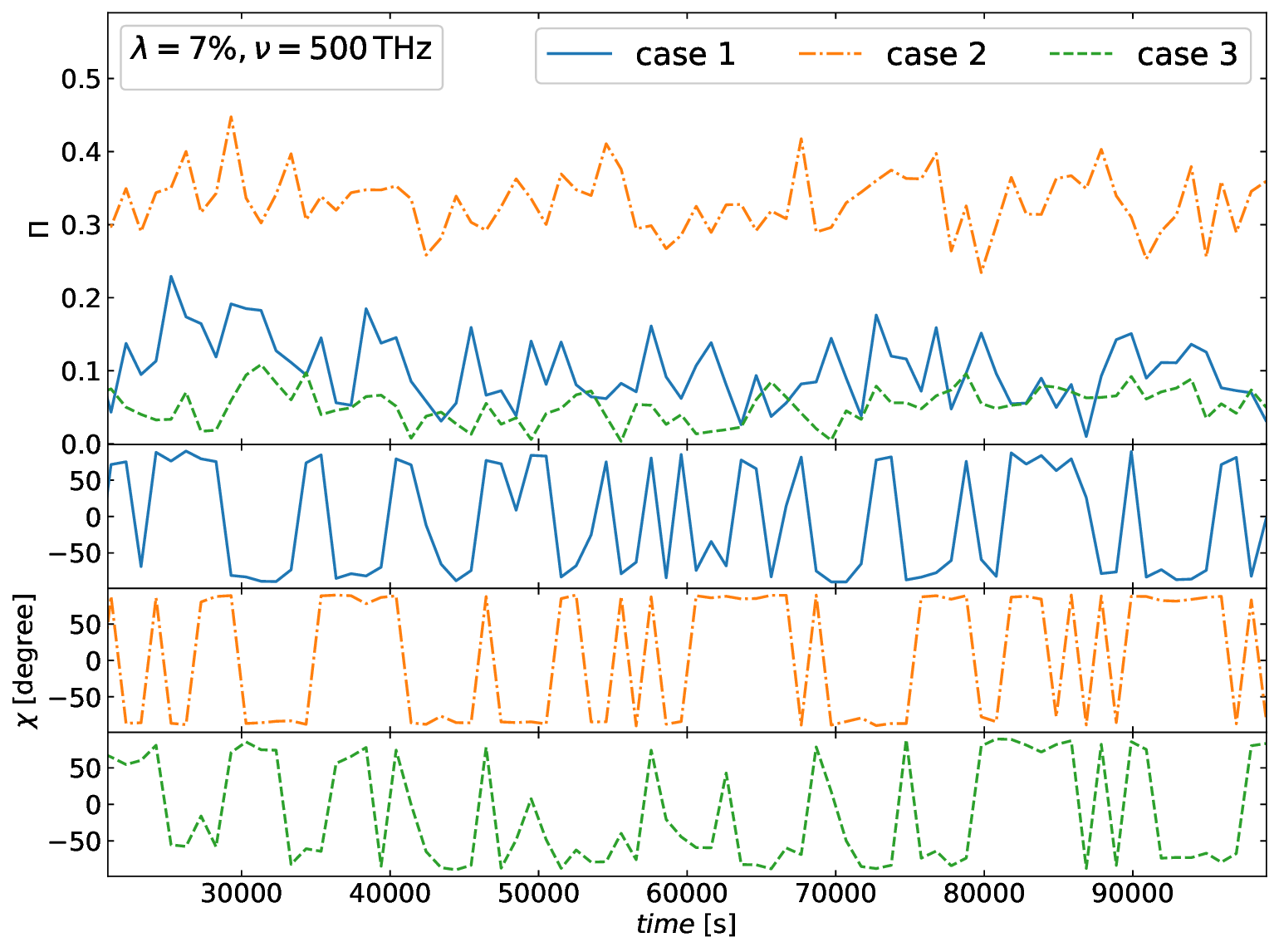}
}
\subfigure{
\includegraphics[width=1\columnwidth]{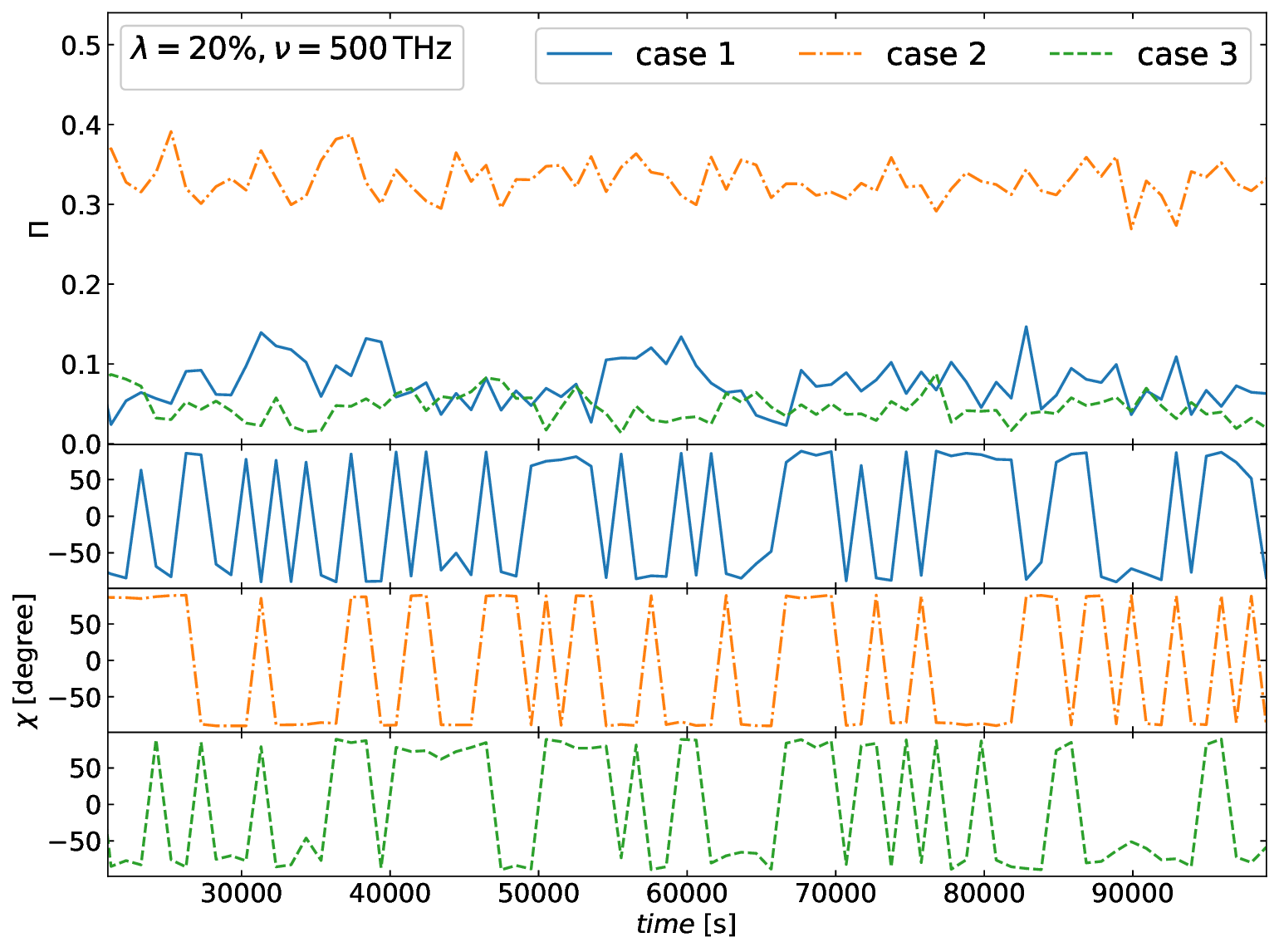}
}
\caption{{Temporal evolution of the DoP and the PA at 500\,THz for the HBL assuming different filling factor $\lambda$. Case 1 (blue curves) shows the result under the benchmark parameters as shown in Table \ref{parameters} with setting the pitch angle to be $\psi'=45^{\circ}$. Parameters used in other two cases are the same as those in case 1, except $\psi'=5^{\circ}$ for the case 2 (orange curves) and $\theta_{\rm obs}=10^{\circ}$ for the case 3 (green curves).
\label{fig:PA}}}
\end{figure}

Substituting $\boldsymbol{v},\,\boldsymbol{n},\,\boldsymbol{l}$ into Eq.~(\ref{cosPA}), we obtain:
\begin{align}\label{PA:full}
     \sin{\chi_{i,j}}&=\frac{\xi B'_{\rm x}-\zeta B'_{\rm z}}{\sqrt{\xi^{2}{B'_{\rm x}}^{2}+\Gamma^2\delta_{\rm D}^2{B'_{\rm y}}^{2}+\zeta^{2}{B'_{\rm z}}^{2}-2\zeta\xi\,B'_{\rm x}B'_{\rm z}}},\\
    \cos\chi_{i,j}&= \frac{\Gamma\delta_{\rm D}B'_{\rm y}}{\sqrt{\xi^{2}{B'_{\rm x}}^{2}+\Gamma^2\delta_{\rm D}^2{B'_{\rm y}}^{2}+\zeta^{2}{B'_{\rm z}}^{2}-2\zeta\xi\,B'_{\rm x}B'_{\rm z}}},
\end{align}    
where $\xi=\cos{\theta_{\rm obs}}-\beta$ and $\zeta=\left(1-\frac{\beta^2\Gamma}{\Gamma+1}\right)\sin\theta_{\rm obs}$.


Fig.~\ref{fig:PA} presents the temporal evolution of the expected DoP (the top panel of each subfigure) and PA (three lower panels) with different jet filling factors $\lambda$. Similar to LCs shown in Fig.~\ref{fig:LCr}, both the DoP and the PA present erratic variations, which is consistent with observations. The smaller $\lambda$ is, the more intense variations are likely to occur. Observations show that some blazars occasionally undergo large-amplitude swings of the PA sometimes even exceeding $180^\circ$ \citep{Marscher08, Marscher10, Abdo10}. This phenomenon may be explained in our model if an intense dissipation event occurs at some point dominating the overall polarization of the blazar.
The amplitude of the DoP depends on the viewing angle $\theta_{\rm obs}$ and the bulk Lorentz factor of the jet $\Gamma$. Also, it is proportional to the ratio of the regular magnetic field to the turbulent magnetic field, and hence its measurement may shed some light on the magnetic topology of the dissipation zones.

\section{Discussion and Conclusion}\label{DC}
Various observational phenomena of blazars are difficult to be explained by the conventional one-zone models, implying a more complex physical picture of the blazar jet. Great efforts have been made to push forward our understanding of blazars' radiation. In addition to the BK conical jet model as already introduced, a decelerating jet model \citep{2003ApJ...594L..27G} has been suggested to account for the additional seed photons needed in modeling the TeV spectral peak of some BL Lacs without invoking high Doppler factors; the spine-layer model \citep{2005A&A...432..401G} has been proposed as an alternative, motivated by the limb brightening morphology of some blazar jets; the ``jet-in-a-jet'' model \citep{2009MNRAS.395L..29G, 2010MNRAS.402.1649G} and jet-star/cloud collision models \citep{2012ApJ...749..119B} can help to explain the minute-scale gamma-ray variability; the ``synchrotron mirror'' model \citep{2005ApJ...621..176B} and the ``ring of fire'' model \citep{2017ApJ...850...87M} may interpret orphan flares; {the "leading blob" model \citep{2011ApJ...743L..19L} aims to reproduce the hardening of the observed gamma-ray spectra}; the inner-outer blob models \citep{2019PhRvD..99f3008L, 2019ApJ...886...23X, 2021ApJ...906...51X} and the neutral beam model \citep{2003ApJ...586...79A, 2020ApJ...889..118Z} may help to reproduce the possible associations between high-energy neutrinos and blazars, etc. 

We would like to point out that the physical picture of the stochastic dissipation model and other models is not mutually exclusive. In fact, our model can provide a framework compatible with others. For example, the inner-outer blob model can be seen as a simplified version of the stochastic dissipation model, with only two emitting blobs considered. On the other hand, the velocity of the jet may not be homogeneous. We may also introduce a decelerating velocity profile for the jet, which can provide an additional radiation field for the IC emission in the inner part of the jet. In the case that the outer boundary of the jet is slower than the jet's central region, the situation is similar to the spine-layer model. If blobs in the jet have their own internal motions relative to the jet's bulk motion, the treatment of the ``jet-in-a-jet'' model may be referred to. One may also combine the ``synchrotron mirror'' model and/or the ``ring-of-fire'' model with our model by assuming the existence of some external structures such as a reflecting cloud or a static synchrotron-emitting region. {In addition, if we consider that one or a few blobs could become much more active than the rest of the blobs, the situation would become similar to the ``leading blob'' model, and some special features predicted in the model such as a hard gamma-ray spectrum, may appear.} Many other models can be integrated into this framework as well. 

We note that our model has four more free parameters with respect to conventional one-zone models. While introducing additional parameters would make the model more flexible, it is challenging to constrain those parameters. Therefore, involving observational data of various aspects in the modelling, such as multiwavelength SED (including the radio band which is usually neglected in one-zone models), temporal variability, polarisation measurements, morphological measurements, multi-messenger observations would be useful to test the model and are necessary to acquire a comprehensive understanding of blazar jets.

In summary, we presented a time-dependent modeling on the multiwavelength emission of blazar jets in the framework of the stochastic dissipation model, aiming to describe features of blazars in the {low} state. In this model, the jet with a conical structure contains numerous spherical radiation zones (or blobs) at different distances along the jet. Relativistic electrons are injected into each blob, losing energy through non-thermal radiative cooling and adiabatic cooling processes. We assume that the probability of a dissipation event occurring at unit time (in the rest frame) and unit jet length at a distance $r$ follows a power law distribution $p(r)=Ar^{-\alpha}$. We then derived the average spectral energy distribution and light curves in various energy bands through summing up the emission from all the blobs that occurred in each segment, and calculated the expected temporal properties of the linear polarization of blazars. The model can reproduce basic features of a blazar's emission such as the double-humped SED, the flat radio spectrum, the erratic temporal variation in flux and DoP. Combined with other models, it can also account for phenomena in the flaring state of blazars that are sometimes difficult to be understood in the conventional one-zone models, such as the minute-scale GeV-TeV gamma-ray variability, orphan flares, and large swings of the polarisation angle. Application of the model to observational data will be explored in upcoming papers.

\section*{Acknowledgements}
We would like to thank the anonymous referee for constructive comments. This study is supported by the National Scientific Foundation of China (NSFC), under grants No.~U2031105. R.X. acknowledges the support by NSFC under Grant No. 12203043. Z.R.W. acknowledges the support by the NSFC under Grant No. 12203024 and the support by the Department of Science $\&$ Technology of Shandong Province under Grant No. ZR2022QA071.
\section*{Data Availability}

No new data were generated or analysed in support of this research.

\bibliographystyle{mnras}

\begin{thebibliography}{}
\bibitem[\protect\citeauthoryear{Abdo et al.}{2010}]{2010ApJ...716...30A} Abdo A.~A., Ackermann M., Agudo I., Ajello M., Aller H.~D., Aller M.~F., Angelakis E., et al., 2010, ApJ, 716, 30. doi:10.1088/0004-637X/716/1/30
\bibitem[\protect\citeauthoryear{Abdo et al.}{2010}]{Abdo10} Abdo A.~A., Ackermann M., Ajello M., Axelsson M., Baldini L., Ballet J., Barbiellini G., et al., 2010, Natur, 463, 919. doi:10.1038/nature08841
\bibitem[\protect\citeauthoryear{Abdo et al.}{2011}]{2011ApJ...736..131A} Abdo A.~A., Ackermann M., Ajello M., Baldini L., Ballet J., Barbiellini G., Bastieri D., et al., 2011, ApJ, 736, 131. doi:10.1088/0004-637X/736/2/131
\bibitem[\protect\citeauthoryear{Abeysekara et al.}{2020}]{2020ApJ...890...97A} Abeysekara A.~U., Benbow W., Bird R., Brill A., Brose R., Buchovecky M., Buckley J.~H., et al., 2020, ApJ, 890, 97. doi:10.3847/1538-4357/ab6612
\bibitem[\protect\citeauthoryear{Ackermann et al.}{2016}]{2016ApJ...824L..20A} Ackermann M., Anantua R., Asano K., Baldini L., Barbiellini G., Bastieri D., Becerra Gonzalez J., et al., 2016, ApJL, 824, L20. doi:10.3847/2041-8205/824/2/L20
\bibitem[\protect\citeauthoryear{Aharonian}{2000}]{2000NewA....5..377A} Aharonian F.~A., 2000, NewA, 5, 377. doi:10.1016/S1384-1076(00)00039-7
\bibitem[\protect\citeauthoryear{Aharonian et al.}{2007}]{2007ApJ...664L..71A} Aharonian F., Akhperjanian A.~G., Bazer-Bachi A.~R., Behera B., Beilicke M., Benbow W., Berge D., et al., 2007, ApJL, 664, L71. doi:10.1086/520635
\bibitem[Aharonian et al.(2017)]{2017ApJ...841...61A} Aharonian, F.~A., Barkov, M.~V., \& Khangulyan, D.\ 2017, \apj, 841, 61. doi:10.3847/1538-4357/aa7049
\bibitem[\protect\citeauthoryear{Albert et al.}{2007}]{2007ApJ...669..862A} Albert J., Aliu E., Anderhub H., Antoranz P., Armada A., Baixeras C., Barrio J.~A., et al., 2007, ApJ, 669, 862. doi:10.1086/521382
\bibitem[\protect\citeauthoryear{Aller, Aller, \& Hughes}{1992}]{1992ApJ...399...16A} Aller M.~F., Aller H.~D., Hughes P.~A., 1992, ApJ, 399, 16. doi:10.1086/171898
\bibitem[\protect\citeauthoryear{Angel \& Stockman}{1980}]{1980ARA&A..18..321A} Angel J.~R.~P., Stockman H.~S., 1980, ARA\&A, 18, 321. doi:10.1146/annurev.aa.18.090180.001541
\bibitem[\protect\citeauthoryear{Atoyan \& Dermer}{2003}]{2003ApJ...586...79A} Atoyan A.~M., Dermer C.~D., 2003, ApJ, 586, 79. doi:10.1086/346261
\bibitem[Barkov et al.(2012)]{2012ApJ...749..119B} Barkov, M.~V., Aharonian, F.~A., Bogovalov, S.~V., et al.\ 2012, \apj, 749, 119. doi:10.1088/0004-637X/749/2/119
\bibitem[\protect\citeauthoryear{Begelman, Fabian, \& Rees}{2008}]{2008MNRAS.384L..19B} Begelman M.~C., Fabian A.~C., Rees M.~J., 2008, MNRAS, 384, L19. doi:10.1111/j.1745-3933.2007.00413.x
\bibitem[\protect\citeauthoryear{Bennett et al.}{2014}]{2014ApJ...794..135B} Bennett C.~L., Larson D., Weiland J.~L., Hinshaw G., 2014, ApJ, 794, 135. doi:10.1088/0004-637X/794/2/135
\bibitem[\protect\citeauthoryear{Blandford \& Rees}{1978}]{1978PhyS...17..265B} Blandford R.~D., Rees M.~J., 1978, PhyS, 17, 265. doi:10.1088/0031-8949/17/3/020
\bibitem[\protect\citeauthoryear{Blandford \& K{\"o}nigl}{1979}]{1979ApJ...232...34B} Blandford R.~D., K{\"o}nigl A., 1979, ApJ, 232, 34. doi:10.1086/157262
\bibitem[\protect\citeauthoryear{B{\"o}ttcher \& Chiang}{2002}]{2002ApJ...581..127B} B{\"o}ttcher M., Chiang J., 2002, ApJ, 581, 127. doi:10.1086/344155
\bibitem[\protect\citeauthoryear{B{\"o}ttcher \& Reimer}{2004}]{2004ApJ...609..576B} B{\"o}ttcher M., Reimer A., 2004, ApJ, 609, 576. doi:10.1086/421320
\bibitem[B{\"o}ttcher(2005)]{2005ApJ...621..176B} B{\"o}ttcher, M.\ 2005, \apj, 621, 176. doi:10.1086/427430
\bibitem[\protect\citeauthoryear{B{\"o}ttcher et al.}{2013}]{2013ApJ...768...54B} B{\"o}ttcher M., Reimer A., Sweeney K., Prakash A., 2013, ApJ, 768, 54. doi:10.1088/0004-637X/768/1/54
\bibitem[\protect\citeauthoryear{B{\"o}ttcher}{2019}]{2019Galax...7...20B} B{\"o}ttcher M., 2019, Galax, 7, 20. doi:10.3390/galaxies7010020
\bibitem[\protect\citeauthoryear{Boula \& Mastichiadis}{2022}]{2022A&A...657A..20B} Boula S., Mastichiadis A., 2022, A\&A, 657, A20. doi:10.1051/0004-6361/202142126
\bibitem[\protect\citeauthoryear{B{\l}a{\.z}ejowski et al.}{2000}]{2000ApJ...545..107B} B{\l}a{\.z}ejowski M., Sikora M., Moderski R., Madejski G.~M., 2000, ApJ, 545, 107. doi:10.1086/317791
\bibitem[\protect\citeauthoryear{Bromberg \& Levinson}{2009}]{2009ApJ...699.1274B} Bromberg O., Levinson A., 2009, ApJ, 699, 1274. doi:10.1088/0004-637X/699/2/1274
\bibitem[\protect\citeauthoryear{Cerruti et al.}{2013}]{2013ApJ...771L...4C} Cerruti M., Dermer C.~D., Lott B., Boisson C., Zech A., 2013, ApJL, 771, L4. doi:10.1088/2041-8205/771/1/L4
\bibitem[\protect\citeauthoryear{Chatterjee et al.}{2019}]{2019MNRAS.490.2200C} Chatterjee K., Liska M., Tchekhovskoy A., Markoff S.~B., 2019, MNRAS, 490, 2200. doi:10.1093/mnras/stz2626
\bibitem[\protect\citeauthoryear{Chen \& Zhang}{2021}]{2021ApJ...906..105C} Chen L., Zhang B., 2021, ApJ, 906, 105. doi:10.3847/1538-4357/abc42d
\bibitem[\protect\citeauthoryear{Christie et al.}{2019}]{2019MNRAS.482...65C} Christie I.~M., Petropoulou M., Sironi L., Giannios D., 2019, MNRAS, 482, 65. doi:10.1093/mnras/sty2636
\bibitem[\protect\citeauthoryear{Clausen-Brown et al.}{2013}]{Clausen-Brown13} Clausen-Brown E., Savolainen T., Pushkarev A.~B., Kovalev Y.~Y., Zensus J.~A., 2013, A\&A, 558, A144. doi:10.1051/0004-6361/201322203
\bibitem[\protect\citeauthoryear{Cleary et al.}{2007}]{2007ApJ...660..117C} Cleary K., Lawrence C.~R., Marshall J.~A., Hao L., Meier D., 2007, ApJ, 660, 117. doi:10.1086/511969
\bibitem[\protect\citeauthoryear{Costamante et al.}{2018}]{2018MNRAS.477.4749C} Costamante L., Cutini S., Tosti G., Antolini E., Tramacere A., 2018, MNRAS, 477, 4749. doi:10.1093/mnras/sty887
\bibitem[\protect\citeauthoryear{Deng et al.}{2021}]{2021MNRAS.506.5764D} Deng X.-J., Xue R., Wang Z.-R., Xi S.-Q., Xiao H.-B., Du L.-M., Xie Z.-H., 2021, MNRAS, 506, 5764. doi:10.1093/mnras/stab2095
\bibitem[\protect\citeauthoryear{Dermer \& Schlickeiser}{1993}]{1993ApJ...416..458D} Dermer C.~D., Schlickeiser R., 1993, ApJ, 416, 458. doi:10.1086/173251
\bibitem[\protect\citeauthoryear{Dermer et al.}{2009}]{2009ApJ...692...32D} Dermer C.~D., Finke J.~D., Krug H., B{\"o}ttcher M., 2009, ApJ, 692, 32. doi:10.1088/0004-637X/692/1/32
\bibitem[\protect\citeauthoryear{Ding et al.}{2017}]{2017MNRAS.464..599D} Ding N., Zhang X., Xiong D.~R., Zhang H.~J., 2017, MNRAS, 464, 599. doi:10.1093/mnras/stw2347
\bibitem[\protect\citeauthoryear{Fabricius et al.}{2016}]{2016A&A...595A...3F} Fabricius C., Bastian U., Portell J., Casta{\~n}eda J., Davidson M., Hambly N.~C., Clotet M., et al., 2016, A\&A, 595, A3. doi:10.1051/0004-6361/201628643
\bibitem[\protect\citeauthoryear{Finke, Dermer, \& B{\"o}ttcher}{2008}]{2008ApJ...686..181F} Finke J.~D., Dermer C.~D., B{\"o}ttcher M., 2008, ApJ, 686, 181. doi:10.1086/590900
\bibitem[\protect\citeauthoryear{Finke}{2019}]{2019ApJ...870...28F} Finke J.~D., 2019, ApJ, 870, 28. doi:10.3847/1538-4357/aaf00c
\bibitem[\protect\citeauthoryear{Fromm et al.}{2016}]{2016A&A...588A.101F} Fromm C.~M., Perucho M., Mimica P., Ros E., 2016, A\&A, 588, A101. doi:10.1051/0004-6361/201527139
\bibitem[\protect\citeauthoryear{Gao, Pohl, \& Winter}{2017}]{2017ApJ...843..109G} Gao S., Pohl M., Winter W., 2017, ApJ, 843, 109. doi:10.3847/1538-4357/aa7754
\bibitem[\protect\citeauthoryear{Georganopoulos \& Kazanas}{2003}]{2003ApJ...594L..27G} Georganopoulos M., Kazanas D., 2003, ApJL, 594, L27. doi:10.1086/378557
\bibitem[\protect\citeauthoryear{Ghisellini, Tavecchio, \& Chiaberge}{2005}]{2005A&A...432..401G} Ghisellini G., Tavecchio F., Chiaberge M., 2005, A\&A, 432, 401. doi:10.1051/0004-6361:20041404
\bibitem[\protect\citeauthoryear{Ghisellini, Tavecchio, \& Ghirlanda}{2009}]{2009MNRAS.399.2041G} Ghisellini G., Tavecchio F., Ghirlanda G., 2009, MNRAS, 399, 2041. doi:10.1111/j.1365-2966.2009.15397.x
\bibitem[\protect\citeauthoryear{Ghisellini et al.}{2014}]{2014Natur.515..376G} Ghisellini G., Tavecchio F., Maraschi L., Celotti A., Sbarrato T., 2014, Natur, 515, 376. doi:10.1038/nature13856
\bibitem[\protect\citeauthoryear{Ghisellini et al.}{2010}]{2010MNRAS.402..497G} Ghisellini G., Tavecchio F., Foschini L., Ghirlanda G., Maraschi L., Celotti A., 2010, MNRAS, 402, 497. doi:10.1111/j.1365-2966.2009.15898.x
\bibitem[\protect\citeauthoryear{Ghisellini \& Tavecchio}{2008}]{2008MNRAS.387.1669G} Ghisellini G., Tavecchio F., 2008, MNRAS, 387, 1669. doi:10.1111/j.1365-2966.2008.13360.x
\bibitem[\protect\citeauthoryear{Giannios, Uzdensky, \& Begelman}{2009}]{2009MNRAS.395L..29G} Giannios D., Uzdensky D.~A., Begelman M.~C., 2009, MNRAS, 395, L29. doi:10.1111/j.1745-3933.2009.00635.x
\bibitem[\protect\citeauthoryear{Giannios, Uzdensky, \& Begelman}{2010}]{2010MNRAS.402.1649G} Giannios D., Uzdensky D.~A., Begelman M.~C., 2010, MNRAS, 402, 1649. doi:10.1111/j.1365-2966.2009.16045.x
\bibitem[\protect\citeauthoryear{Hada et al.}{2011}]{2011Natur.477..185H} Hada K., Doi A., Kino M., Nagai H., Hagiwara Y., Kawaguchi N., 2011, Natur, 477, 185. doi:10.1038/nature10387
\bibitem[\protect\citeauthoryear{Hayashida et al.}{2012}]{2012ApJ...754..114H} Hayashida M., Madejski G.~M., Nalewajko K., Sikora M., Wehrle A.~E., Ogle P., Collmar W., et al., 2012, ApJ, 754, 114. doi:10.1088/0004-637X/754/2/114
\bibitem[\protect\citeauthoryear{Hayashida et al.}{2015}]{2015ApJ...807...79H} Hayashida M., Nalewajko K., Madejski G.~M., Sikora M., Itoh R., Ajello M., Blandford R.~D., et al., 2015, ApJ, 807, 79. doi:10.1088/0004-637X/807/1/79
\bibitem[\protect\citeauthoryear{Hovatta et al.}{2009}]{2009A&A...494..527H} Hovatta T., Valtaoja E., Tornikoski M., L{\"a}hteenm{\"a}ki A., 2009, A\&A, 494, 527. doi:10.1051/0004-6361:200811150
\bibitem[\protect\citeauthoryear{Impey \& Tapia}{1988}]{1988ApJ...333..666I} Impey C.~D., Tapia S., 1988, ApJ, 333, 666. doi:10.1086/166775
\bibitem[\protect\citeauthoryear{Jones, O'dell, \& Stein}{1974}]{1974ApJ...188..353J} Jones T.~W., O'dell S.~L., Stein W.~A., 1974, ApJ, 188, 353. doi:10.1086/152724
\bibitem[\protect\citeauthoryear{Jorstad et al.}{2005}]{2005AJ....130.1418J} Jorstad S.~G., Marscher A.~P., Lister M.~L., Stirling A.~M., Cawthorne T.~V., Gear W.~K., G{\'o}mez J.~L., et al., 2005, AJ, 130, 1418. doi:10.1086/444593
\bibitem[\protect\citeauthoryear{Kaiser}{2006}]{2006MNRAS.367.1083K} Kaiser C.~R., 2006, MNRAS, 367, 1083. doi:10.1111/j.1365-2966.2006.10030.x
\bibitem[\protect\citeauthoryear{Komissarov et al.}{2007}]{2007MNRAS.380...51K} Komissarov S.~S., Barkov M.~V., Vlahakis N., K{\"o}nigl A., 2007, MNRAS, 380, 51. doi:10.1111/j.1365-2966.2007.12050.x
\bibitem[\protect\citeauthoryear{Kovalev et al.}{2007}]{2007ApJ...668L..27K} Kovalev Y.~Y., Lister M.~L., Homan D.~C., Kellermann K.~I., 2007, ApJL, 668, L27. doi:10.1086/522603
\bibitem[\protect\citeauthoryear{Lefa, Aharonian, \& Rieger}{2011}]{2011ApJ...743L..19L} Lefa E., Aharonian F.~A., Rieger F.~M., 2011, ApJL, 743, L19. doi:10.1088/2041-8205/743/1/L19
\bibitem[\protect\citeauthoryear{Li \& Kusunose}{2000}]{2000ApJ...536..729L} Li H., Kusunose M., 2000, ApJ, 536, 729. doi:10.1086/308960
\bibitem[\protect\citeauthoryear{Li, Yuan, \& Wang}{2017}]{2017MNRAS.468.2552L} Li Y.-P., Yuan F., Wang Q.~D., 2017, MNRAS, 468, 2552. doi:10.1093/mnras/stx655
\bibitem[\protect\citeauthoryear{Liodakis et al.}{2019}]{2019ApJ...880...32L} Liodakis I., Romani R.~W., Filippenko A.~V., Kocevski D., Zheng W., 2019, ApJ, 880, 32. doi:10.3847/1538-4357/ab26b7
\bibitem[\protect\citeauthoryear{Liu et al.}{2019}]{2019PhRvD..99f3008L} Liu R.-Y., Wang K., Xue R., Taylor A.~M., Wang X.-Y., Li Z., Yan H., 2019, PhRvD, 99, 063008. doi:10.1103/PhysRevD.99.063008
\bibitem[Lyutikov et al.(2003)]{2003ApJ...597..998L} Lyutikov, M., Pariev, V.~I., \& Blandford, R.~D.\ 2003, \apj, 597, 998. doi:10.1086/378497
\bibitem[\protect\citeauthoryear{Lyutikov, Pariev, \& Gabuzda}{2005}]{2005MNRAS.360..869L} Lyutikov M., Pariev V.~I., Gabuzda D.~C., 2005, MNRAS, 360, 869. doi:10.1111/j.1365-2966.2005.08954.x
\bibitem[\protect\citeauthoryear{Lyutikov \& Lister}{2010}]{2010ApJ...722..197L} Lyutikov M., Lister M., 2010, ApJ, 722, 197. doi:10.1088/0004-637X/722/1/197
\bibitem[MacDonald et al.(2017)]{2017ApJ...850...87M} MacDonald, N.~R., Jorstad, S.~G., \& Marscher, A.~P.\ 2017, \apj, 850, 87. doi:10.3847/1538-4357/aa92c8
\bibitem[\protect\citeauthoryear{MAGIC Collaboration et al.}{2019}]{2019A&A...623A.175M} MAGIC Collaboration, Acciari V.~A., Ansoldi S., Antonelli L.~A., Arbet Engels A., Baack D., Babi{\'c} A., et al., 2019, A\&A, 623, A175. doi:10.1051/0004-6361/201834010
\bibitem[\protect\citeauthoryear{MAGIC Collaboration et al.}{2021}]{2021A&A...655A..89M} MAGIC Collaboration, Acciari V.~A., Ansoldi S., Antonelli L.~A., Arbet Engels A., Artero M., Asano K., et al., 2021, A\&A, 655, A89. doi:10.1051/0004-6361/202141004
\bibitem[\protect\citeauthoryear{Marcowith et al.}{2020}]{2020LRCA....6....1M} Marcowith A., Ferrand G., Grech M., Meliani Z., Plotnikov I., Walder R., 2020, LRCA, 6, 1. doi:10.1007/s41115-020-0007-6
\bibitem[\protect\citeauthoryear{Marscher et al.}{2003}]{2003heba.conf..173M} Marscher A.~P., Jorstad S.~G., McHardy I.~M., Aller M.~F., Balonek T.~J., Villata M., Raiteri C.~M., et al., 2003, heba.conf, 299, 173
\bibitem[\protect\citeauthoryear{Marscher \& Gear}{1985}]{1985ApJ...298..114M} Marscher A.~P., Gear W.~K., 1985, ApJ, 298, 114. doi:10.1086/163592
\bibitem[\protect\citeauthoryear{Marscher et al.}{2008}]{Marscher08} Marscher A.~P., Jorstad S.~G., D'Arcangelo F.~D., Smith P.~S., Williams G.~G., Larionov V.~M., Oh H., et al., 2008, Natur, 452, 966. doi:10.1038/nature06895
\bibitem[\protect\citeauthoryear{Marscher}{2010}]{Marscher10} Marschern A.~P., 2010, LNP, 173. doi:10.1007/978-3-540-76937-8\_7
\bibitem[\protect\citeauthoryear{Marscher \& Jorstad}{2011}]{2011ApJ...729...26M} Marscher A.~P., Jorstad S.~G., 2011, ApJ, 729, 26. doi:10.1088/0004-637X/729/1/26
\bibitem[Marscher(2014)]{2014ApJ...780...87M} Marscher, A.~P.\ 2014, \apj, 780, 87. doi:10.1088/0004-637X/780/1/87
\bibitem[\protect\citeauthoryear{Mastichiadis \& Kirk}{1997}]{1997A&A...320...19M} Mastichiadis A., Kirk J.~G., 1997, A\&A, 320, 19
\bibitem[Medina-Torrej{\'o}n et al.(2021)]{2021ApJ...908..193M} Medina-Torrej{\'o}n, T.~E., de Gouveia Dal Pino, E.~M., Kadowaki, L.~H.~S., et al.\ 2021, \apj, 908, 193. doi:10.3847/1538-4357/abd6c2
\bibitem[\protect\citeauthoryear{Meng, Lin, \& Yuan}{2015}]{2015RAA....15..207M} Meng Y., Lin J., Yuan F., 2015, RAA, 15, 207-214. doi:10.1088/1674-4527/15/2/005
\bibitem[\protect\citeauthoryear{Mertens et al.}{2016}]{2016A&A...595A..54M} Mertens F., Lobanov A.~P., Walker R.~C., Hardee P.~E., 2016, A\&A, 595, A54. doi:10.1051/0004-6361/201628829
\bibitem[\protect\citeauthoryear{Nathanail et al.}{2020}]{2020MNRAS.495.1549N} Nathanail A., Fromm C.~M., Porth O., Olivares H., Younsi Z., Mizuno Y., Rezzolla L., 2020, MNRAS, 495, 1549. doi:10.1093/mnras/staa1165
\bibitem[\protect\citeauthoryear{O'Sullivan \& Gabuzda}{2009}]{2009MNRAS.400...26O} O'Sullivan S.~P., Gabuzda D.~C., 2009, MNRAS, 400, 26. doi:10.1111/j.1365-2966.2009.15428.x
\bibitem[\protect\citeauthoryear{Paliya et al.}{2017}]{2017ApJ...851...33P} Paliya V.~S., Marcotulli L., Ajello M., Joshi M., Sahayanathan S., Rao A.~R., Hartmann D., 2017, ApJ, 851, 33. doi:10.3847/1538-4357/aa98e1
\bibitem[Patel et al.(2021)]{2021JHEAp..29...31P} Patel, S.~R., Bose, D., Gupta, N., et al.\ 2021, Journal of High Energy Astrophysics, 29, 31. doi:10.1016/j.jheap.2020.12.001
\bibitem[\protect\citeauthoryear{Petropoulou}{2014}]{2014A&A...571A..83P} Petropoulou M., 2014, A\&A, 571, A83. doi:10.1051/0004-6361/201424603
\bibitem[\protect\citeauthoryear{Petropoulou, Vasilopoulos, \& Giannios}{2017}]{2017MNRAS.464.2213P} Petropoulou M., Vasilopoulos G., Giannios D., 2017, MNRAS, 464, 2213. doi:10.1093/mnras/stw2453
\bibitem[\protect\citeauthoryear{Petropoulou, Giannios, \& Sironi}{2016}]{2016MNRAS.462.3325P} Petropoulou M., Giannios D., Sironi L., 2016, MNRAS, 462, 3325. doi:10.1093/mnras/stw1832
\bibitem[\protect\citeauthoryear{Planck Collaboration et al.}{2011}]{2011A&A...536A..15P} Planck Collaboration, Aatrokoski J., Ade P.~A.~R., Aghanim N., Aller H.~D., Aller M.~F., Angelakis E., et al., 2011, A\&A, 536, A15. doi:10.1051/0004-6361/201116466
\bibitem[\protect\citeauthoryear{Plavin, Kovalev, \& Petrov}{2019}]{2019ApJ...871..143P} Plavin A.~V., Kovalev Y.~Y., Petrov L.~Y., 2019, ApJ, 871, 143. doi:10.3847/1538-4357/aaf650
\bibitem[\protect\citeauthoryear{Potter \& Cotter}{2013}]{2013MNRAS.429.1189P} Potter W.~J., Cotter G., 2013, MNRAS, 429, 1189. doi:10.1093/mnras/sts407
\bibitem[\protect\citeauthoryear{Potter \& Cotter}{2015}]{2015MNRAS.453.4070P} Potter W.~J., Cotter G., 2015, MNRAS, 453, 4070. doi:10.1093/mnras/stv1657
\bibitem[\protect\citeauthoryear{Potter \& Cotter}{2012}]{2012MNRAS.423..756P} Potter W.~J., Cotter G., 2012, MNRAS, 423, 756. doi:10.1111/j.1365-2966.2012.20918.x
\bibitem[\protect\citeauthoryear{Rees}{1967}]{1967MNRAS.137..429R} Rees M.~J., 1967, MNRAS, 137, 429. doi:10.1093/mnras/137.4.429
\bibitem[\protect\citeauthoryear{Rybicki \& Lightman}{1986}]{1986rpa..book.....R} Rybicki G.~B., Lightman A.~P., 1986, rpa..book, 400
\bibitem[\protect\citeauthoryear{Schlickeiser \& Ruppel}{2010}]{2010NJPh...12c3044S} Schlickeiser R., Ruppel J., 2010, NJPh, 12, 033044. doi:10.1088/1367-2630/12/3/033044
\bibitem[\protect\citeauthoryear{Shukla et al.}{2018}]{2018ApJ...854L..26S} Shukla A., Mannheim K., Patel S.~R., Roy J., Chitnis V.~R., Dorner D., Rao A.~R., et al., 2018, ApJL, 854, L26. doi:10.3847/2041-8213/aaacca
\bibitem[\protect\citeauthoryear{Sironi, Giannios, \& Petropoulou}{2016}]{2016MNRAS.462...48S} Sironi L., Giannios D., Petropoulou M., 2016, MNRAS, 462, 48. doi:10.1093/mnras/stw1620
\bibitem[\protect\citeauthoryear{Sokolovsky et al.}{2011}]{2011A&A...532A..38S} Sokolovsky K.~V., Kovalev Y.~Y., Pushkarev A.~B., Lobanov A.~P., 2011, A\&A, 532, A38. doi:10.1051/0004-6361/201016072
\bibitem[\protect\citeauthoryear{Tan et al.}{2020}]{2020ApJS..248...27T} Tan C., Xue R., Du L.-M., Xi S.-Q., Wang Z.-R., Xie Z.-H., 2020, ApJS, 248, 27. doi:10.3847/1538-4365/ab8cc6
\bibitem[\protect\citeauthoryear{Tavecchio \& Ghisellini}{2008}]{2008MNRAS.386..945T} Tavecchio F., Ghisellini G., 2008, MNRAS, 386, 945. doi:10.1111/j.1365-2966.2008.13072.x
\bibitem[\protect\citeauthoryear{Tramacere et al.}{2009}]{2009A&A...501..879T} Tramacere A., Giommi P., Perri M., Verrecchia F., Tosti G., 2009, A\&A, 501, 879. doi:10.1051/0004-6361/200810865
\bibitem[\protect\citeauthoryear{Tramacere et al.}{2021}]{2021arXiv211203941T} Tramacere A., Sliusar V., Walter R., Jurysek J., Balbo M., 2021, arXiv, arXiv:2112.03941
\bibitem[\protect\citeauthoryear{Urry \& Padovani}{1995}]{1995PASP..107..803U} Urry C.~M., Padovani P., 1995, PASP, 107, 803. doi:10.1086/133630
\bibitem[\protect\citeauthoryear{Wang, Li, \& Xue}{2004}]{2004ApJ...617..113W} Wang J., Li H., Xue L., 2004, ApJ, 617, 113. doi:10.1086/425354
\bibitem[\protect\citeauthoryear{Wang et al.}{2017}]{2017Ap&SS.362..189W} Wang Z., Xue R., Xie Z., Du L., Yi T., Xu Y., Liu W., 2017, Ap\&SS, 362, 189. doi:10.1007/s10509-017-3164-2
\bibitem[\protect\citeauthoryear{Wang et al.}{2022}]{2022PhRvD1053005W} Wang Z.-R., Liu R.-Y., Petropoulou M., Oikonomou F., Xue R., Wang X.-Y., 2022, PhRvD, 105, 023005. doi:10.1103/PhysRevD.105.023005
\bibitem[\protect\citeauthoryear{Xue et al.}{2019a}]{2019ApJ...871...81X} Xue R., Liu R.-Y., Wang X.-Y., Yan H., B{\"o}ttcher M., 2019, ApJ, 871, 81. doi:10.3847/1538-4357/aaf720
\bibitem[\protect\citeauthoryear{Xue et al.}{2019b}]{2019ApJ...886...23X} Xue R., Liu R.-Y., Petropoulou M., Oikonomou F., Wang Z.-R., Wang K., Wang X.-Y., 2019, ApJ, 886, 23. doi:10.3847/1538-4357/ab4b44
\bibitem[\protect\citeauthoryear{Xue et al.}{2021}]{2021ApJ...906...51X} Xue R., Liu R.-Y., Wang Z.-R., Ding N., Wang X.-Y., 2021, ApJ, 906, 51. doi:10.3847/1538-4357/abc886
\bibitem[\protect\citeauthoryear{Yan, Zeng, \& Zhang}{2014}]{2014MNRAS.439.2933Y} Yan D., Zeng H., Zhang L., 2014, MNRAS, 439, 2933. doi:10.1093/mnras/stu146
\bibitem[\protect\citeauthoryear{Yuan et al.}{2009}]{2009MNRAS.395.2183Y} Yuan F., Lin J., Wu K., Ho L.~C., 2009, MNRAS, 395, 2183. doi:10.1111/j.1365-2966.2009.14673.x
\bibitem[\protect\citeauthoryear{Yuan \& Zhang}{2012}]{2012ApJ...757...56Y} Yuan F., Zhang B., 2012, ApJ, 757, 56. doi:10.1088/0004-637X/757/1/56
\bibitem[\protect\citeauthoryear{Zabalza}{2015}]{2015ICRC...34..922Z} Zabalza V., 2015, ICRC, 34, 922
\bibitem[\protect\citeauthoryear{Zdziarski, Stawarz, \& Sikora}{2019}]{2019MNRAS.485.1210Z} Zdziarski A.~A., Stawarz {\L}., Sikora M., 2019, MNRAS, 485, 1210. doi:10.1093/mnras/stz475
\bibitem[\protect\citeauthoryear{Zhang et al.}{2012}]{2012ApJ...752..157Z} Zhang J., Liang E.-W., Zhang S.-N., Bai J.~M., 2012, ApJ, 752, 157. doi:10.1088/0004-637X/752/2/157
\bibitem[\protect\citeauthoryear{Zhang et al.}{2020}]{2020ApJ...889..118Z} Zhang B.~T., Petropoulou M., Murase K., Oikonomou F., 2020, ApJ, 889, 118. doi:10.3847/1538-4357/ab659a
\bibitem[\protect\citeauthoryear{Zhang et al.}{2020}]{2020PASJ...72...44Z} Zhang H.-M., Wang Z.-J., Zhang J., Yi T.-F., Chen L., Lu R.-J., Liang E.-W., 2020, PASJ, 72, 44. doi:10.1093/pasj/psaa029
\end{thebibliography}

\appendix
\section{Dependence of the low-energy SED shape on the dissipation probability distribution}\label{appA}
In Section \ref{sec:sed} we mentioned that we need $\alpha=2$ in order to reproduce the flat radio spectra. This can be understood by considering the emission in the {low} state as follows. Note that the high-frequency IC spectral shape can be studied in this way as well, but due to the contribution from various target radiation fields and the KN effect, the dependence is not straightforward, and hence we do not study it here.

\begin{figure*}
\centering
\subfigure{
\includegraphics[width=0.9\columnwidth]{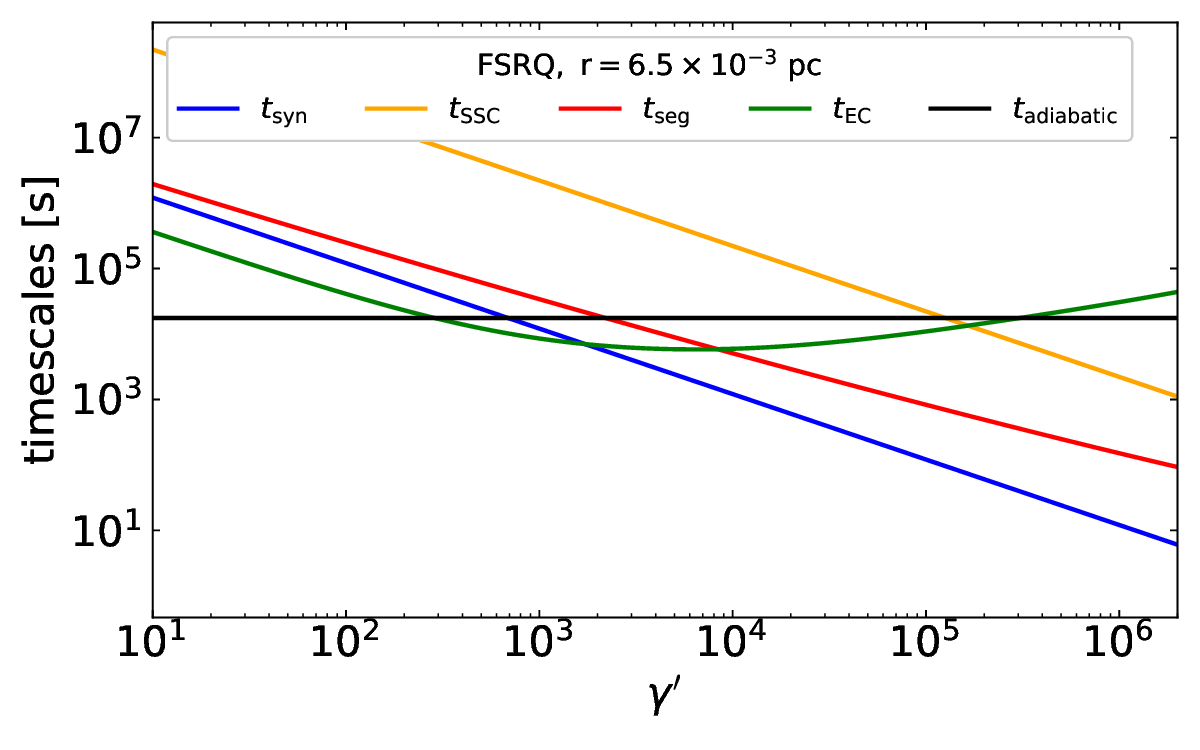}
}\hspace{-5mm}
\quad
\subfigure{
\includegraphics[width=0.9\columnwidth]{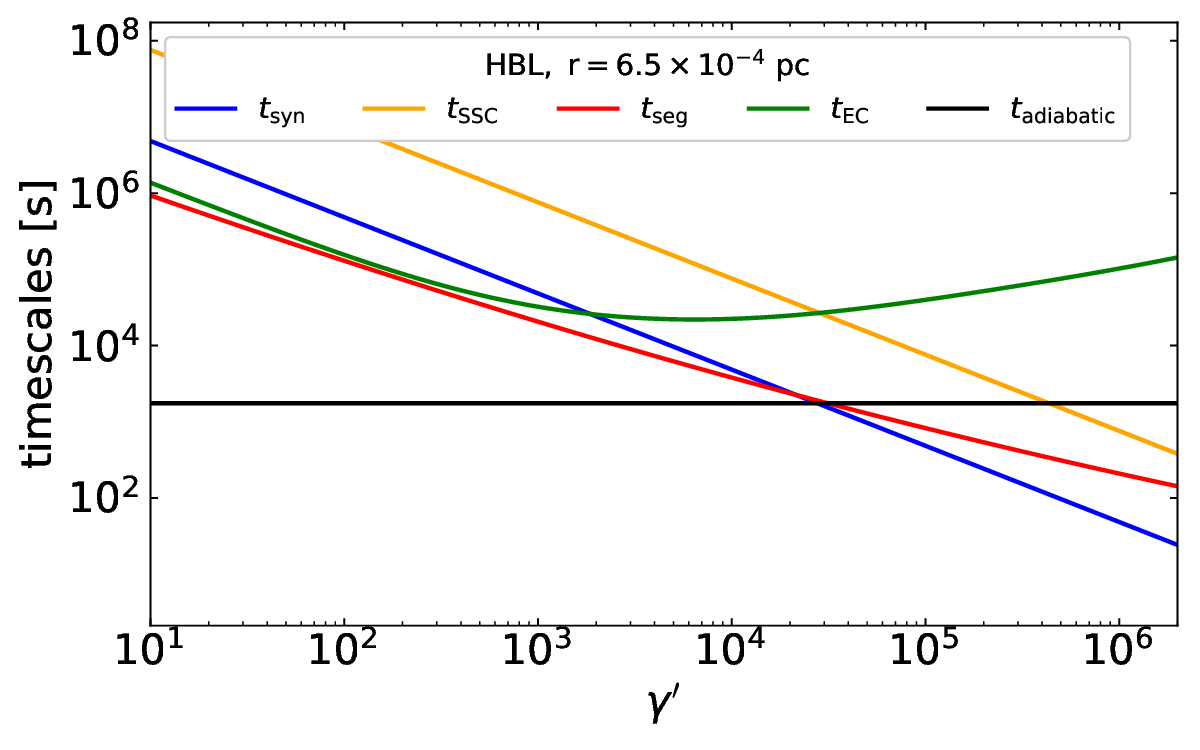}
}\hspace{-5mm}
\quad
\subfigure{
\includegraphics[width=0.9\columnwidth]{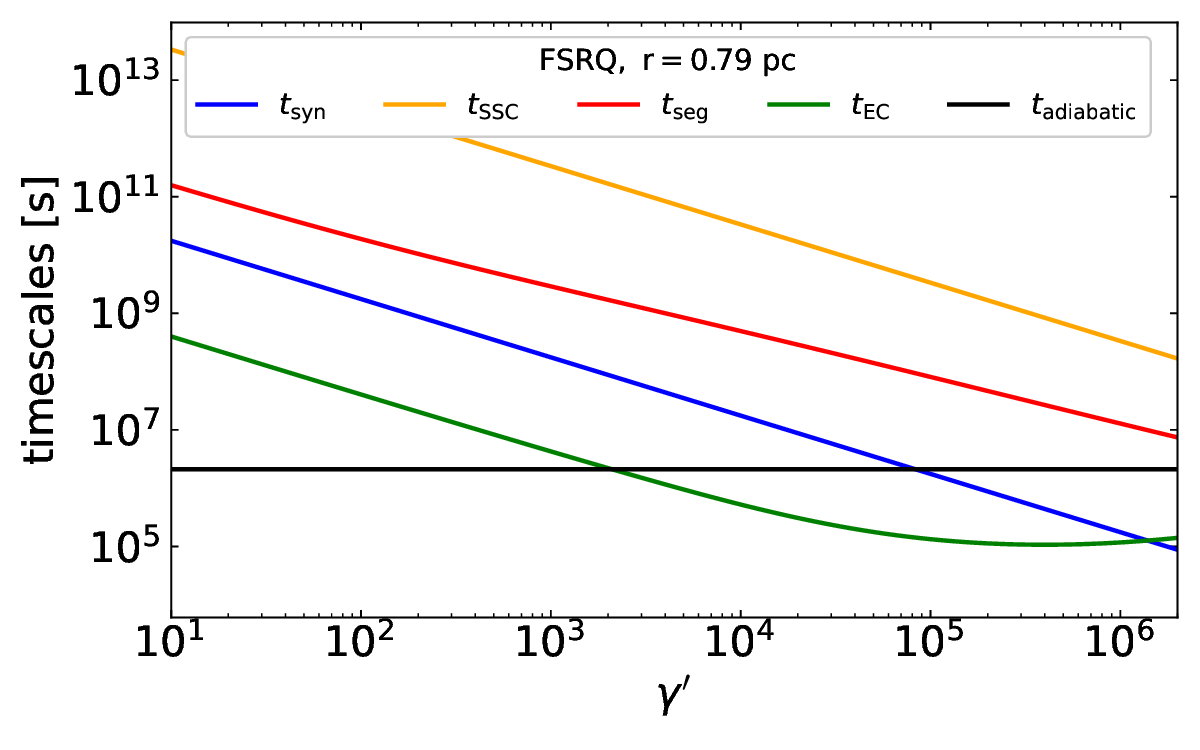}
}\hspace{-5mm}
\quad
\subfigure{
\includegraphics[width=0.9\columnwidth]{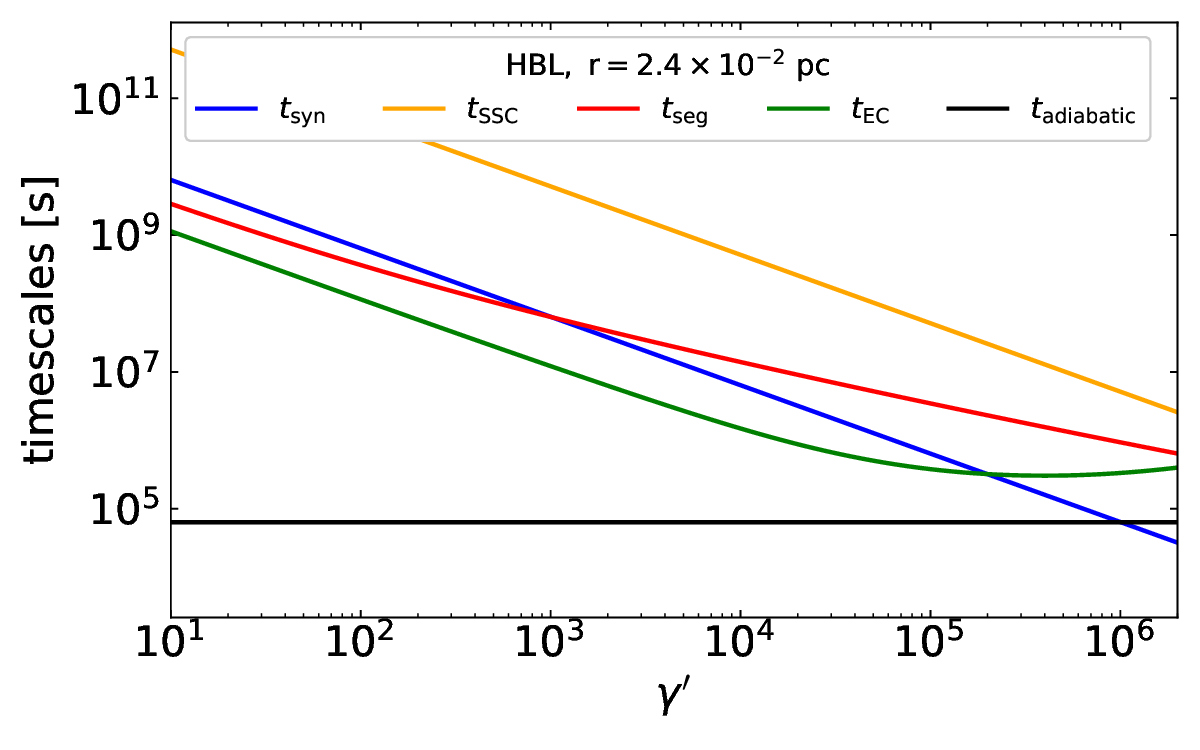}
}\hspace{-5mm}
\quad
\subfigure{
\includegraphics[width=0.9\columnwidth]{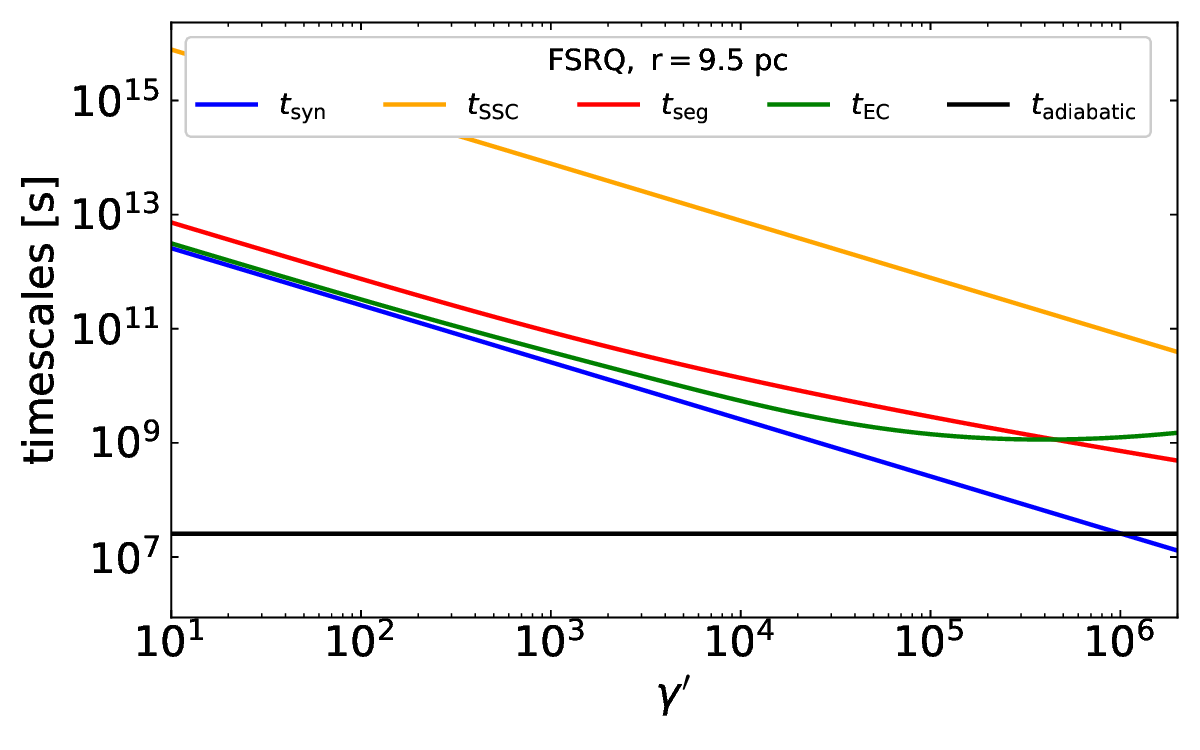}
}\hspace{-5mm}
\quad
\subfigure{
\includegraphics[width=0.9\columnwidth]{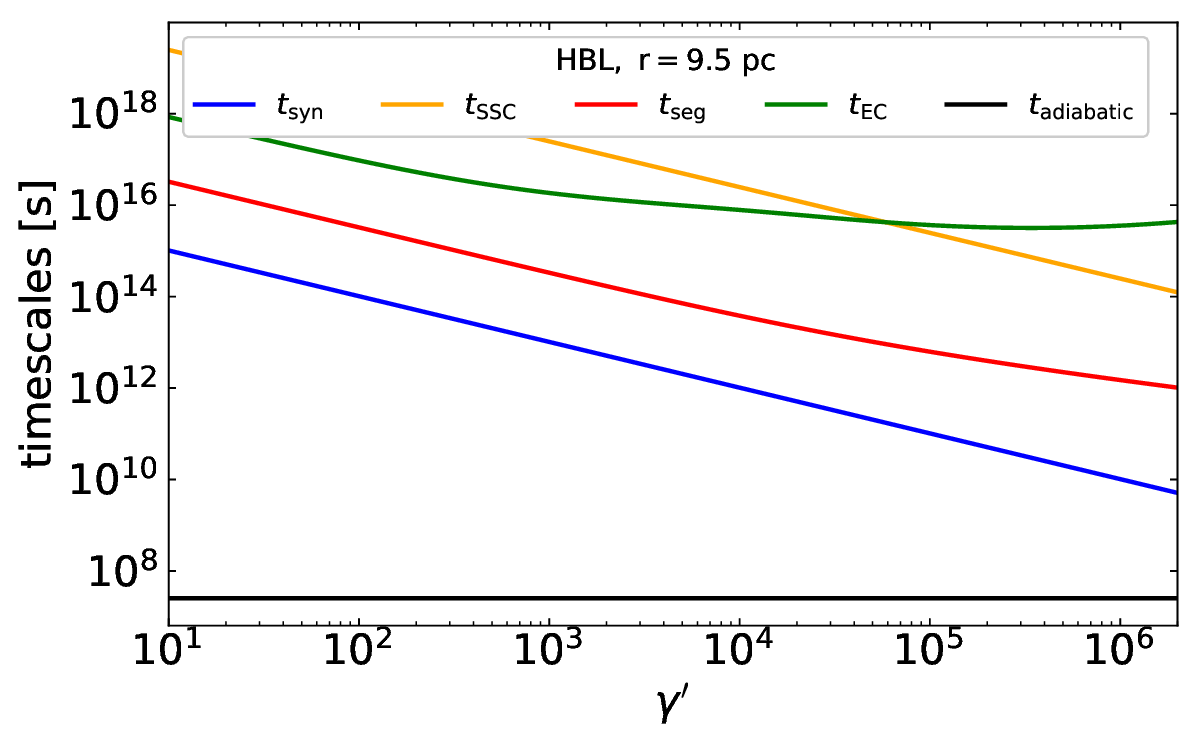}
}
\caption{Timescales of various cooling processes for electrons in the blobs at jet base, $r_{\rm DT}$, and beyond DT for FSRQ (three panels on the left) and HBL (three panels on the right), respectively. Timescale are measured in the comoving frame. The parameters are the same as those used in Table~\ref{parameters}. The meaning of all curves is explained in the inset legend.
\label{tcool}}
\end{figure*}

Since the duration of each dissipation event can last roughly $\sim R(r)/c$, we may obtain the number of blobs (or dissipation zones) in unit distance arond $r$ as $dN/dr=p(r) R(r)/c\propto r^{1-\alpha}$. For each blob, we assumed identical injection differential luminosity of electrons, i.e., $L'_{\rm inj}(\gamma')=Q_0'\gamma'^{2-s}$. The synchrotron radiation efficiency of electrons with Lorentz factor $\gamma'$ in a blob at a distance $r$ can be given by
\begin{equation}
    \begin{split}
    k_{\rm syn}(\gamma',r)&={t'}_{\rm syn}^{-1}(\gamma',r)\\
    &\times \left[{t'}_{\rm syn}^{-1}(\gamma', r)+{t'}_{\rm IC}^{-1}(\gamma', r)+{t'}_{\rm esc}^{-1}(r)+{t'}_{\rm ad}^{-1}(r) \right]^{-1}
    \end{split}
\end{equation}
For parameters shown in Table~\ref{parameters}, we found that the denominator of the above equation is dominated by the term $t'_{\rm ad}$  for low-energy electrons (see Fig.~\ref{tcool}). Therefore, except for the highest-end of the superimposed synchrotron spectrum, we may simply consider $k_{\rm syn}={t'}_{\rm syn}^{-1}/{t'}_{\rm ad}^{-1}$, i.e., the slow-cooling scenario (in other words, the total number of emitting electrons in each blob is $L'_{\rm inj}(\gamma')t_{\rm ad}$). Given ${t'}_{\rm syn}\propto {\gamma'}^{-1}B^{-2}$, $B\propto r^{-1}$, and ${t'}_{\rm ad}\propto r$ we have $k_{\rm syn}\propto \gamma' r^{-1} $. Hence, we have
\begin{equation}\label{eq:dLdr}
\begin{split}
    \frac{dL_{\rm syn}}{dr}&=k_{\rm syn}L'_{\rm inj}\frac{dN}{dr}\propto r^{-\alpha}{\gamma'}^{3-s}\\
    &\propto \nu^{(3-s)/2}r^{(3-s)/2-\alpha}\exp\left({-\frac{\nu}{\nu_{\rm max}(r_0)}\frac{r}{r_0}}\right),
\end{split}
\end{equation}
where we have assumed a monochromatic synchrotron spectrum for an electron with the relation $\nu\propto {\gamma'}^2B$. This is consistent with the flux profile $\overline{\nu F_\nu(E,r)}$, which is approximately proportional to $rdL/dr$,  at the frequency dominated by synchrotron radiation while not affected by the synchrotron self-absorption (SSA) process (e.g., in the infrared and the optical band), as shown in Fig.~\ref{fig:LCr}. We can see $\overline{\nu F_\nu(E,r)}\propto r^{-0.4}$ for $\alpha=1.9$ and $s=2$, and $\propto r^{0.2}$ for $\alpha=1.3$ and $s=2$

Let us now also take into account the SSA effect at low frequencies. Assuming injected electrons distributed homogeneously in each dissipation zone, we need to multiply a factor $(1-\exp(-\tau_{\rm ssa}))/\tau_{\rm ssa}$ to Eq.~(\ref{eq:dLdr}) with $\tau_{\rm ssa}$ being the SSA opacity. It is dependent on the frequency, the magnetic field strength and the column density of the electrons in the blob \citep{1986rpa..book.....R}, i.e.,
\begin{equation}\label{eq:ssa}
\begin{split}
\tau_{\rm ssa}&=\frac{(p+2)c^2}{8\pi\nu^2}\int \frac{Q_0t_{\rm ad}}{4\pi R(r)^2}{\gamma'}^{-s-1} P(\nu,\gamma')d\gamma'\\
&\propto r^{-1}B^{(s+2)/2}\nu^{-(s+4)/2}\propto r^{-(s+4)/2}\nu^{-(s+4)/2}
\end{split}
\end{equation}
For simplicity, we normalize the opacity at $\nu_{\rm max}(r_0)$ so that we have $\tau_{\rm ssa}=\tau_0(r\nu/r_0\nu_{\rm max}(r_0))^{-(s+4)/2}$.
Then, we can sum up the contribution at different distance $r$ to get the overall SED of the jet, i.e.,
\begin{equation}
\overline{\nu F_\nu}=\int \frac{dL_{\rm syn}}{dr}\frac{1-\exp(-\tau_{\rm ssa})}{\tau_{\rm ssa}}dr
\end{equation}
By defining $x\equiv r\nu/r_0\nu_{\rm max}(r_0)$ with $\nu_{\rm max}(r_0)=3\delta_D{\gamma'}_{\rm max}^2 e B_0/4\pi m_e c$, we can rewrite the above equation as
\begin{equation}
\overline{\nu F_\nu}\propto \nu^{\alpha-1}\int_{x_0}^{x_1} x^{\frac{7}{2}-\alpha}\exp(-x)\left[1-\exp{(-\tau_0 x^{-\frac{s+4}{2}})}\right]dx,
\end{equation}
where $x_0=\nu/\nu_{\rm max}(r_0)$ and $x_1=(L/r_0)x_0$.

For $\tau_0x^{-(s+4)/2}\gg 1$, the SSA effect is important and we have
\begin{equation}\label{eq:taugg1}
\begin{split}
\overline{\nu F_\nu}&\propto \nu^{\alpha-1}\int_{x_0}^{x_1} x^{\frac{7}{2}-\alpha}\exp(-x)dx\\
&\propto \nu^{\alpha-1}\left[\Gamma\left(\frac{9}{2}-\alpha, x_0\right)-\Gamma\left(\frac{9}{2}-\alpha, x_1\right)\right]
\end{split}
\end{equation}
where $\Gamma$ is the upper incomplete Gamma function. For $x_1<1$ or $\nu <(r_0/L)\nu_{\rm max}(r_0)$, the innermost blob dominates, and we have $\Gamma(\frac{9}{2}-\alpha, x_0)-\Gamma(\frac{9}{2}-\alpha, x_1)\propto \nu^{9/2-\alpha}$, leading to $\overline{\nu F_\nu} \propto \nu^{7/2}$.
For $(r_0/L)\nu_{\rm max}(r_0)<\nu<\nu_{\rm max}(r_0)$, $\Gamma(\frac{9}{2}-\alpha, x_0)-\Gamma(\frac{9}{2}-\alpha, x_1)\propto\,$Const., so that $\overline{\nu F_\nu}\propto \nu^{\alpha-1}$. This explains why we obtained $\overline{F_\nu} \propto \nu^{-0.1}$ for $\alpha=1.9$ and $\overline{F_\nu} \propto \nu^{0}$ for $\alpha=2$ in Section \ref{sec:sed}.

\begin{figure}
\centering
\subfigure{
\includegraphics[width=0.45\columnwidth]{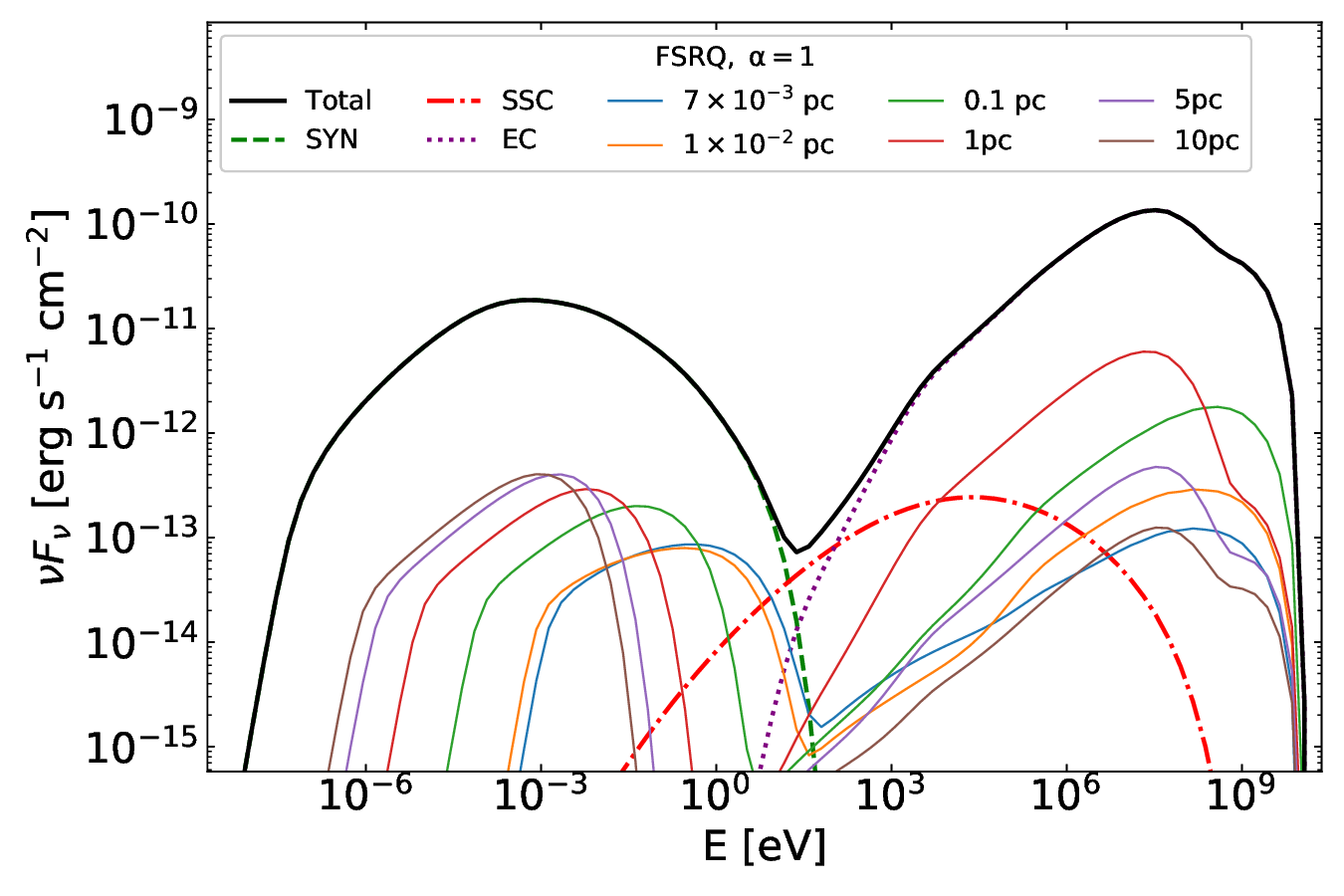}
}
\subfigure{
\includegraphics[width=0.45\columnwidth]{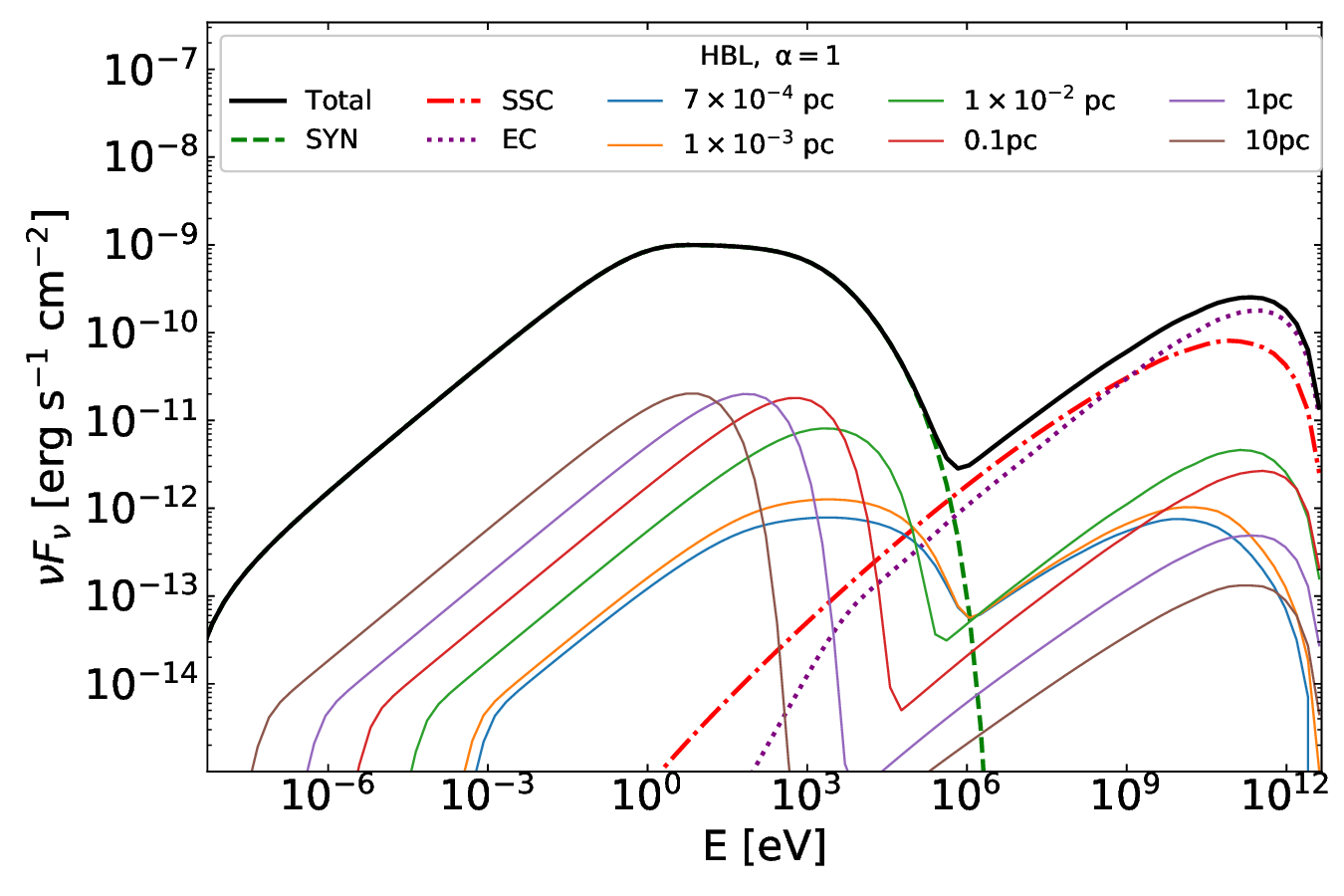}
}
\subfigure{
\includegraphics[width=0.45\columnwidth]{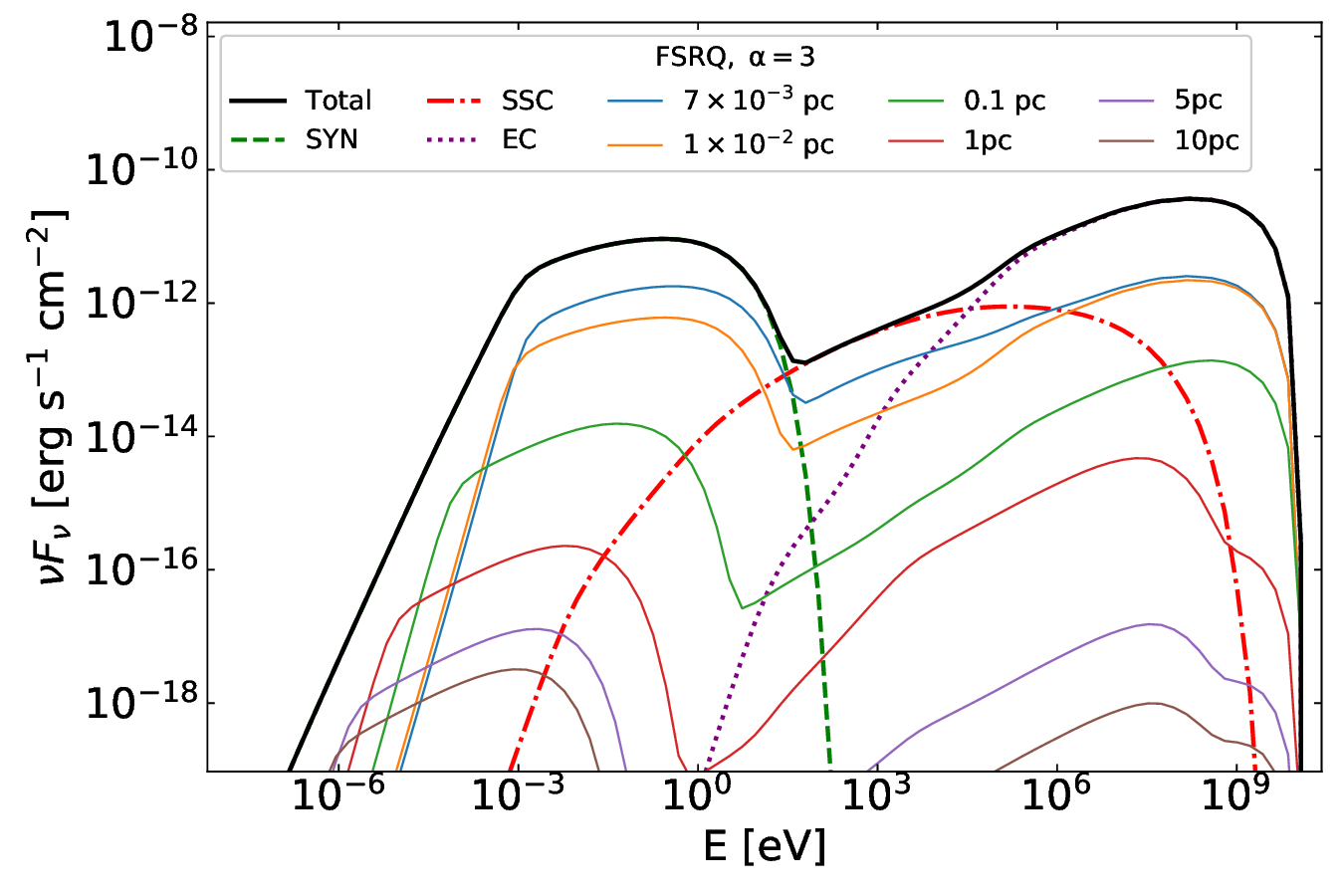}
}
\subfigure{
\includegraphics[width=0.45\columnwidth]{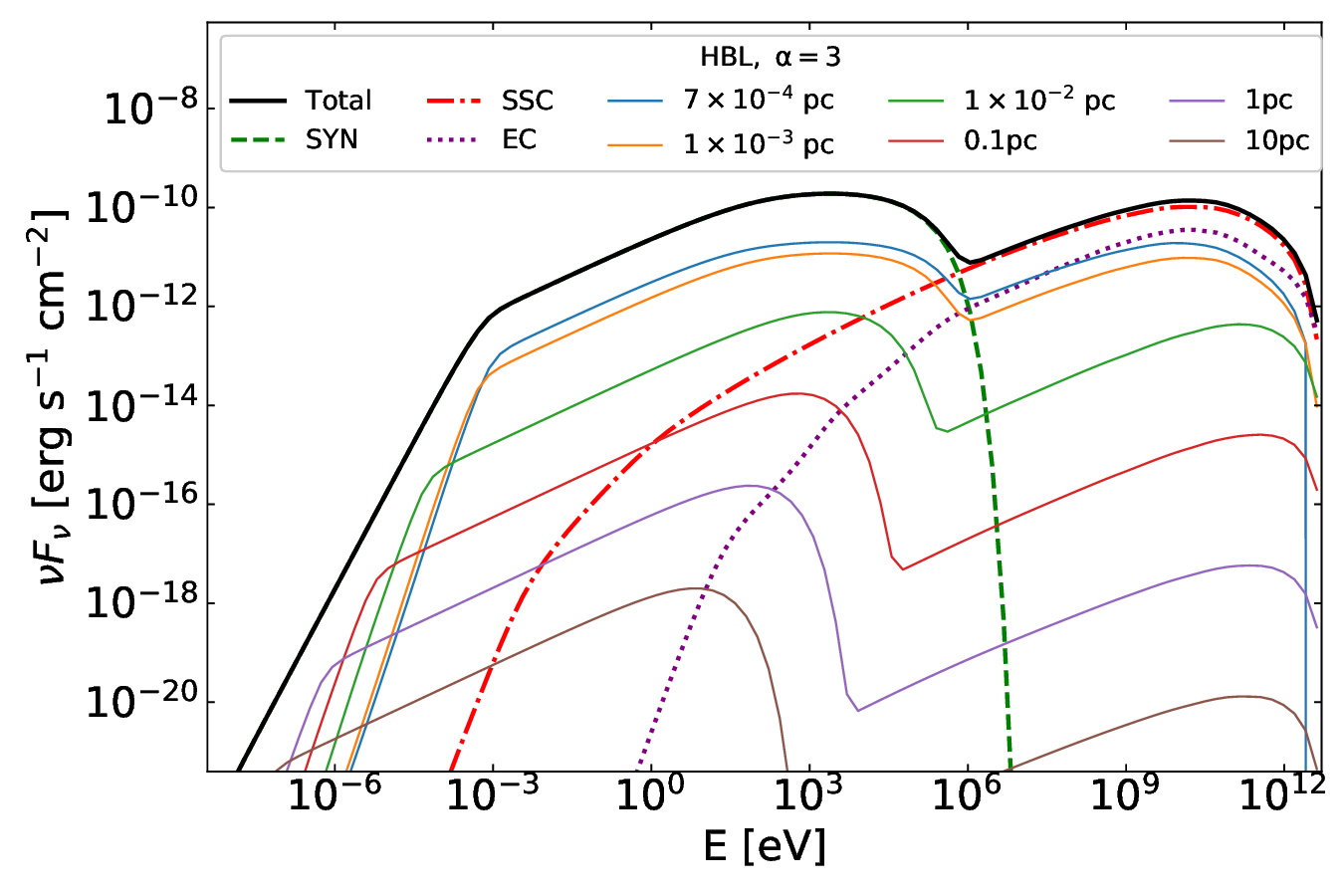}
}
\caption{Sum of the average of FSRQ (left panels) and HBL (right panels) when setting $\alpha=1$ and $\alpha=3$. The line styles have the same meaning as in Fig.~\ref{ave}.
\label{fig:sed_alpha}}
\end{figure}
\begin{figure}
\centering
\subfigure{
\includegraphics[width=0.45\columnwidth]{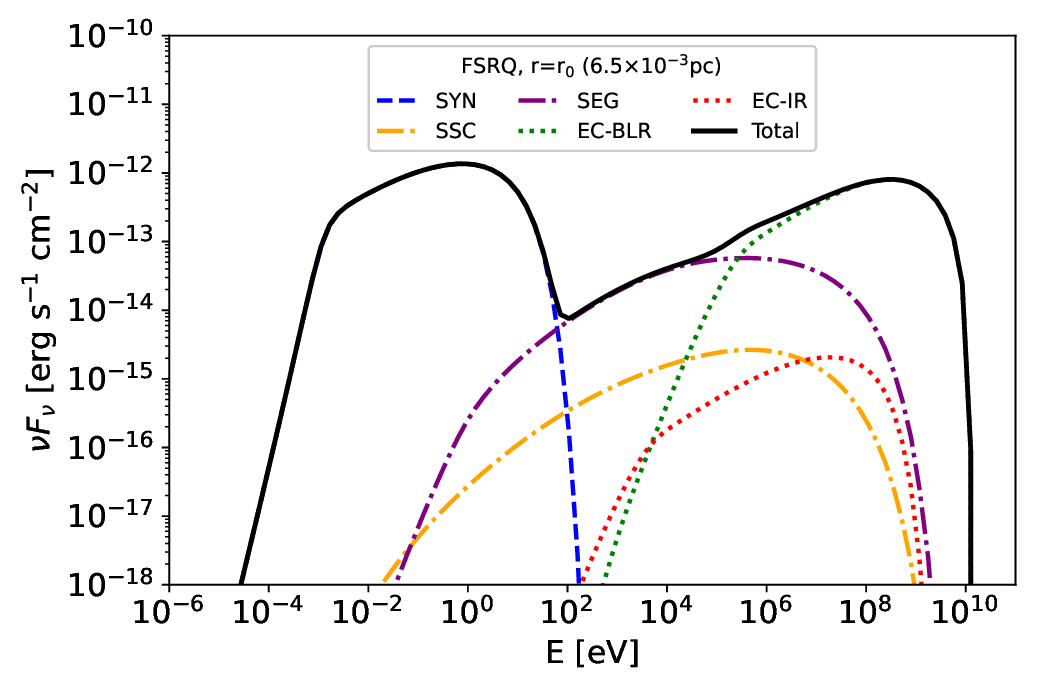}
}
\subfigure{
\includegraphics[width=0.45\columnwidth]{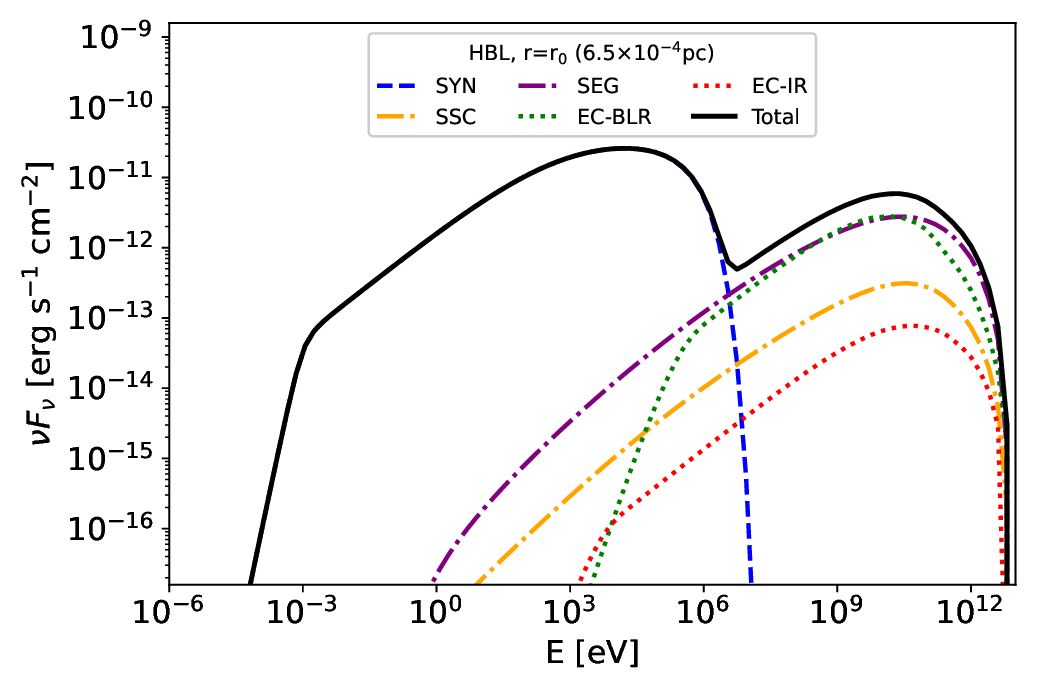}
} 
\subfigure{
\includegraphics[width=0.45\columnwidth]{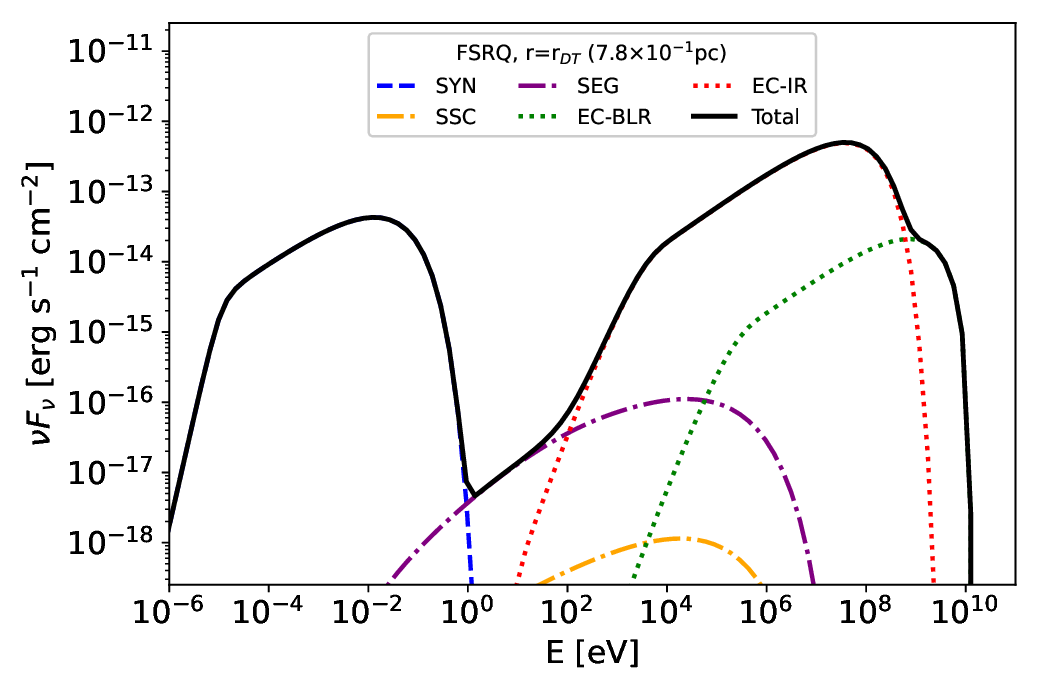}
}
\subfigure{
\includegraphics[width=0.45\columnwidth]{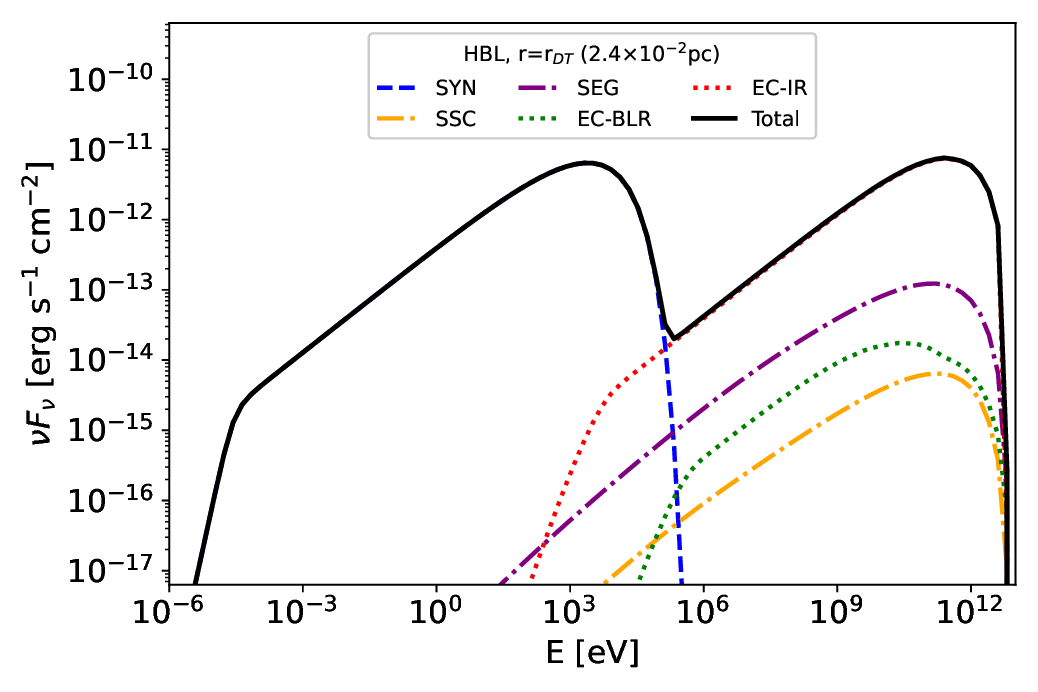}
}
\subfigure{
\includegraphics[width=0.45\columnwidth]{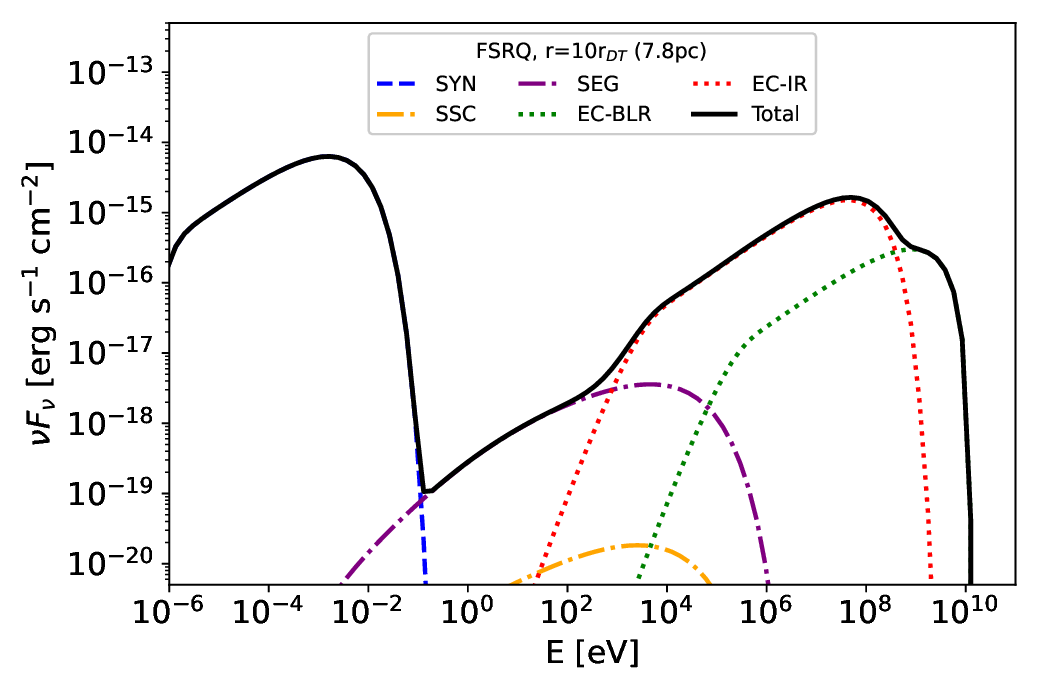}
}
\subfigure{
\includegraphics[width=0.45\columnwidth]{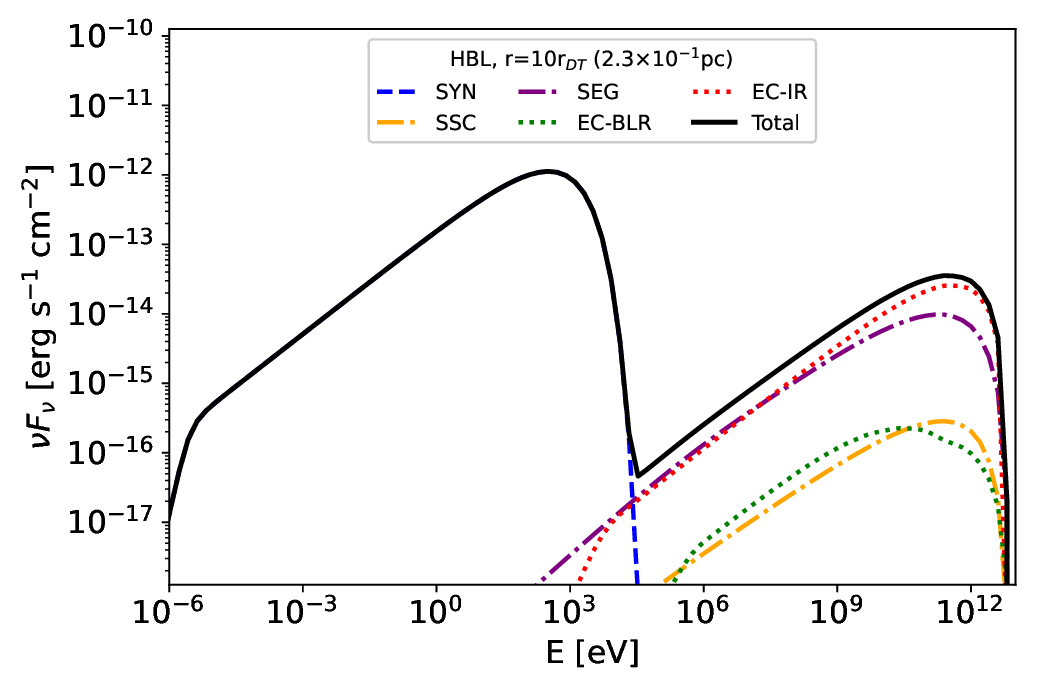}
}
\caption{{The average SEDs produced by blobs at jet base, $r_{\rm DT}$, and $10r_{\rm DT}$ for FSRQ (three panels on the left) and HBL (three panels on the right) respectively when setting $\alpha=2$. The dashed blue curve is the synchrotron emission. The high energy band is dominated by IC emission, and different line styles represent different target photons. The dot-dashed yellow and purple curves show the IC emission by scattering the synchrotron photons in the same blob and the other blobs in the same segment respectively. The dotted green and red curves represent the photons from BLR and DT. The thick black curve is the total emission. The parameters are the same as those shown in Table~\ref{parameters}.} 
\label{fig:sed_single_blob}}
\end{figure}

Lastly, we show the superimposed average SEDs of the FSRQ and the HBL when setting $\alpha=1$ and $\alpha=3$ in Fig.~\ref{fig:sed_alpha}, respectively, {while the SED of a blob at different distance is shown in Fig.~\ref{fig:sed_single_blob} for reference.} When setting $\alpha=1$, the number of blobs that far from the jet base (such as within DT) will be more than that in the $\alpha=2$ case. Since the energy densities $u'_{\rm BLR}$ and  $u'_{\rm DT}$ within the characteristic distances of BLR and DT remains unchanged while the magnetic field decreases, more blobs will contribute considerable EC radiation, making sum of the average flux of EC radiation higher. On the contrary, when setting $\alpha=3$, sum of the average flux of EC radiation would be lower. It implies that a precise measurement of the blazar spectrum can be used to study the dissipation probability distribution in the jets.

\onecolumn

\section{Influence of the electron injection timescale}\label{appB}
\begin{figure}
    \centering
    \includegraphics[width=0.8\textwidth]{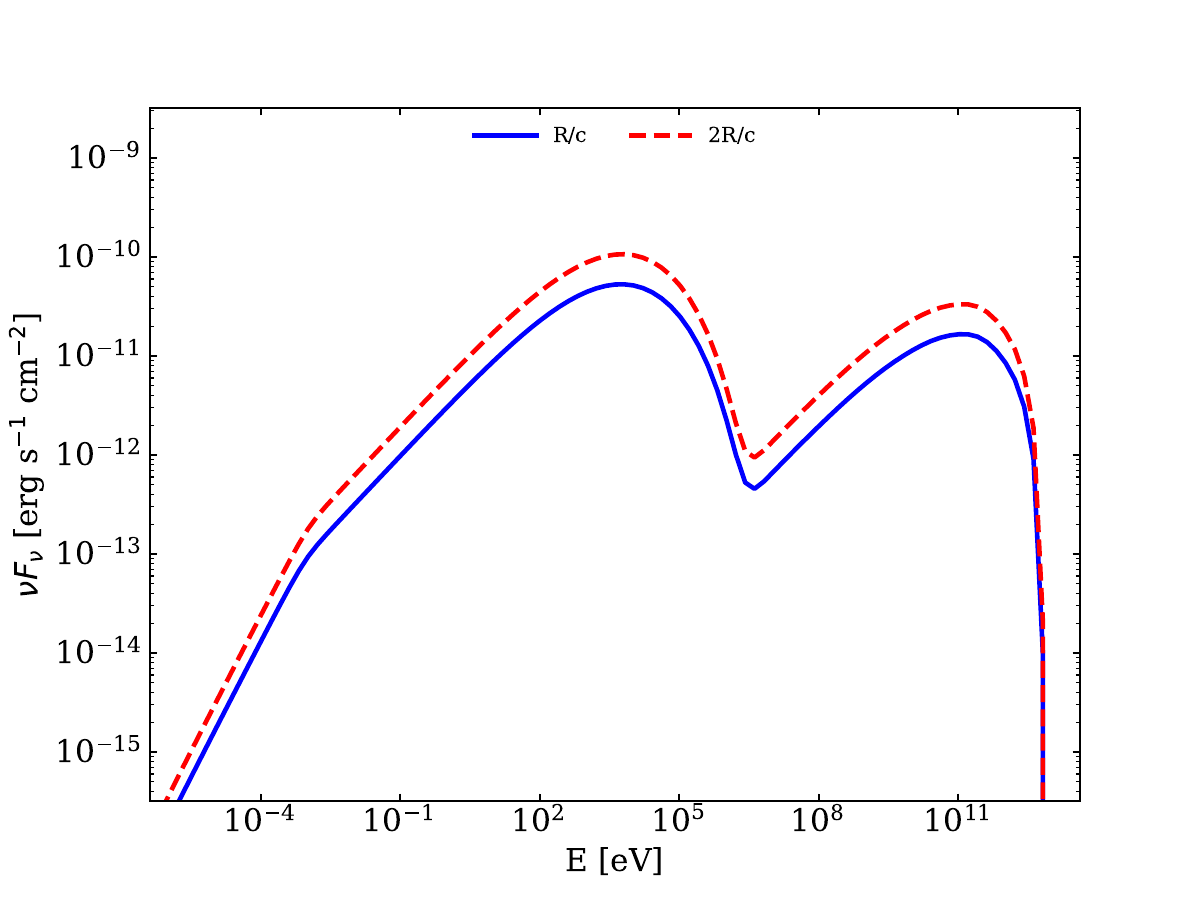}
    \caption{Comparison of the SED generated with $t_{\rm inj}'=R/c$ (solid blue curve) and $t_{\rm inj}'=2R/c$ (dashed red curve).}
    \label{fig:sed_tinjcompare}
\end{figure}

\begin{figure}
    \centering
    \includegraphics[width=1\textwidth]{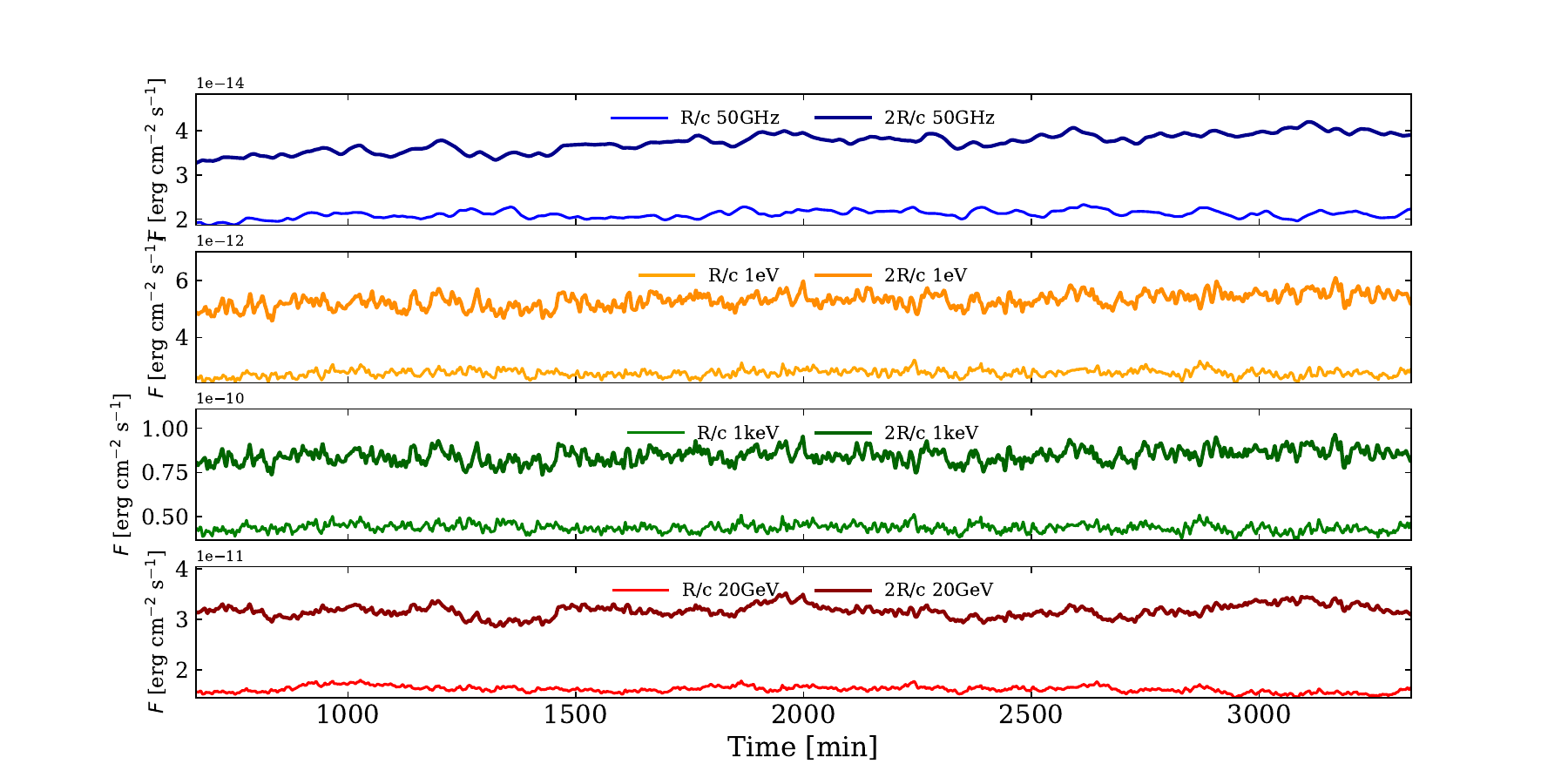}
    \caption{Comparison of the multiwavelength LCs generated with $t_{\rm inj}'=R/c$ (thin curves) and $t_{\rm inj}'=2R/c$ (thick curves).}
    \label{fig:LC_tinjcompare}
\end{figure}

{We compare the SED and multiwavelength LCs of the blazar with $t_{\rm inj}'=2R'/c$ to the benchmark case with $t_{\rm inj}'=R'/c$ in Fig.~\ref{fig:sed_tinjcompare} and \ref{fig:LC_tinjcompare}. Other parameters are all the same as the benchmark parameters. The flux in the case of $t_{\rm inj}'=2R'/c$ is about two times higher than that in the case of $t_{\rm inj}'=R'/c$, because the amount of injected electrons in the former case is twice that in the latter case. We also see that the LC in the former case is less variable than that in the latter case, which is consistent with expectation. }

\bsp	
\label{lastpage}
\end{document}